\documentclass[11pt]{physmath_art}

\usepackage{cite} 
\usepackage{url}  
\usepackage{ifthen}  
\usepackage{multicol}   
\usepackage[utf8]{inputenc} 
\usepackage{graphicx}
\usepackage{dcolumn}
\usepackage{bm}
\usepackage{epsfig}

\def\Dstar{\ifmmode {{ D}^*} \else {${ D}^*$} \fi}
\def\Dzero{\ifmmode {{ D}^0} \else {${ D}^0$} \fi}
\def\md0c{M_{D^0}^{cand}}
\def\etal{{\it et al.}}

\def\bbar{$\bar {B^0}$}

\def\dsp{$D^{*+}$}

\def\vcb{\mbox{$|V_{cb}|$}}

\def\fvcb{${\cal F}(1)|V_{ cb}|$}
\def\btods{$\overline{B}^0\to D^{*+}\ell^-{\bar \nu}_\ell$}

\def\taup{\tau^+}
\def\mum{\mu^-}
\def\mup{\mu ^+}
\def\Bs{B_s}
\def\bep{\mbox{\boldmath $\epsilon$}}

\def\ra{\rightarrow}

\def\be{\begin{equation}}
\def\ee{\end{equation}}
\def\bea{\begin{eqnarray}}
\def\eea{\end{eqnarray}}

\def\etal{{\it et al}}

\newcommand{\Dz}{D^0}

\newcommand{\pim}{\pi ^-}

\newcommand{\Bz}{B^0}
\newcommand{\Bp}{B^+}

\newcommand{\Ds}{D_s}
\newcommand{\epm}{e^+e^-}

\newcommand{\vub}{|V_{ub}|}

\newcommand{\qsq}{q^2}
\newcommand{\nub}{\bar{\nu}_\ell}
\newcommand{\Ufs}{\Upsilon (4S)}
\newcommand{\mbstar}{M_{B^\star}}
\newcommand{\bstarpr}{B^{\star \prime}}
\newcommand{\aerr}[4]   {\mbox{${{#1}^{+ #2}_{- #3}\pm #4}$}}
\newcommand{\berr}[4]   {\mbox{${{#1}\pm #2^{+ #3}_{- #4}}$}}
\newcommand{\cerr}[3]   {\mbox{${{#1}^{+ #2}_{- #3}}$}}

\newcommand{\err}[3]   {\mbox{${{#1}\pm{#2}\pm{#3}}$}}

\def\babar{\mbox{\slshape B\kern-0.1em{\smaller A}\kern-0.1em B\kern-0.1em{\smaller A\kern-0.2em R}}}
\def\etal{{\it et al.}}

\def\lim{\mathop{\rm lim}}

\newboolean{publ}

\newenvironment{physmath}{\begin{raggedright}\baselineskip16pt\setboolean{publ}{false}}{\end{raggedright}}

\begin{document}
\begin{physmath}


\title{B Meson Decays}

\author{Marina Artuso$^{1}$%
       \email{Marina Artuso - artuso@physics.syr.edu}%
       \and
         Elisabetta Barberio$^2$%
         \email{Elisabetta Barberio - barberio@unimelb.edu.au}
       and
         Sheldon Stone\correspondingauthor$^1$%
         \email{Sheldon Stone\correspondingauthor - stone@physics.syr.edu}%
      }


\address{%
    \iid(1)Department of Physics, Syracuse University,
  Syracuse, N. Y. 13244, USA\\
	\iid(2)School of Physics, University of Melbourne,
  Victoria 3010, Australia
}

\maketitle

\begin{abstract}
We discuss the most important Physics thus far extracted from studies
of $B$ meson decays. Measurements of the four CP violating angles accessible in $B$ decay are reviewed as well
as direct CP violation. A detailed discussion of the  measurements of the CKM elements
$V_{cb}$ and $V_{ub}$ from semileptonic decays is given, and the differences between resulting values using
inclusive decays versus exclusive decays is discussed. Measurements of ``rare" decays are also reviewed.   We
point out where CP violating and rare decays could lead to observations of physics beyond that of the Standard Model
in future experiments. If such physics is found  by directly observation of new particles, e.g. in LHC experiments, $B$ decays can play a decisive role in interpreting the nature of these particles.

\end{abstract}

\newpage
\tableofcontents
\section{Introduction}
The forces of nature generally reveal their properties by how they act on matter. On
the most fundamental material scale that we are aware of matter is formed from
fermions. These take two forms, leptonic matter and quark matter. The former do not
have any strong interactions, that is they lack a property called ``color charge",
which allow quarks to bind together either mesonically or
baryonically. New and therefore as yet unknown forces could effect both leptons and
quarks. Here, we concentrate on how such forces effect quarks, especially the $b$
quark.

Light matter consists mostly of $u$, $d$ and $s$ quarks. In 1963 Cabibbo showed that
weak interactions of mesons and baryons containing $s$ quarks were suppressed with
respect to those without $s$ quarks by an amount $\tan\theta_C$, where the ``Cabibbo"
angle $\theta_C$ must be determined experimentally \cite{Cabibbo}. The $s$ was
further shown to have an important and at that time a mystifying role, by the
discovery of CP violation in $K_L^0$ decays in 1964 \cite{CP,Bigi-Sanda}. When the
$c$ quark was discovered in Nov. 1974 \cite{TR}(though its existence was speculated earlier \cite{Glashow}), it became clear that $\theta_C$ was
the mixing angle between two quark doublets, $(c,~s)$ and $(u,~d)$. However,
it is not possible to generate a CP violating phase with only two quark
doublets.

This was recognized even before the discovery of the charm quark by Kobayashi and
Maskawa, who postulated the existence of yet another quark doublet $(b,~t)$ \cite{KM}, in work for which they were awarded the Nobel Prize in 2008.  While the $t$ quark is the heaviest, having a mass of 173 GeV, they are difficult to produce and decay before they can form a hadron, thus excluding many otherwise possible studies.  Much interesting work has been done with the $s$ and $c$ quarks, but in this article we discuss the physics of the $b$ quark, which turns out to be the most interesting of the six quarks to study.

\subsection{How {\it\textbf B}'s Fit Into the Standard Model}
First we will discuss how particles formed with $b$-quarks fit into current paradigm of
particle physics, the ``Standard Model" (SM) \cite{SM}. The SM has at its basis
the gauge group SU(3)xSU(2)xU(1). The first term corresponds to the strong interaction
and SU(3) describes the octet of colored gluons which are the strong force carriers of
quantum chromodynamics. SU(2)xU(1) describes the weak interaction and is the product
of weak isospin and hypercharge. We speak of the fundamental objects being spin-1/2
quarks and leptons and the force carriers generally spin-1 objects. The spin-0 Higgs
boson, yet to be discovered, is necessary for generating mass \cite{Higgs}.

Particles containing $b$ quarks can be $B^0$, $B^-$, $B_s$, or $B_c$ mesons,
depending on whether the light anti-quark that it pairs with is $\bar{d}$, $\bar{u}$,
$\bar{s}$, or $\bar{c}$, or a baryon containing two other quarks. Mesons containing
$b$ and $\bar{b}$ quarks are also interesting especially for studies of quantum
chromodynamics, but will not be discussed further in this article, as we will
concentrate on weak decays and discuss strong interactions as an important and
necessary complication that must be understood in many cases to extract information on
fundamental $b$ quark couplings.

The quarks come in three repetitions called generations, as do the leptons. The first
generation is $d~u$, the second $s~c$ and the third $b~t$. In the second and third
generations the charge +2/3 quark is heavier than the charge -1/3; the first
generation has two very light quarks on the order of a few MeV with the $d$ thought to
be a bit heavier.\footnote{Isospin invariance is related to the equality of $u$ and
$d$ quark masses. The PDG \cite{PDG} gives the $u$ quark mass between 1.5-3.3 MeV and
the $d$ mass between 3.5-6.0 MeV, where the large range indicates the
considerable uncertainties.}  Decays usually proceed within generations, so the $c$
decays predominantly to the $s$ quark via the quark level process $c\to W^+s$, though
some decays do go to the first generation as $c\to W^+d$. The ratio of these
amplitudes approximate the Cabibbo angle discussed earlier.

The mixing matrix proposed by Kobayshi and Maskawa \cite{KM} parameterizes the mixing
between the mass eigenstates and weak eigenstates as couplings between the charge +2/3
and -1/3 quarks. We use here the Wolfenstein approximation \cite{Wolf} good to order
$\lambda^4$: 
\begin{eqnarray}\label{eq:CKM}
V_{CKM}&=& \left(\begin{array}{ccc} V_{ud} & V_{us}  & V_{ub} \\ V_{cd} & V_{cs}  &
V_{cb} \\ V_{td} & V_{ts}  & V_{tb}
\end{array}\right) \\\nonumber
&=& \left(\begin{array}{ccc} 1-\lambda^2/2 &  \lambda & A\lambda^3(\rho-i\eta) \\
-\lambda &  1-\lambda^2/2-\lambda^4(1+4A^2)/8 & A\lambda^2 \\ A\lambda^3(1-\rho-i\eta)
& -A\lambda^2+A\lambda^4(1/2-(\rho+i\eta))& 1-A^2\lambda^4/2
\end{array}\right).
\end{eqnarray}

In the Standard Model $A$, $\lambda$, $\rho$ and $\eta$ are fundamental constants of
nature like $G$, or $\alpha_{EM}$; $\eta$ multiplies the imaginary $i$ and is
responsible for all Standard Model CP violation. We know $\lambda$=0.226, $A \sim$0.8
and we have constraints on $\rho$ and $\eta$. Often the variables $\overline{\rho}$
and $\overline{\eta}$ are used where
\begin{equation}
\overline{\rho}+i\overline{\eta}=(\rho+i\eta)(1-\lambda^2/2)
=-\frac{V_{ud}V_{ub}^*}{V_{cd}V_{cb}^*},
\end{equation}
where the definition in terms of the CKM matrix elements is correct to all orders in
$\lambda$ \cite{CKMfitter}.

Applying unitarity constraints allows us to construct the six independent triangles
shown in Figure~\ref{six_tri}. Another basis for the CKM matrix are four angles labeled
as $\chi$, $\chi'$ and any two of $\alpha$, $\beta$ and $\gamma$ since $\alpha +\beta
+\gamma =\pi$ \cite{akl}. (These angles are also shown in Figure~\ref{six_tri}.) CP
violation measurements make use of these angles.\footnote{The Belle collaboration
defines $\phi_2\equiv\alpha$, $\phi_1\equiv\beta$, and $\phi_3\equiv\gamma$.}

\begin{figure}
\centerline{\epsfig{figure= 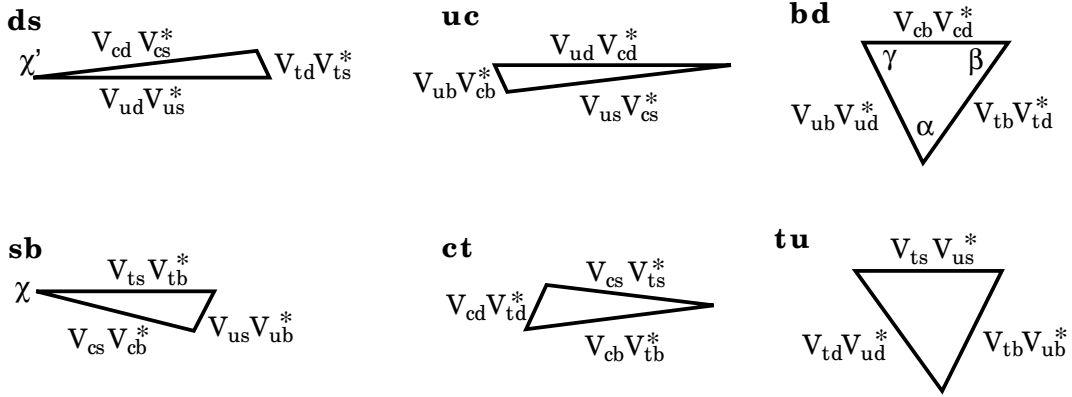,height=2.1in}}
\caption{The 6 CKM triangles resulting from applying unitarity
constraints to the indicated row and column. The CP violating angles are also shown.}
\label{six_tri}       
\end{figure}

$B$ meson decays can occur through various processes. Some decay diagrams with
intermediate charged vector bosons are shown in Figure~\ref{Bdiagrams2}. The simple
spectator diagram shown Figure~\ref{Bdiagrams2}(a) has by far the largest rate.
Semileptonic decays proceed through this diagram, and allow us to measure the CKM
couplings $V_{cb}$ and $V_{ub}$ by considering only the hadronic uncertainties due
to the spectator quark. The color suppressed diagram Figure~\ref{Bdiagrams2}(b) exists
only for hadronic decays. It can occur only when the colors of the quarks from the
virtual $W^-$ decay match those of the initial $B$ meson. Since this happens only 1/3
of the time in amplitude, the rate is down by almost an order of magnitude from the
spectator decays. The annihilation Figure~\ref{Bdiagrams2}(c) describes the important
decay $B^-\to \tau^-\overline{\nu}$ and will be discussed in detail later. The $W$
exchange Figure~\ref{Bdiagrams2}(d) diagram is small. The box diagram
Figure~\ref{Bdiagrams2}(e) is the source of mixing in the $B$ system. An analogous
diagram exists also for the $B_s$ meson. Finally the Penguin diagram
Figure~\ref{Bdiagrams2}(f) is an example of one of several loop diagrams leading to ``rare
decays" that will be discussed in detail in a subsequent section.

\begin{figure}
\centerline{\epsfig{figure=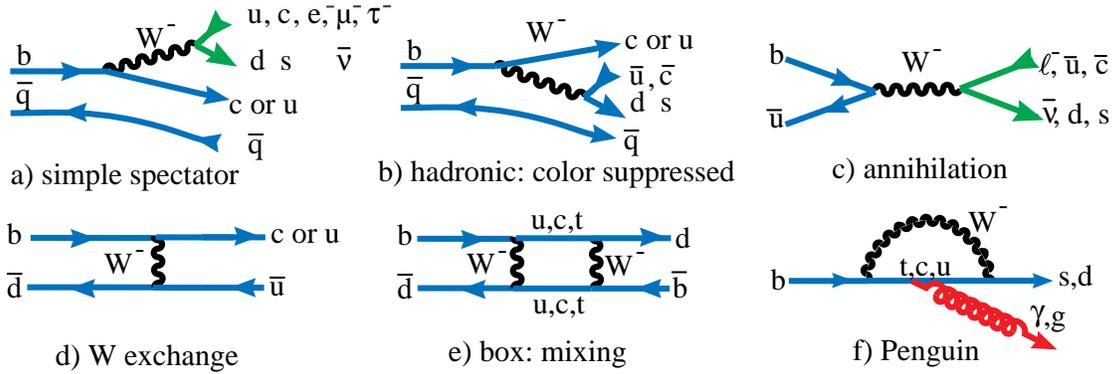,height=2.0in}}
\caption{Some $B$ decay diagrams.} \label{Bdiagrams2}
\end{figure}

\subsubsection{Dark Matter}

``Dark Matter" was first shown to exist by Zwicky studying rotation curves of galaxies
\cite{Zwicky}. The motion could only be explained if there was massive cloud of matter
that was not luminous. We still do not know what composes this dark matter, though hopes
are it will be discovered at the LHC. An even more mysterious phenomena called ``Dark Energy" may also have a
connection to particle physics experiments \cite{Trodden}, perhaps via ``Extra
Dimensions" \cite{ExtraD}.

\subsubsection{Baryogenesis}

When the Universe began with the Big Bang, there was an equal amount of matter and
antimatter. Now we have mostly matter. How did it happen? Sakharov gave three
necessary conditions: Baryon (${\cal B}$) number violation, departure from thermal equilibrium,
and C and CP violation \cite{Sakh}. (The operation of Charge Conjugation (C) takes
particle to anti-particle and Parity (P) takes a vector $\overrightarrow{r}$ to
$-\overrightarrow{r}$.)

These criteria are all satisfied by the Standard Model. ${\cal B}$ is violated in Electroweak
theory at high temperature, though baryon minus lepton number is conserved; in
addition we need quantum tunneling, which is powerfully suppressed at the low
temperatures that we now have. Non-thermal equilibrium is provided by the electroweak
phase transition. C and CP are violated by weak interactions. However the violation is
too small. The ratio of the number of baryons to the entropy in the observed part of the Universe
needs to be $\sim 5\times 10^{-11}$, while the SM provides many orders of magnitude less.
Therefore, there must be new physics \cite{Gavela}.

\subsubsection{The Hierarchy Problem}

Our worry is why the Planck scale at $\sim 10^{19}$ GeV is so much higher than the
scale at which we expect to find the Higgs Boson, $\sim$100 GeV. As Lisa Randall said
\cite{Lisa} ``The gist of it is that the universe seems to have two entirely different
mass scales, and we don't understand why they are so different. There's what's called
the Planck scale, which is associated with gravitational interactions. It's a huge
mass scale, but because gravitational forces are proportional to one over the mass
squared, that means gravity is a very weak interaction. In units of GeV, which is how
we measure masses, the Planck scale is 10$^{19}$ GeV. Then there's the
electroweak scale, which sets the masses for the W and Z bosons. These are particles
that are similar to the photons of electromagnetism and which we have observed and
studied well. They have a mass of about 100 GeV. So the hierarchy problem, in its
simplest manifestation, is how can you have these particles be so light when the other
scale is so big."  We expect the explanation lies in physics beyond the Standard Model
\cite{Lykken}.

\subsection{{\it\textbf B} Decays as Probes for New Physics}

When we make measurements on $B$ decays we observe the contributions of SM processes as
well as any other processes that may be due to physics beyond the SM or New Physics
(NP). Other diagrams would appear adding to those in Figure~\ref{Bdiagrams2} with new
intermediate particles. Thus, when it is declared by those who say that there isn't
any evidence of NP in $B$ decays, we have to be very careful that we have not absorbed
such new evidence into what we declare to be SM physics. There are several approaches that can be followed.

One approach is to simply predict the decay rate of a single process in the SM with
known couplings and compare to the measurements. The classical case here is $b\to
s\gamma$ and we will discuss this and other specific examples later. Another approach
is make different measurements of the CKM parameters in different ways and see if they
agree. This is normally done by measuring both angles and sides of the CKM triangle,
but other quantities can also be used. This is the approach used by the CKM fitter
\cite{CKMfitter} and UT fit groups \cite{UTfit}. In yet a third approach, the exact
same quantity can be measured in several ways, even if cannot be predicted in the SM.
An example here is measuring the CP violating angle $\beta$ using $B^0\to J/\psi K_S$
decays that proceed through the diagram in Figure~\ref{Bdiagrams2}(b), at least in the
SM, and another process that uses the ``Penguin" diagram in Figure~\ref{Bdiagrams2}(f), e.g. $B^0\to \phi K_S$.

The punch line is that if new, more massive particles exist in a mass
range accessible to the LHC then they MUST contribute to rare and CP violating
$B$ decays! Even if measurements are precise enough only to limit the size of these
effects, the properties of these new particles will be much better understood. This is the
raison d'$\hat{\rm e}$tre for the further study of $B$ decays.

\section{Measurements of Mixing \& CP Violation}

\subsection{Neutral {\it\textbf B} Meson Mixing}
Neutral $B$ mesons can transform into their anti-particles before they decay. The
diagrams for this process are shown in Figure~\ref{bmix} for the $B_d$. There is a
similar diagram for the $B_s$. Although $u$, $c$ and $t$ quark exchanges are all
shown, the $t$ quark plays a dominant role mainly due to its mass, as the amplitude of
this process is proportional to the mass of the exchanged fermion.
\begin{figure}[thb]
\centerline{\epsfig{figure=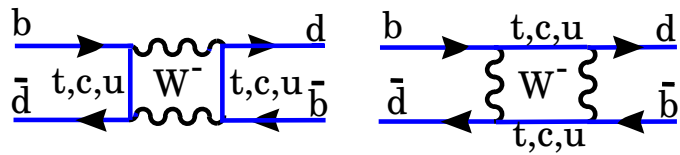,width=3.5in}} 
\caption{\label{bmix}The two diagrams for $B_d$ mixing.}
\end{figure}

Under the weak interactions the eigenstates of flavor degenerate in pure QCD can mix.
Let the quantum mechanical basis vectors be $\{|1\rangle,|2\rangle\} \equiv
\{|B^0\rangle,|\overline{B}^0\rangle\}$. The Hamiltonian is then

\begin{equation}
{\cal H}=M-{i\over 2}\Gamma=\left(\begin{array}{cc} M & M_{12}\\ M_{12}^* & M
\end{array}
\right) -{i\over 2}\left(\begin{array}{cc} \Gamma & \Gamma_{12} \\ \Gamma_{12}^* &
\Gamma
\end{array}\right)  .
\end{equation}

The Schr\"odinger equation is
\begin{equation}
i\frac{d}{dt}\left(\begin{array}{c} \mid B^0(t)\rangle \\ \mid
\overline{B}^0(t)\rangle
\end{array}\right)
={\cal H}\left(\begin{array}{c} \mid B^0(t)\rangle \\ \mid \overline{B}^0(t)\rangle
\end{array}\right)
\end{equation}

Diagonalizing we have
\begin{eqnarray}
\Delta m &=& m_{B_H}-m_{B_L}=2\left|M_{12}\right|\\ \Delta\Gamma
&=&\Gamma_L-\Gamma_H=2\left|\Gamma_{12}\right|\cos\phi ~
\end{eqnarray}
where $H$ refers to the heavier and $L$ the lighter of the two weak eigenstates, and
$\phi = {\rm arg}\left(-M_{12}/\Gamma_{12}\right)$. We expect that $\Delta\Gamma$ is
very small for $B^0$ mesons but should be significant for $B_s$ mesons.

$B_d$ mixing was first discovered by the ARGUS experiment \cite{Albrecht1983} (There
was a previous measurement by UA1 indicating mixing for a mixture of $B_d^0$ and
$B_s^0$ \cite{Albajar1987}). At the time it was quite a surprise, since the top-quark mass, $m_t$, was
thought to be in the 30 GeV range. Since $b$-flavored hadrons are produced in pairs,
it is possible to observe a mixed event by having both $B$'s decay semileptonically.
Thus the number of events where both $B$s decay into leptons of the same sign is
indicative of mixing. We can define  $R$ as the ratio of events with same-sign leptons
to the sum of same-sign plus opposite-sign dilepton events. This is related to the
mixing probability.  The OPAL data for $R$ are shown in Figure~\ref{opal_mix}
\cite{Akers1995}.

\begin{figure}[thb]
\centerline{\epsfig{figure=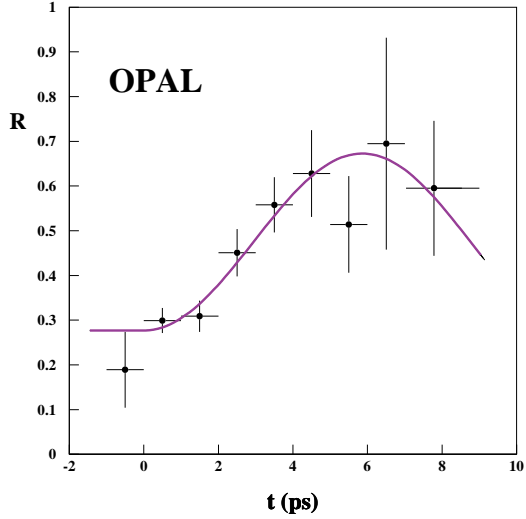,width=2.7in}}
\caption{\label{opal_mix} The ratio, R, of same-sign to total dilepton
events as a function of proper decay time, for selected $B\to D^{*+}X\ell^-\bar{\nu}$
events. The jet charge in the opposite hemisphere is used to determine the sign
correlation. The curve is the result of a fit to the mixing parameter.}
\end{figure}

Data from many experiments has been combined by ``The Heavy Flavor Averaging Group,"
(HFAG) to obtain an average value $\Delta m_d = (0.507\pm0.004)\times 10^{12}$
ps$^{-1}$ \cite{HFAG}.

 The probability of mixing is related to the CKM matrix elements as
 \cite{Gaillard1974,Bigi-Sanda}
\begin{equation}
x_d\equiv \frac{\Delta m}{\Gamma}={G_F^2\over
6\pi^2}B_{B_d}f_B^2m_B\tau_B|V^*_{tb}V_{td}|^2m_t^2 F{\left(m^2_t\over
M^2_W\right)}\eta_{QCD}, \label{eq:Bdmix}
\end{equation}
where $B_B$ is a parameter related to the probability of the $d$ and $\bar{b}$ quarks
forming a hadron and must be estimated theoretically, $F$ is a known function which
increases approximately as $m^2_t$, and $\eta_{QCD}$ is a QCD correction, with value
about 0.8. By far the largest uncertainty arises from the decay constant,
$f_B$.

In principle $f_B$ can be measured. The decay rate of the annihilation process
$B^-\to\ell^-\overline{\nu}$ is proportional to the product of $f_B^2|V_{ub}|^2$. Of
course there is a substantial uncertainty associated with $|V_{ub}|$. The experimental
evidence for $B^-\to\tau^-\overline{\nu}$ is discussed in Section \ref{sec:tau_nu}, and substantial
uncertainty exists in the branching ratio measurement. Thus, we need to rely on theory for a value of $f_B$.

The ratio of $B_s$ to $B_d$ mixing frequency, however, provides a better situation in
terms of reducing model dependent errors. Using Eq.~\ref{eq:Bdmix} for $B_d$ mixing
and an analogous relation for $B_s$ mixing and then dividing them results in

\begin{equation}
\frac{x_d}{x_s}=\frac{B_{B}}{B_{B_s}}\frac{f_B^2}{f^2_{B_s}}\frac{m_B}{m_{B_s}}
\frac{\tau_B}{\tau_{B_s}}\frac{|V^*_{tb}V_{td}|^2}{|V^*_{tb}V_{ts}|^2}~.
\label{eq:mix}
\end{equation}

The CKM terms are
\begin{eqnarray}
 |V^*_{tb}V_{td}|^2&=&A\lambda^3|(1-\rho-i\eta)|^2=A\lambda^3(\rho-1)^2+\eta^2
 ~{\rm and}\\\nonumber
 |V^*_{tb}V_{ts}|^2&=&A\lambda^2~,
 \label{eq:mixrhoeta}
\end{eqnarray}
ignoring the higher order term in $V_{ts}$. Solving the ratio of the two above
equations for $\left(1\over\lambda\right)\frac{|V_{td}|}{|V_{ts}|}$ gives a circle
centered at (1,0) in the $\rho - \eta$ plane whose radius depends on $x_d/x_s$. The
theoretical errors are now due to the difference between having a light quark versus a
strange quark in the $B$ meson, called SU(3) splitting.

For many years experiments at the $Z^0$ using $e^+e^-$ colliders at both LEP and the
SLC had set lower limits on $B_s^0$ mixing \cite{PDG}. In 2006 $B_s$ mixing was
measured by the CDF collaboration \cite{CDF-Bs}. The D0 collaboration had previously
presented both a lower and an upper limit \cite{D0-Bs}.

We now discuss the CDF measurement. The probability, ${\cal{P}}(t)$ for a $B_s$ to
oscillate into a $\overline{B}_s$ is given as
\begin{equation}
{\cal{P}}(t)\left(B_s\to\overline{B}_s\right)={1\over 2}\Gamma_s e^{-\Gamma_s t}
\left[1+\cos\left(\Delta m_s t\right)\right]~~,
\end{equation}
where $t$ is the proper time.

An amplitude $A$ for each test frequency $\omega$, is defined as \cite{MOSER}
\begin{equation}
{\cal{P}}(t)={1\over 2}\Gamma_s e^{-\Gamma_s t} \left[1+A\cos\left(\omega
t\right)\right]~~. \label{eq:Bs}
\end{equation}
For each frequency the expected result is either zero for no mixing, or one for mixing.
No other value is physical, although measurement errors admit other values.
Figure~\ref{fig:amplitudeScan} shows the CDF results.
\begin{figure}[ht]
\begin{center}
\includegraphics[width=5in]{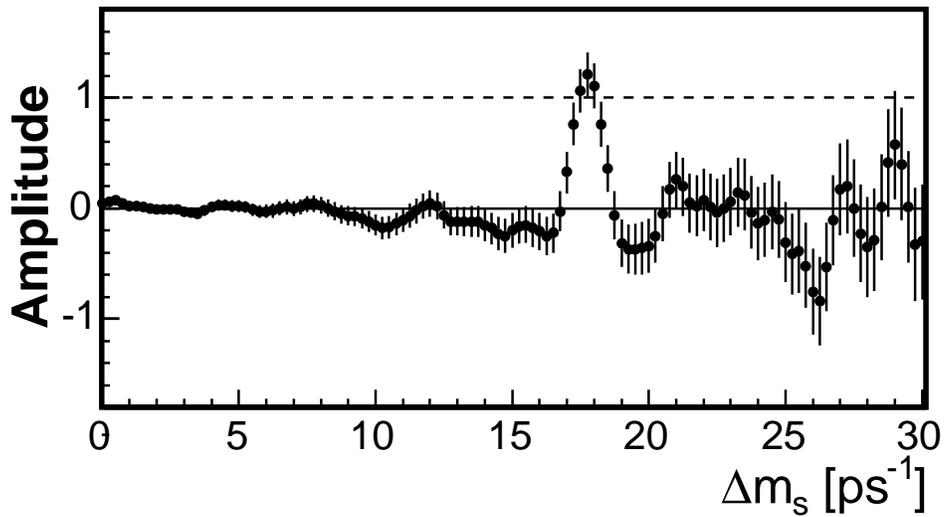}
\caption[] { 
 The measured amplitude values and uncertainties versus the $B_s\overline{B}_s$
 oscillation frequency $\Delta m_s$ for a combination of semileptonic and
 hadronic decay modes from CDF.
} \label{fig:amplitudeScan}
\end{center}
\end{figure}

 At $\Delta m_s = 17.75$ ps$^{-1}$, the observed amplitude
 ${A} = 1.21\pm 0.20~({\rm stat.})$
 is consistent with unity, indicating that the data
 are compatible with $B_s\overline{B}_s$ oscillations with that frequency, while the
 amplitude is inconsistent with zero: ${A}/\sigma_{A}=6.05$,
 where $\sigma_{A}$ is the statistical uncertainty on ${A}$,
 the ratio has negligible systematic uncertainties. (This then is called a ``6$\sigma$ effect.)
The measured value is
\begin{equation}
\Delta m_s = 17.77\pm 0.010~\pm{0.07}~ {\rm ps}^{-1}~,
\end{equation}
where the first error is statistical and the second systematic.

In order to translate the mixing measurements to constraints on the CKM parameters
$\rho$ and $\eta$, we need to use theoretical values for the ratios
$\frac{B_{B}}{B_{B_s}}$ and $\frac{f_B}{f_{B_s}}$ in Eq.~\ref{eq:mix}. It is
interesting that in practice it is not possible to measure either of these numbers
directly. It is usually assumed, however, that $f_{B^+}$, which could in principle be
measured via the process $B^-\to \tau^-\nu$ is the same as $f_B$. There is no way to
measure $f_{B_s}$. The charm system, on the other hand, provides us with both
$D^+\to\ell^+\nu$ and $D_s^+\to\ell^+\nu$, and both rates have been measured. They
also have been calculated in unquenched lattice quantum electrodynamics (QCD).

The combined efforts of the HPQCD and UKQCD collaborations predict $f_{D^+}=(207\pm
4$) MeV \cite{Follana}, while the CLEO measurement is in astonishing agreement:
(205.8$\pm$8.5$\pm$2.5) MeV \cite{CLEOfDp}. Furthermore, the measurements of
$f_{D_s^+}$ are not in such good agreement with the Follana \etal ~calculation of
(241$\pm$3)~MeV. The average of CLEO and Belle results as determined by Rosner and
Stone is (273$\pm$10)~MeV \cite{Rosner-Stone}. The discrepancy is at the 3 standard
deviation level and not yet statistically significant, though it bears watching.

Unfortunately, the group that has calculated $f_{D^+}$ is not the same group that has
determined $f_{B^+}$. The theoretical calculations used for the $B_B$ terms and the
$f_B$ terms are summarized by Tantalo \cite{Tantalo}. He suggests using values of
$f_{B_s}=(268\pm 17\pm 20)$ MeV, $f_{B_s}/f_{B}=1.20\pm 0.02\pm 0.05$, $B_{B_s}=0.84\pm 0.03 \pm 0.055$ and $B_{B}=0.83\pm 0.01 \pm 0.06$. These numbers allow us to measure the length of one
side of the CKM triangle using Eq.~\ref{eq:mix}. Use of this measurement will
be discussed in more detail in Section~\ref{sec:CKMfits}.

\subsection{CP Violation in the {\it\textbf B} System}

We have two quantum mechanical operators: Charge Conjugation, $C$, and Parity, $P$.
When applied to a particle wavefunction $C$ changes particle to antiparticle and
vice-versa. Applying $P$ to a wavefunction $\psi(\bf r)$ we have ${ P}\psi({\bf
r})=\psi(-{\bf r})$. The $P$ operator can be thought of changing the natural
coordinate system from right-handed to left-handed. If nature was blind to handedness,
then $P$ would be always conserved. By applying the $P$ operator twice we end up with
${P}^2\psi(\bf r)=\psi(\bf r)$, so the eigenvalues of $P$ are $\pm$1. Therefore
wave-functions, or particles represented by such wave-functions, have either intrinsic
positive parity $+1$ (right-handed) or $-1$ (left-handed).

Weak interactions, characterized by a combination of vector minus axial-vector
currents, are known to be left-handed. Therefore, handedness matters and its well known
that Parity is maximally violated in weak decays \cite{Bigi-Sanda}. Since $C$ changes
left-handed particles to right-handed anti-particles, the product $CP$ symmetry could have been
preserved, but nature decided otherwise. Different particle transitions involve the
different $CP$ violating angles shown in Figure~\ref{six_tri}.  Measuring these
independently allows comparisons with measurements of the sides of the triangle and
any differences in constraints in $\rho$ and $\eta$ can be due to the presence of new
physics.

Consider the case of a process $B\to f$ that goes via two amplitudes, $\cal{A}$ and
$\cal{B}$ each of which has a strong part e. g. $s_{\cal{A}}$ and a weak part
$w_{\cal{A}}$. Then we have
\begin{eqnarray}
\Gamma(B\to f)&=&\left(\left|{\cal{A}}\right|e^{i(s_{\cal{A}}+w_{\cal{A}})}
+\left|{\cal{B}}\right|e^{i(s_{\cal{B}}+w_{\cal{B}})}\right)^2 \\
\Gamma(\overline{B}\to \overline{f})&=&
\left(\left|{\cal{A}}\right|e^{i(s_{\cal{A}}-w_{\cal{A}})}
+\left|{\cal{B}}\right|e^{i(s_{\cal{B}}-w_{\cal{B}})}\right)^2 \\ \Gamma(B\to
f)-\Gamma(\overline{B}\to \overline{f})&=&2\left|{\cal{AB}}\right|
\sin(s_{\cal{A}}-s_{\cal{B}})\sin(w_{\cal{A}}-w_{\cal{B}})~~.
\end{eqnarray}

Any two amplitudes will do, though its better that they be of approximately equal
size. Thus charged $B$ decays can exhibit CP violation as well as neutral $B$ decays.
In some cases, we will see that it is possible to guarantee that
$\left|\sin(s_{\cal{A}}-s_{\cal{B}})\right|$ is unity, so we can get information on
the weak phases. In the case of neutral $B$ decays, mixing serves as the second
amplitude.

\subsubsection{Formalism of CP Violation in Neutral {\it\textbf B} Decays}
Consider the operations of Charge Conjugation, $C$,  and Parity, $P$:
\begin{eqnarray}
&C|B(\overrightarrow{p})\big>=|\overline{B}(\overrightarrow{p})\big>,~~~~~~~~~
&C|\overline{B}(\overrightarrow{p})\big>=|{B}(\overrightarrow{p})\big> \\
&P|B(\overrightarrow{p})\big>=-|{B}(-\overrightarrow{p})\big>,~~~~~
&P|\overline{B}(\overrightarrow{p})\big>=-|\overline{B}(-\overrightarrow{p})\big> \\
&CP|B(\overrightarrow{p})\big>=-|\overline{B}(-\overrightarrow{p})\big>,~~~
&CP|\overline{B}(\overrightarrow{p})\big>=-|{B}(-\overrightarrow{p})\big> ~~.
\end{eqnarray}
For neutral mesons we can construct the $CP$ eigenstates
\begin{eqnarray}
\big|B^0_1\big>&=&{1\over \sqrt{2}}\left(\big|B^0\big>-
\big|\overline{B}^0\big>\right)~~,\\ \big|B^0_2\big>&=&{1\over
\sqrt{2}}\left(\big|B^0\big>+\big|\overline{B}^0\big>\right)~~,
\end{eqnarray}
where
\begin{eqnarray}
CP\big|B^0_1\big>&=&\big|B^0_1\big>~~, \\ CP\big|B^0_2\big>&=&-\big|B^0_2\big>~~.
\end{eqnarray}
Since $B^0$ and $\overline{B}^0$ can mix, the mass eigenstates are a superposition of
$a\big|B^0\big> + b\big|\overline{B}^0\big>$ which obey the Schrodinger equation
\begin{equation}
i{d\over dt}\left(\begin{array}{c}a\\b\end{array}\right)= {\cal
H}\left(\begin{array}{c}a\\b\end{array}\right)= \left(M-{i\over
2}\Gamma\right)\left(\begin{array}{c}a\\b\end{array}\right). \label{eq:schrod}
\end{equation}
If $CP$ is not conserved then the eigenvectors, the mass eigenstates $\big|B_L\big>  $
and  $\big|B_H\big>$, are not the $CP$ eigenstates but are
\begin{equation}
\big|B_L\big> = p\big|B^0\big>+q\big|\overline{B}^0\big>,~~\big|B_H\big> =
p\big|B^0\big>-q\big|\overline{B}^0\big>,
\end{equation}
where
\begin{equation}
p={1\over \sqrt{2}}{{1+\epsilon_B}\over {\sqrt{1+|\epsilon_B|^2}}},~~ q={1\over
\sqrt{2}}{{1-\epsilon_B}\over {\sqrt{1+|\epsilon_B|^2}}}.
\end{equation}
$CP$ is violated if $\epsilon_B\neq 0$, which occurs if $|q/p|\neq 1$.

The time dependence of the mass eigenstates is
\begin{eqnarray}
\big|B_L(t)\big> &= &e^{-\Gamma_Lt/2}e^{im_Lt/2} \big|B_L(0)\big> \\
 \big|B_H(t)\big> &= &e^{-\Gamma_Ht/2}e^{im_Ht/2} \big|B_H(0)\big>,
\end{eqnarray}
leading to the time evolution of the flavor eigenstates as
\begin{eqnarray}
\big|B^0(t)\big>&=&e^{-\left(im+{\Gamma\over 2}\right)t} \left(\cos{\Delta mt\over
2}\big|B^0(0)\big>+i{q\over p}\sin{\Delta mt\over 2}\big|\overline{B}^0(0)\big>\right)
\\ \big|\overline{B}^0(t)\big>&=&e^{-\left(im+{\Gamma\over 2}\right)t} \left(i{p\over
q}\sin{\Delta mt\over 2}\big|B^0(0)\big>+ \cos{\Delta mt\over
2}\big|\overline{B}^0(0)\big>\right),
\end{eqnarray}
where $m=(m_L+m_H)/2$, $\Delta m=m_H-m_L$ and $\Gamma=\Gamma_L\approx \Gamma_H$, and
$t$ is the decay time in the $B^0$ rest frame, the so-called ``proper time''. Note that
the probability of a $B^0$ decay as a function of $t$ is given by
$\big<B^0(t)\big|B^0(t)\big>^*$, and is a pure exponential, $e^{-\Gamma t/2}$, in the
absence of $CP$ violation.

\subsubsection{{\it\textbf CP} Violation for {\it\textbf B} Via the Interference of Mixing and Decay}

Here we choose a final state $f$ which is accessible to both $B^0$ and
$\overline{B}^0$  decays \cite{Bigi-Sanda}. The second amplitude necessary for interference
is provided by mixing.  Figure~\ref{eigen_CP} shows the decay into $f$ either directly
or indirectly via  mixing.
\begin{figure}[htb]
\centerline{\epsfig{figure=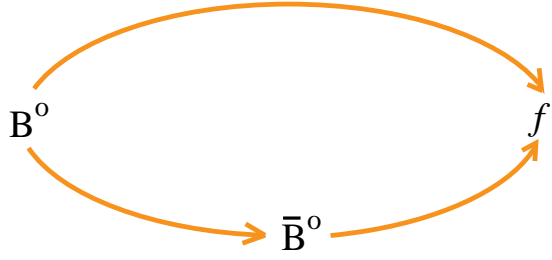,height=1.5in}}
\caption{\label{eigen_CP}Two interfering ways for a $B^0$ to decay into a final state
$f$.} 
\end{figure}
It is  necessary only that $f$ be accessible directly from either state; however if
$f$ is a $CP$  eigenstate the situation is far simpler. For $CP$ eigenstates
\begin{equation} CP\big|f_{CP}\big>=\pm\big|f_{CP}\big>. \end{equation}

It is useful to define the amplitudes
\begin{equation}
A=\big<f_{CP}\big|{\cal H}\big|B^0\big>,~~ \overline{A}=\big<f_{CP}\big|{\cal
H}\big|\overline{B}^0\big>.
\end{equation}
If $\left|{\overline{A}\over A}\right|\neq 1$, then we have ``direct" $CP$ violation in
the decay amplitude, which we will discuss in detail later. Here $CP$ can be violated by
having
\begin{equation}
\lambda = {q\over p}\cdot {\overline{A}\over A}\neq 1,
\end{equation}
which requires only that $\lambda$  acquire a non-zero phase, i.e. $|\lambda|$ could
be unity and $CP$ violation can occur.

Other useful variables, that are independent of any phase convention are
\begin{eqnarray}\label{eq:phi}
&&\phi_{12}={\rm arg}\left(-\frac{M_{12}}{\Gamma_{12}}\right),{~\rm and} \\\nonumber
&&{\rm{Im}}\frac{\Gamma_{12}}{M_{12}}=\frac{1-|q/p|^4}{1+|q/p|^4}\equiv A_{SL}(t).
\end{eqnarray}
The first quantity can be related to CKM angles, while the second can be measured by
the ``semileptonic asymmetry," or for that matter in any flavor specific decay \cite{Beneke}:
\begin{equation}
\label{eq:asl}
A_{SL}(t)=\frac{\Gamma[\overline{B}^0(t)\to \ell^-X]-\Gamma[{B}^0(t)\to \ell^+X]}
{\Gamma[\overline{B}^0(t)\to \ell^-X]+\Gamma[{B}^0(t)\to \ell^+X]}=
\frac{\Delta\Gamma}{\Delta M}\tan\phi_{12}~,
\end{equation}
for either $B^0$ or for $B_s$ mesons, separately.

A comment on neutral $B$ production at $e^+e^-$ colliders is in order. At the
$\Upsilon (4S)$ resonance there is coherent production of $B^0\bar{B}^0$ pairs. This
puts the $B$'s in a $C=-1$ state. In hadron colliders, or at $e^+e^-$ machines
operating at the $Z^0$, the $B$'s are produced incoherently. For the rest of this
article we will assume incoherent production except where explicitly noted.

The asymmetry, in the case of $CP$ eigenstates, is defined as
\begin{equation}
a_{f_{CP}}={{\Gamma\left(B^0(t)\to f_{CP}\right)- \Gamma\left(\overline{B}^0(t)\to
f_{CP}\right)}\over {\Gamma\left(B^0(t)\to f_{CP}\right)+
\Gamma\left(\overline{B}^0(t)\to f_{CP}\right)}},
\end{equation}
which for $|q/p|=1$ gives
\begin{equation}
\label{eq:bothamp} a_{f_{CP}}={{\left(1-|\lambda|^2\right)\cos\left(\Delta
mt\right)-2{\rm Im}\lambda \sin(\Delta mt)}\over {1+|\lambda|^2}}.
\end{equation}
For the cases where there is only one decay amplitude $A$, $|\lambda |$ equals 1, and
we have
\begin{equation}
a_{f_{CP}}=-{\rm Im}\lambda \sin(\Delta mt).
\end{equation}
Only the factor ${\rm -Im}\lambda$ contains information about the level of CP
violation, the sine term is determined by $B$ mixing. In fact, the time
integrated asymmetry is given by
\begin{equation}
a_{f_{CP}}=-{x \over {1+x^2}}{\rm Im}\lambda = -0.48 {\rm Im}\lambda ~~.
\label{eq:aint}
\end{equation}
This is quite lucky for the study of $B_d$ mesons, as the maximum size of the coefficient for any $x$ is $-0.5$, close to the measured value of $x_d$.

Let us now find out how ${\rm Im}\lambda$ relates to the CKM parameters. Recall
$\lambda={q\over p}\cdot {\overline{A}\over A}$. The first term is the part that comes
from mixing:
\begin{equation}
{q\over p}={{\left(V_{tb}^*V_{td}\right)^2}\over {\left|V_{tb}V_{td}\right|^2}}
={{\left(1-\rho-i\eta\right)^2}\over {\left(1-\rho+i\eta\right)\left(1-\rho-
i\eta\right)}} =e^{-2i\beta}{\rm~~and}
\end{equation}
\begin{equation}
{\rm Im}{q\over p}= -{{2(1-\rho)\eta}\over
{\left(1-\rho\right)^2+\eta^2}}=\sin(2\beta).
\end{equation}

\subsubsection{Measurements of {$\sin(2\beta)$}}
To evaluate the decay part we need to consider specific final states. For example,
consider the final state $J/\psi K_s$. The decay diagram is shown in
Figure~\ref{psi_ks}. In this case we do not get a phase from the decay part because
\begin{equation}
{\overline{A}\over A} = {{\left(V_{cb}V_{cs}^*\right)^2}\over
{\left|V_{cb}V_{cs}\right|^2}}
\end{equation}
is real to order $1/\lambda^4$.

In this case the final state is a state of negative $CP$, i.e. $CP\big|J/\psi
K_s\big>=-\big|J/\psi K_s\big>$. This introduces an additional minus sign in the
result for ${\rm Im}\lambda$. Before finishing discussion of this final state we need
to consider in more detail the presence of the $K_s$ in the final state. Since neutral
kaons can mix, we pick up another mixing phase (similar diagrams as for $B^0$, see
Figure~\ref{bmix}). This term creates a phase given by
\begin{equation}
\left({q\over p}\right)_K={{\left(V_{cd}^*V_{cs}\right)^2}\over
{\left|V_{cd}V_{cs}\right|^2}},
\end{equation}
which is real to order $\lambda^4$. It necessary to include this term, however, since
there are other  formulations of the CKM matrix than Wolfenstein, which have the phase
in a  different location. It is important that the physics predictions not depend on
the  CKM convention.\footnote{Here we don't include $CP$ violation in the neutral kaon
since it is much smaller than what is expected in the $B$ decay.}

In summary, for the case of $f=J/\psi K_s$, ${\rm Im}\lambda=-\sin(2\beta)$. The angle
$\beta$ is the best measured $CP$ violating angle measured in $B$ meson decays. The
process used is $B^0\to J/\psi K_S$, although there is some data used with the
$\psi(2S)$ or with $K_L$ (where the phase is $+\sin(2\beta)$).

\begin{figure}[ht]
\begin{center}
\includegraphics[width=3in]{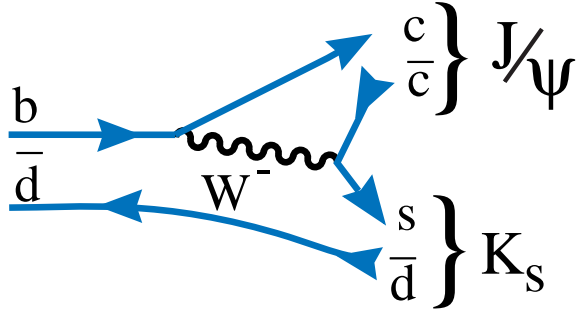}
\caption[] { 
 The Feynman diagram for the decay $B^0\to J/\psi K^0$.
} \label{psi_ks}
\end{center}
\end{figure}

Although it is normally thought that only one decay amplitude contributes here, in
fact one can look for the presence of another source of $CP$ violation, presumably in
the decay amplitude, by not assuming $|\lambda|$ equals one in Eq.~\ref{eq:bothamp}.
Then the time dependence of the decay rate is given by
\begin{equation}
\label{eq:timedep} a_{f_{CP}}=-\frac{2{\rm Im}\lambda}{1+|\lambda|^2} \sin(\Delta mt)
-\frac{1-|\lambda|^2}{1+|\lambda|^2}\cos(\Delta mt).
\end{equation}
Thus $a_{f_{CP}}(t)$ has both $\sin(\Delta mt)$ and $\cos(\Delta mt)$ terms, and the
coefficients of these terms can be measured. Let us assign the labels
\begin{equation}
{\cal{S}}=-\frac{2{\rm
Im}\lambda}{1+|\lambda|^2},~~~~~~~~~~~~{\cal{C}}=\frac{1-|\lambda|^2}{1+|\lambda|^2}.
\end{equation}
(Note that the sign of the ${\cal{S}}$ term changes depending on the $CP$ of the final state,
but not the ${\cal{C}}$ term.)

The most precise measurements of $\sin(2\beta)$ have been made by the BaBar and Belle
experiments \cite{sin2beta}. These measurements are made at the $\Upsilon(4S)$
resonance using $e^+e^-\to \Upsilon(4S)\to B^0\overline{B}^0$, and with one of the
neutral $B$'s decaying into $J/\psi K^0$. Because the $\Upsilon(4S)$ is a definite
quantum state with $C=-1$, the formulae given above have to modified somewhat. One important
change is the definition of $t$. The time used here is the difference between the
decay time of the $J/\psi K^0$ and the other $B^0$, also called the ``tagging" $B$
because we need to determine its flavor, whether $B^0$ or $\overline{B}^0$, in order to
make the asymmetry measurement.

While we will not discuss flavor tagging in general, it is an important part of $CP$
violation measurements. At the $\Upsilon(4S)$ once such very useful tag is that of high
momentum lepton as a $b$-quark, part of a $\overline{B}^0$ meson decays
semileptonically into an $\ell^-$, while a $\overline{b}$-quark decays into an $e^+$.
Other flavor tagging observables include the charge of kaons and fast pions. Modern
experiments combine these in a ``neural network," to maximize their tagging
efficiencies \cite{neural}. The BaBar data are shown in Figure~\ref{BaBar-psiKs}, Belle has similar
results.

\begin{figure}[htb]
\centerline{\epsfig{figure= 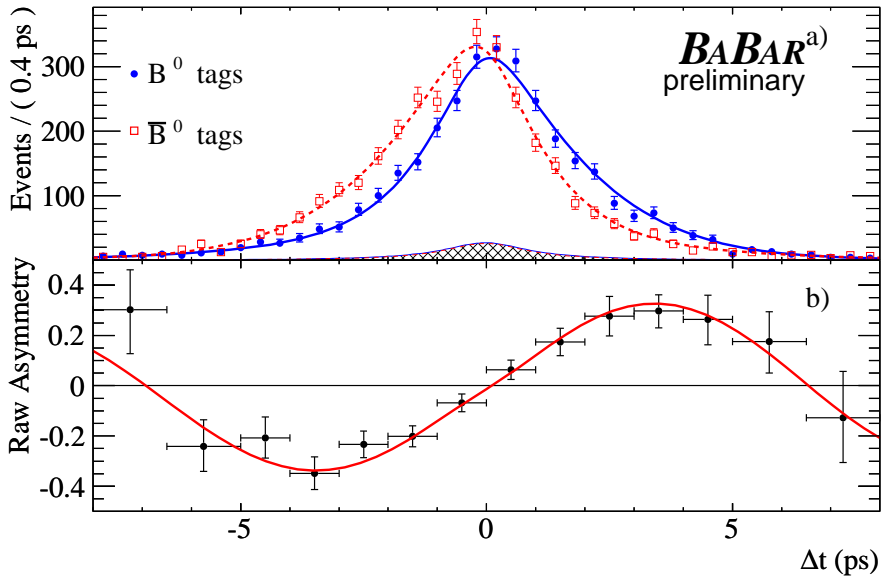,height=3.1in}}
\caption{(a) The number of $J/\psi K_s$ candidates in the signal
region with either a $B^0$ tag ($N_{B^0}$), or a $\overline{B}^0$ tag
($N_{\overline{B}^0}$) as a function of $\Delta t$. (b) The raw asymmetry,
$(N_{B^0}-N_{\overline{B}^0})/(N_{B^0}+N_{\overline{B}^0})$. The solid (dashed) curves
represent the fit projections as functions of $\Delta t$ for both  $B^0$ and
$\overline{B}^0$ tags. The shaded regions represent the estimated background
contributions.} \label{BaBar-psiKs}       
\end{figure}

The average value as determined by the Heavy Flavor Averaging Group is
$\sin(2\beta)=0.671\pm0.024$, where the dominant part of the error is statistical. No
evidence is found for a non-zero ${\cal{C}}$ term with the measured average given as 0.005$\pm0.020$~.

Determining the sine of any angle gives a four-fold ambiguity in the angle. The decay
mode $B^0\to J/\psi K^{*0}$, $K^{*0}\to K_s\pi^0$ offers a way of measuring
$\cos{2\beta}$ and resolving the ambiguities. This is a subtle analysis that we will
not go into detail on \cite{psiKstar}. The result is that $\beta=
21.07^{+0.94}_{-0.92}$ degrees.

\subsubsection{Measurements of {$\alpha$}}

The next state to discuss is the $CP$+ eigenstate $f\equiv\pi^+\pi^-$. The simple
spectator decay diagram is shown in Figure~\ref{rhorho}(a). For the moment we will
assume that this is the only diagram, though the Penguin diagram shown in
Figure~\ref{rhorho}(b) could also contribute; its presence can be inferred because it
would induce a non-zero value for ${\cal{C}}$, the coefficient of the cosine term in
Eq.~\ref{eq:timedep}. For this $b\to u\bar{u}d$ process we have
\begin{equation}
{\overline{A}\over A}={{\left(V_{ud}^*V_{ub}\right)^2}\over
{\left|V_{ud}V_{ub}\right|^2}}={{(\rho-i\eta)^2}\over
 {(\rho-i\eta)(\rho+i\eta)}}=e^{-2i\gamma}
\end{equation}
and
\begin{equation}
{\rm Im}(\lambda)={\rm Im}(e^{-2i\beta}e^{-2i\gamma})= {\rm
Im}(e^{2i\alpha})=-\sin(2\alpha)
\end{equation}

\begin{figure}
\centerline{\epsfig{figure=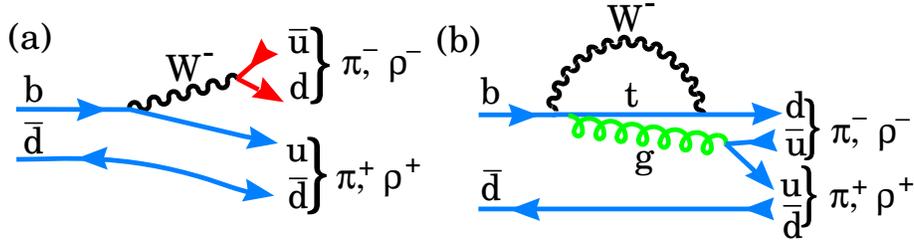,height=1.4in}} 
\caption{Tree (a) and Penguin (b) processes for neutral $B$ decay into either
$\pi^+\pi^-$ or $\rho^+\rho^-$.} \label{rhorho}       
\end{figure}

Time dependent $CP$ violation measurements have been made by both the BaBar and Belle
collaborations \cite{alpha-pipi}. Both groups find a non-zero value of both ${\cal{C}}$  and
${\cal{S}}$, though their values are not in particularly good agreement.
The HFAG average is
shown in Figure~\ref{pi+pi-} along with the experimental results. The value of ${\cal{C}}$
clearly is not zero, thus demonstrating direct $CP$ violation.  (Historically, this was an important observation because it showed that $CP$ violation could not be due to some kind of ``superweak" model ala Wolfenstein \cite{superweak}.)

\begin{figure}
\centerline{\epsfig{figure=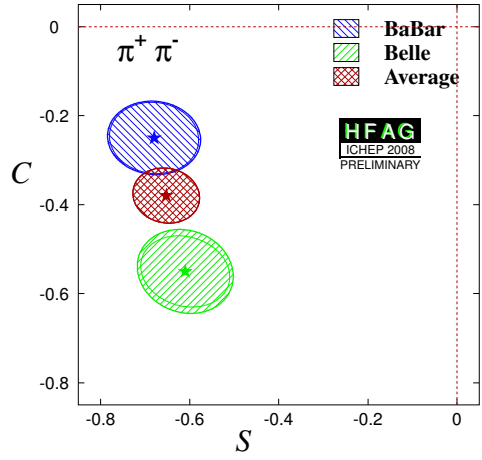,height=2.4in}} 
\caption{Coefficients of the sine term ${\cal{S}}$  and the cosine term ${\cal{C}}$  in time dependent
$CP$ violation for neutral $B$ decay into $\pi^+\pi^-$, showing BaBar and Belle
results, and the HFAG average. Contours are shown at 60.7\% confidence level.}
\label{pi+pi-}
\end{figure}

The non-zero value of ${\cal{C}}$  shows the presence of at least two amplitudes, presumably tree and
penguin, in addition to the mixing amplitude, making extraction of $\alpha$ difficult. All is not lost, however. Gronau and London \cite{GL} showed that $\alpha$ could be
extracted without theoretical error by doing a full isotopic spin analysis of the
$\pi\pi$ final state. The required measurements include: ${\cal{C}}$ , ${\cal{S}}$,  ${\cal B}(B^+\to
\pi^+\pi^0)$, ${\cal B}(B^0\to \pi^0\pi^0)$, and  ${\cal B}(\overline{B}^0\to
\pi^0\pi^0)$. The last two items require a flavored tagged analysis that has not yet
been done and whose prospects are bleak. Grossman and Quinn have showed, however, that
an upper limit on $\Gamma(B^0\to\pi^0\pi^0)/\Gamma(B^+\to\pi^+\pi^0)$ can be used to
limit the penguin shift to $\alpha$ \cite{GQ}. Unfortunately, current data show a
relative large ${\cal B}(B^0\to\pi^0\pi^0)=(1.55\pm 0.19)\times 10^{-6}$ rate,
compared with ${\cal B}(B^0\to\pi^+\pi^-)=(5.16\pm 0.22)\times 10^{-6}$, implying a
large penguin contribution \cite{PDG,HFAG}, and the limit is very weak.

Use of the $\rho^+\rho^-$ final state is in principle quite similar to $\pi^+\pi^-$,
with some important caveats. First of all, it is a vector-vector final state and
therefore could have both $CP$+ and $CP$- components. It is however possible, doing a
full angular analysis to measure the $CP$ violating phase separately in each of these
two amplitudes. The best method for this is in the  ``transversity" basis and will be
discussed later \cite{transversity} in Section~\ref{sec:transversity}. It is possible, however, for one polarization
component to be dominant and then the angular analysis might not be necessary. In fact
the longitudinal polarization is dominant in this case.  BaBar measures the fraction as  $0.992\pm
0.024^{+0.026}_{-0.013}$ \cite{BaBar-rhorho}, and Belle measures it as
$0.941^{+0.034}_{-0.040}\pm 0.030$ \cite{Belle-rhorho}. Thus we can treat this final
state without worrying about the angular analysis and just apply a correction for the
amount of transverse amplitude.

In addition, $\rho$ mesons are wide, so non-$B$ backgrounds could be a problem and
even if the proper $B$ is reconstructed, there are non-resonant and possible $a_1\pi$
contributions. Furthermore, it has been pointed out that the large $\rho$ width could
lead to the violation of the isospin constraints and this effect should be
investigated \cite{Falk-rhorho}. The relevant branching ratios are given in
Table~\ref{tab:rhorho}.

\begin{table}
\begin{center}
\caption{Branching ratios $B\to\rho\rho$ modes in units of $10^{-6}$}
\label{tab:rhorho}
\begin{tabular}{lccc}
\hline\hline Mode & BaBar&  Belle & Average \cite{PDG,HFAG} \\\hline $\rho^+\rho^-$ &
$25.5\pm 2.1^{+3.6}_{-3.9}$ & $22.8\pm 3.8^{+2.3}_{-2.6}$ & $24.2\pm 3.1$\\
$\rho^+\rho^0$ & $16.8\pm 2.2\pm 2.3$ & $31.7\pm 7.1^{+3.8}_{-6.7}$ & $18\pm 4$ \\
$\rho^0\rho^0$ &$0.92\pm 0.32 \pm 0.14$& $0.4^{+0.4}_{-0.2}\pm 0.3$ &
$0.74^{+0.30}_{-0.27}$  \\ \hline
\end{tabular}
\end{center}
\end{table}

\begin{figure}[htb]
\centerline{\epsfig{figure=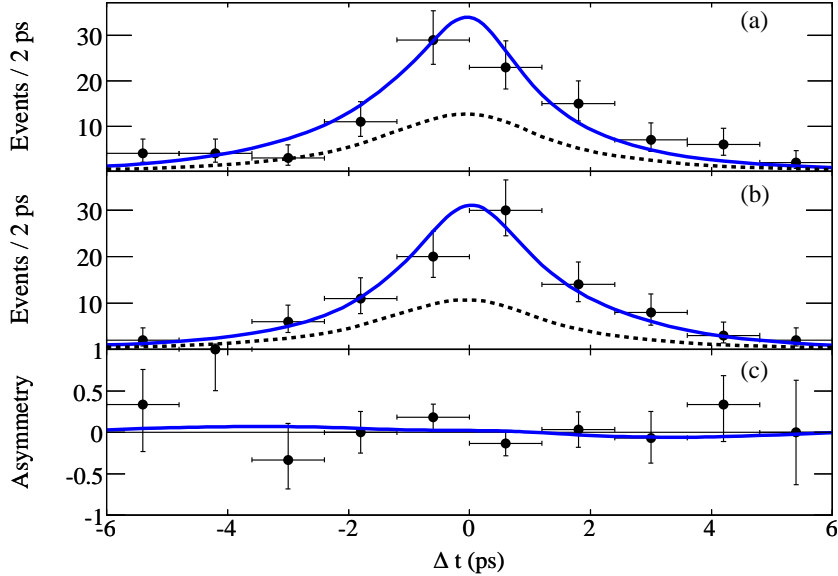,height=3.1in}}
\caption{ The number of $\rho^+\rho^-$ candidates in the signal
region with either a $B^0$ tag (a) ($N_{B^0}$), or a $\overline{B}^0$ tag (b)
($N_{\overline{B}^0}$) as a function of $\Delta t$. (c) The raw asymmetry,
$(N_{B^0}-N_{\overline{B}^0})/(N_{B^0}+N_{\overline{B}^0})$. The dashed curves
represent the estimated backgrounds.} \label{CP-rhorho}
\end{figure}

Nevertheless, the small branching ratio for $\rho^0\rho^0$, if indeed it has been
observed at all, shows that the effects of the Penguin diagram on the extracted value
of $\sin(2\alpha)$ are small, and this may indeed be a good way to extract a value of
$\alpha$. The time dependent decay rates separately for $B^0\to\rho^+\rho^-$ and
$\overline{B}^0\to\rho^+\rho^-$ and their difference are shown in Figure~\ref{CP-rhorho}
from BaBar. In is interesting to compare these results with those in
Figure~\ref{BaBar-psiKs}. We see that the measured asymmetry in the $\rho^+\rho^-$ decay
more or less cancels that from $B^0-\overline{B}^0$ mixing inferring that
$\sin(2\alpha)$ is close to zero.

Results from the time dependent $CP$ violation analysis from both Belle and BaBar are
shown in Figure~\ref{rho+rho-SvsC}.
\begin{figure}[htb]
\centerline{\epsfig{figure=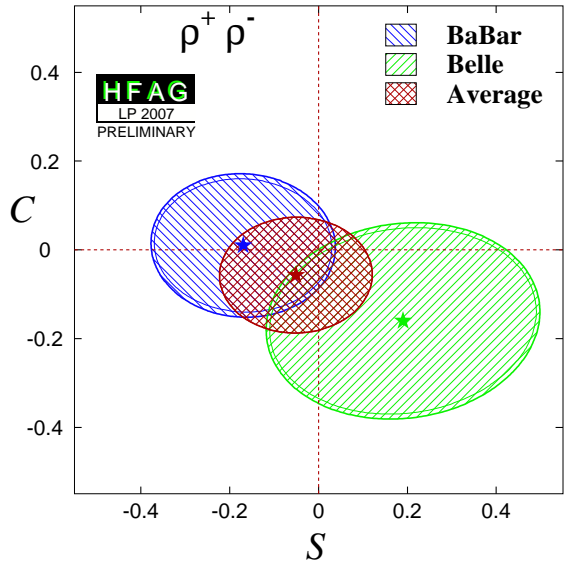,height=3.1in}}
\caption{Coefficients of the sine term ${\cal{S}}$ and the cosine term ${\cal{C}}$  in
time dependent $CP$ violation for neutral $B$ decay into $\rho^+\rho^-$, showing BaBar
and Belle results, and the HFAG average. Contours are shown at 60.7\% confidence
level.} \label{rho+rho-SvsC}
\end{figure}

The data can be averaged and $\alpha$ determined by using the isospin analysis and the
rates listed in Table~\ref{tab:rhorho}. Unfortunately the precision of the data leads
only to constraints. These have been determined by both the CKM fitter group
\cite{CKMfit} and the UT fit group \cite{UTfit}. These groups disagree in some cases.
The CKM fitter group use a frequentist statistical approach, while UT fit uses a Bayseian
approach. The basic difference is the that the Bayesian approach the theoretical
errors are taken as having a Gaussian distribution. Here we show in
Figure~\ref{ckm_alpha_rhorho} the results from CKM fitter.

\begin{figure}[htb]
\centerline{\epsfig{figure=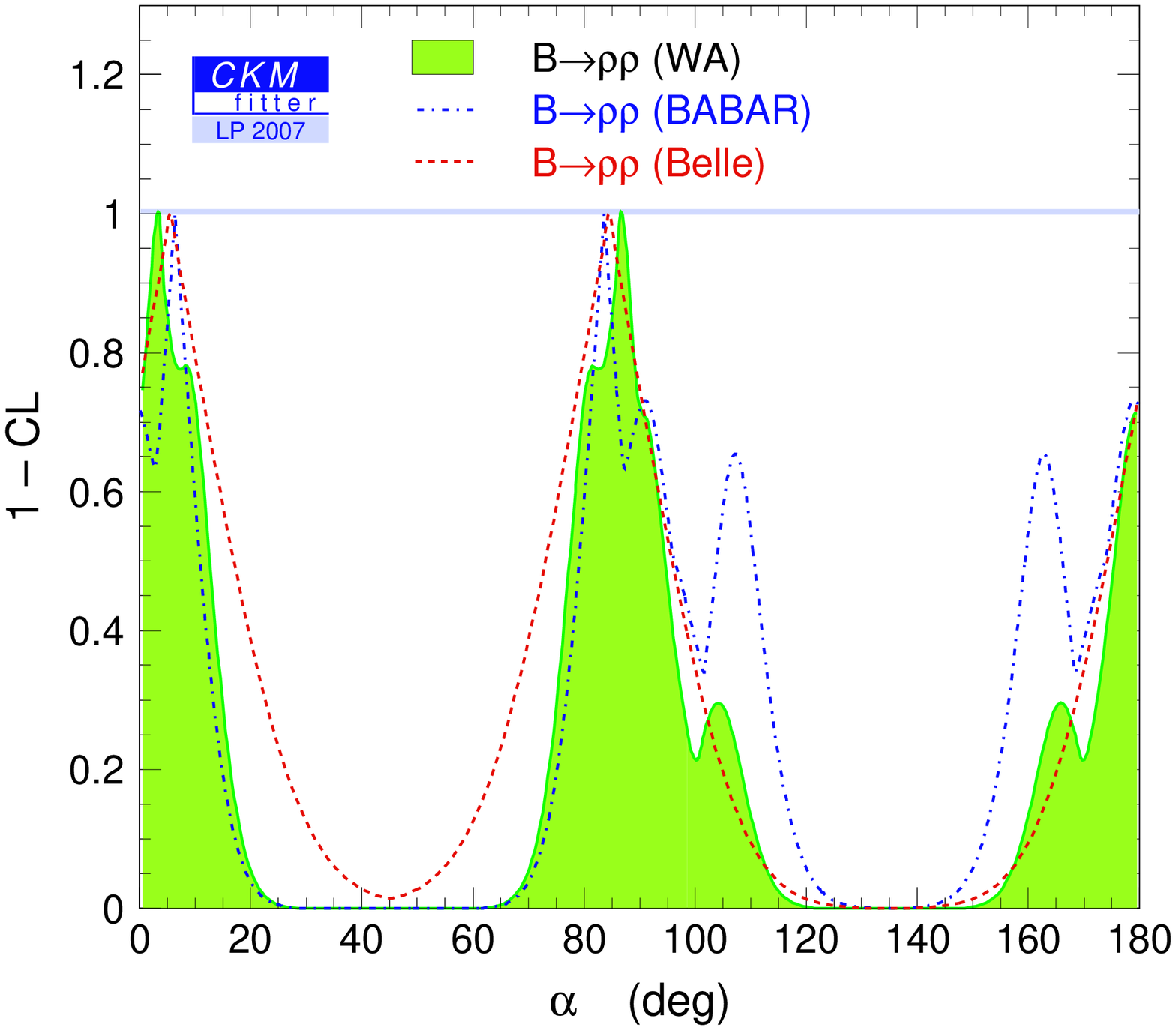,height=3.1in}}
\caption{The probability density, defined as 1 - CL (confidence
level) for the angle $\alpha$ as determined by measurements of $B\to\rho\rho$.}
\label{ckm_alpha_rhorho}
\end{figure}

The final state $\pi^+\pi^-\pi^0$ can also be used to extract $\alpha$. Snyder and
Quinn proposed that a Dalitz plot analysis of $B\to\rho\pi\to\pi^+\pi^-\pi^0$ can be
used to unravel both $\alpha$ and the relative penguin-tree phases \cite{SQ}. The
Dalitz plot for simulated events unaffected by detector acceptance is shown in
Figure~\ref{Dalitz}.
\begin{figure}[htb]
\centerline{\epsfig{figure=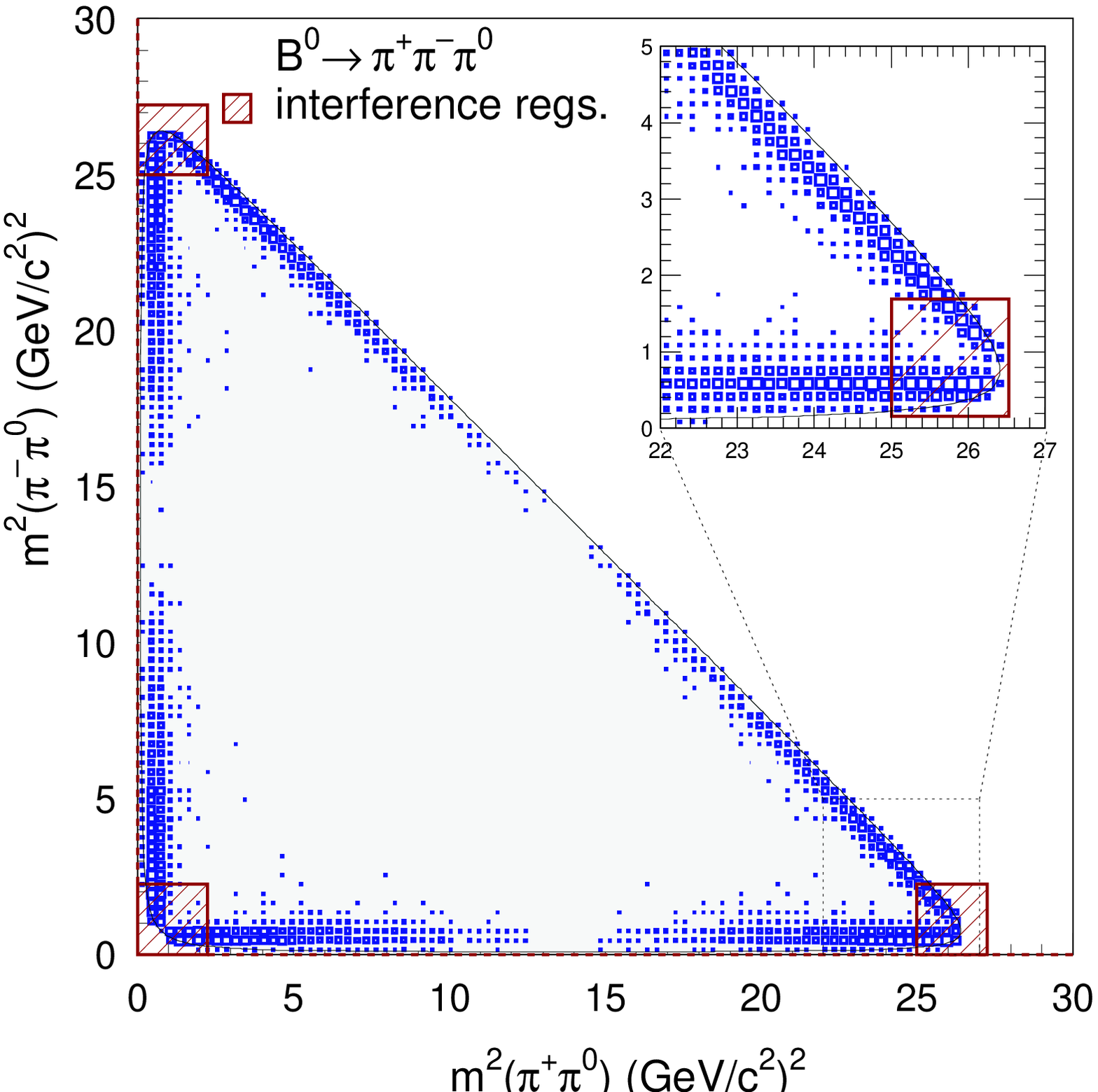,height=3.1in}} 
\caption{Dalitz plot for Monte-Carlo generated $B^0\to\rho\pi\to\pi^+\pi^-\pi^0$
decays. The decays have been simulated without any detector effect and the amplitudes
for $\rho^+\pi^-$, $\rho^-\pi^+$ and $\rho^0\pi^0$ have all been generated with equal
magnitudes in order to have destructive interferences where the $\rho$ bands overlap.
The main overlap regions between the $\rho$ bands are indicated by the hatched areas.
Adapted from \cite{BaBar-rhopi}.} \label{Dalitz}
\end{figure}

The task at hand is to do a time dependent analysis of the magnitude of the decay
amplitudes and phases. The analyses of both collaborations allow for $\rho(1450)$ and
$\rho(1700)$ contributions in addition to the $\rho(770)$. There are a total of 26
free parameters in the fit. The statistics for such an analysis are not overwhelming:
BaBar has about 2100 signal events \cite{BaBar-rhopi}, while Belle has about 1000
\cite{Belle-rhopi}.

The results for the confidence levels of $\alpha$ found by both collaborations are
shown in Figure~\ref{alpha-rhopi}. Note that there is also a mirror solution at
$\alpha$+180$^{\circ}$. The Belle collaboration uses their measured decay rates for
$B^+\to\rho^+\pi^0$ and $B^+\to\rho^0\pi^+$ coupled with isospin relations
\cite{isospin-p} to help constrain $\alpha$.

The LHCb experiment expects to be able to significantly improve on the determination
of $\alpha$ using the $\rho\pi$ mode. They expect 14,000 events in for an integrated
luminosity of 2 fb$^{-1}$, with a signal to background ratio greater than 1
\cite{lhcb-alpha}. It is expected that this amount of data can be accumulated in one
to two years of normal LHC operation.

\begin{figure}[htb]
\centerline{\epsfig{figure=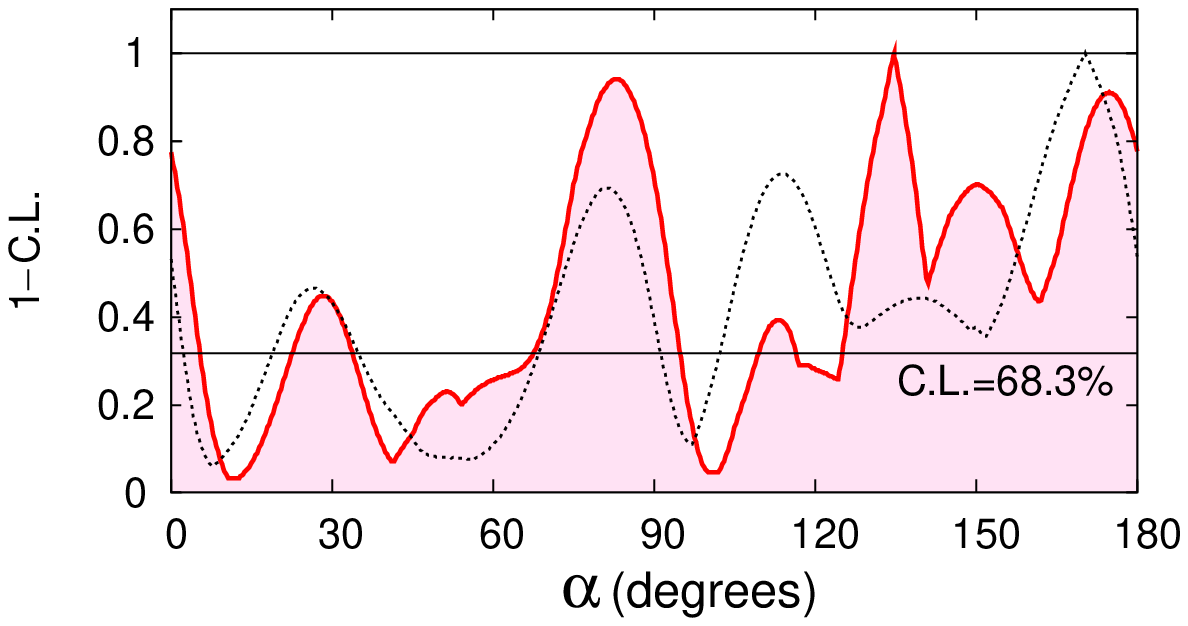,height=1.8in}\hspace
{5mm}\vspace{-5mm} \epsfig{figure=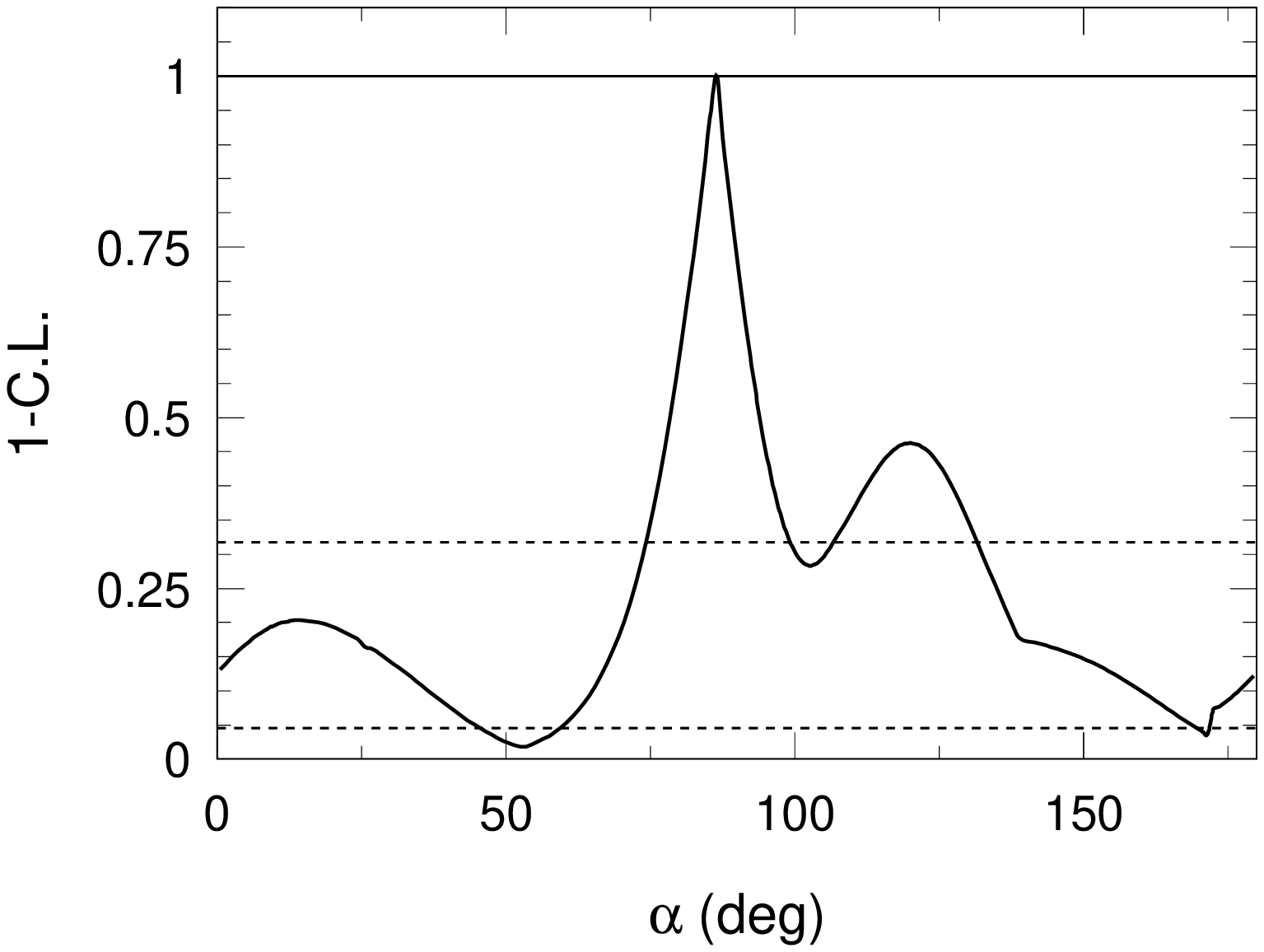,height=1.9in} } 
\caption{The experimental confidence levels for $\alpha$ as determined separately by
the Belle collaboration (left) and the BaBar collaboration (right). The dotted curve
for Belle shows the result without using constraints from charged $B$ to $\rho\pi$
final states.} \label{alpha-rhopi}
\end{figure}

Combining all the data, both the CKM fitter and UT fit groups derive a the confidence
level plot for $\alpha$ shown in Figure~\ref{ckm_alpha_all_08}. There is a clear
disagreement between the two groups, UT fit preferring a solution in the vicinity of
160$^{\circ}$, and CKM fitter a value closer to 90$^{\circ}$. The CKM fitter
group believes that this is due to the UT fit group's use of Bayesian statistics which
they criticize \cite{controversy}.
\begin{figure}[htb]
\centerline{\epsfig{figure=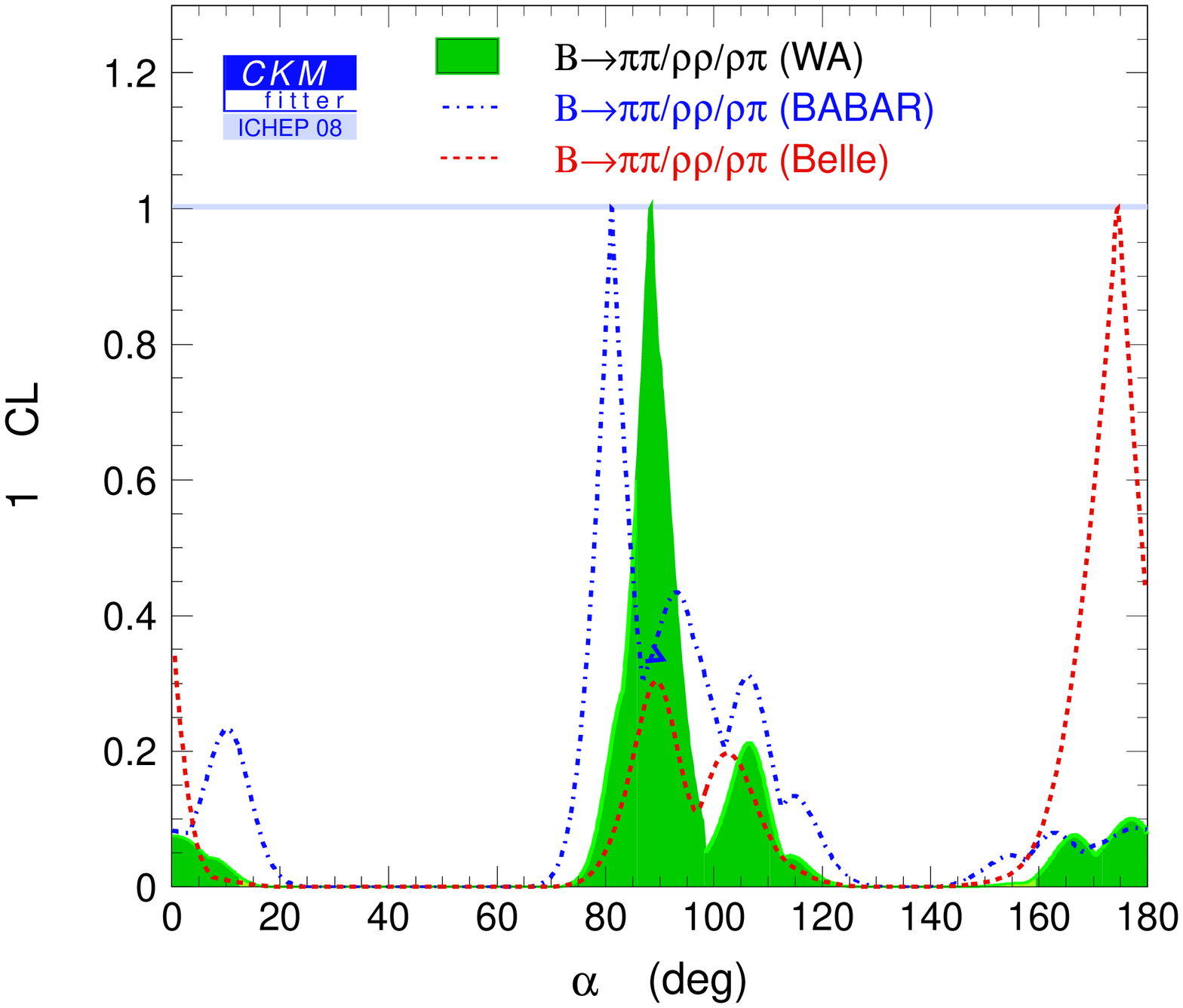,height=2.5in}\hspace
{5mm}\vspace{-0.1mm} \epsfig{figure=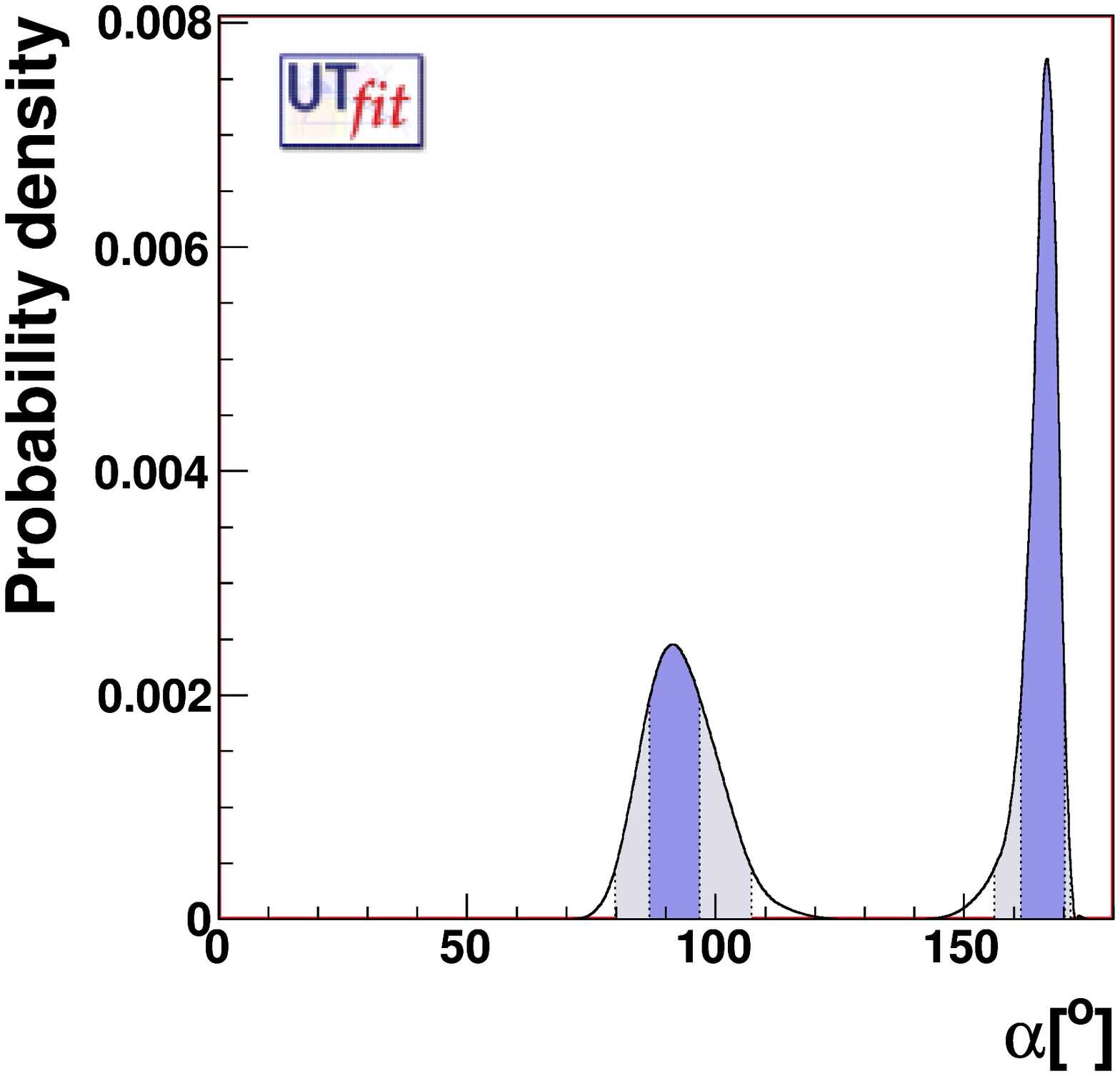,height=2.5in} } \vspace{-0.4cm}
\caption{The experimental confidence levels for $\alpha$ as determined separately by
the CKM fitter and UT fit groups.} \label{ckm_alpha_all_08}
\end{figure}


\subsubsection{The Angle {$\gamma$}}

The angle $\gamma={\rm arg}\left[-\frac{V_{ud}V^*_{ub}}{V_{cd}V^*_{cb}}\right].$ It
can be viewed as the phase of the $|V_{ub}|$ matrix element, with respect to
$|V_{cb}|$.
  Interference measurements are required to determine phases. We can use
  neutral or even charged $B$ decays to measure $\gamma$, and there are several ways
  to do this without using any theoretical assumptions, as is the case for $\alpha$
  and $\beta$. The first relies on the color suppressed tree diagrams shown in
  Figure~\ref{D0K}, and the second on using the interference in $B_s$ mixing. At first
  glance, the diagrams in Figure~\ref{D0K} don't appear to have any interfering
  components. However, if we insist that the final state is common to $D^0$ and
  $\overline{D}^0$, then we have two diagrams which differ in that one has a $b\to c$
  amplitude and the other a $b\to u$ amplitude, the relative weak phase is then
  $\gamma$. Explicitly the amplitudes are defined as
 \begin{eqnarray}\label{eq:gamma}
 A(B^-\to D^0K^-) &\equiv &A_B \\\nonumber
 A(B^-\to \overline{D}^0K^-) &\equiv &A_Br_Be^{i(\delta_B-\gamma)},
 \end{eqnarray}
 where $r_B$ reflects the amplitude suppression for the $b\to u$ mode
 and $\delta_B$ is the relative phase. We have not yet used identical final states for
 $D^0$ and $\overline{D}^0$ decays, yet we can see that there will be
 an ambiguity in our result between $\delta_B$ and $\gamma$ that could be resolved by
 using different $D^0$ decay modes.

\begin{figure}[htb]
\centerline{\epsfig{figure=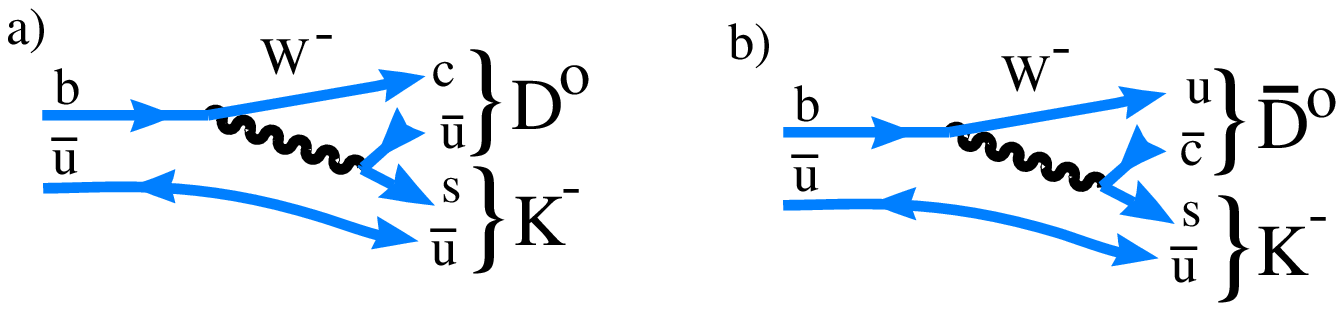,height=2.5in} } \vspace{-2.6cm}
\caption{Two diagrams for a charged $B$ decay into a neutral $D$ meson and a charged
kaon.} \label{D0K}
\end{figure}

There are several suggestions as to different $D^0$ decay modes to use. In the
original paper on this topic Gronau and Wyler \cite{GW} propose using $CP$ eigenstates
for the $D^0$ decay, such as $K^+K^-$, $\pi^+\pi^-$ etc.., combining with charge
specific decays and comparing $B^-$ with $B^+$ decays. In the latter, the sign of the
strong phase is flipped with respect to the weak phase. In fact modes such as $D^{*0}$
or $K^{*-}$ can also be used. (When using $D^{*0}$ there is a difference in $\delta_B$
of $\pi$ between $\gamma D^0$ and $\pi^0 D^0$ decay modes \cite{BG}.)

It is convenient to define the follow variables:
\begin{eqnarray}
A_{CP\pm} &=& \frac{\left[\Gamma(B^- \to D^{0(*)}_{CP\pm}K^{-(*)}) -\Gamma(B^+ \to
D^{0(*)}_{CP\pm}K^{+(*)})\right] }{\left[\Gamma(B^- \to D^{0(*)}_{CP\pm}K^{-(*)})
+\Gamma(B^+ \to D^{0(*)}_{CP\pm}K^{+(*)})\right]},\\\nonumber
R_{CP\pm}&=&\frac{\left[\Gamma(B^- \to D^{0(*)}_{CP\pm}K^{-(*)}) +\Gamma(B^+ \to
D^{0(*)}_{CP\pm}K^{+(*)})\right]}{\left[\Gamma(B^- \to D^{0(*)}K^{-(*)}) +\Gamma(B^+ \to
\overline{D}^{0(*)}K^{+(*)})\right]/2}.
\end{eqnarray}

These are related to the variables defined in Eq.~\ref{eq:gamma} as
\begin{eqnarray}
R_{CP\pm}&=&1+r_B^2\pm 2r \cos\delta_B\cos\gamma, \\\nonumber A_{CP\pm}&=&\pm 2 r
\sin\delta_B\sin\gamma/R_{CP\pm}.
\end{eqnarray}

Measurements have been made by the BaBar, Belle and CDF collaborations
\cite{GLW-gamma}. These data, however, are not statistically powerful enough by
themselves to give a useful measurement of $\gamma$. Atwood, Dunietz and Soni
suggested using double-Cabibbo suppressed decays as an alternative means of generating the
interference between decay amplitudes \cite{ADS}. For example the final state $B^-\to
D^0 K^-$ can proceed via the tree level diagram in Figure~\ref{Bdiagrams2}(a) when the
$W^-\to \bar{u}s$. Then if we use the doubly-Cabibbo suppressed decay $D^0\to K^+\pi^-$, we
get a final state that interferes with the $\overline{D}^0K^-$ final state with the
$\overline{D}^0\to K^+\pi^-$, which is Cabibbo allowed. The advantage of this method,
besides extending the number of useful final states, is that both amplitudes are closer
to being equal than in the above method, which can generate larger interferences.
Any doubly-Cabibbo suppressed decay can be used. Similar equations exist relating the
measured parameters to $\gamma$ and the strong phase shift. Measurements have been
made mostly using the $K^{\pm}\pi^{\mp}$ final state \cite{ADS-gamma}. Again, these
attempts do not yet produce accurate results.

Thus far, the best method for measuring $\gamma$ uses the three-body final state $K_S
\pi^+\pi^-$. Since the final state is accessible by both $D^0$ and $\overline{D}^0$
decays interference results. By its very nature a three-body state is complicated, and
consists of several resonant and non-resonant parts. It is typically analyzed by
examining the Dalitz plot where two of the possible three sub-masses (squared) are
plotted versus one another \cite{Dalitz}. In this way, the phase space is described as
a uniform density and thus any resonant structure is clearly visible. For our
particular case a practical method was suggested by Giri \etal~\cite{Giri}, where the
decay is analyzed in separate regions of the Dalitz plot. Results from this analysis
have been reported by BaBar \cite{babar-dalitz} and Belle \cite{belle-dalitz}. BaBar
finds $\gamma=(76\pm 22\pm 5 \pm 5)^{\circ}$, and Belle finds
$\gamma=(76^{+12}_{-13}\pm 4\pm 5)^{\circ}$.  In both cases the first error is
statistical, the second systematic, and the third refers to uncertainties caused by
the Dalitz plot model. (There is also a mirror solution at $\pm 180^{\circ}$.)

It has been shown \cite{Bondar} that measurement of the amplitude magnitudes and
phases found in the decays of $\psi(3770)\to {\rm (CP\pm~Tag)} (K_S\pi^+\pi^-)_D$
provide useful information that help the narrow model error. The CLEO collaboration is
working on such an analysis, and preliminary results have been reported \cite{Naik}.
(CLEO also is working on incorporating other $D^0$ decay modes.)

Both the CKM fitter and UT fit groups have formed liklihoods for $\gamma$ based on
the measured results, as shown in Figure~\ref{gamma-avg}. In this case CKM fitter and UT
fit agree on the general shape of the liklihood curve.  The CKM fitter plot shows a
small disagreement between the Daltiz method and a combination of the other two
methods.

\begin{figure}[htb]
\centerline{\epsfig{figure=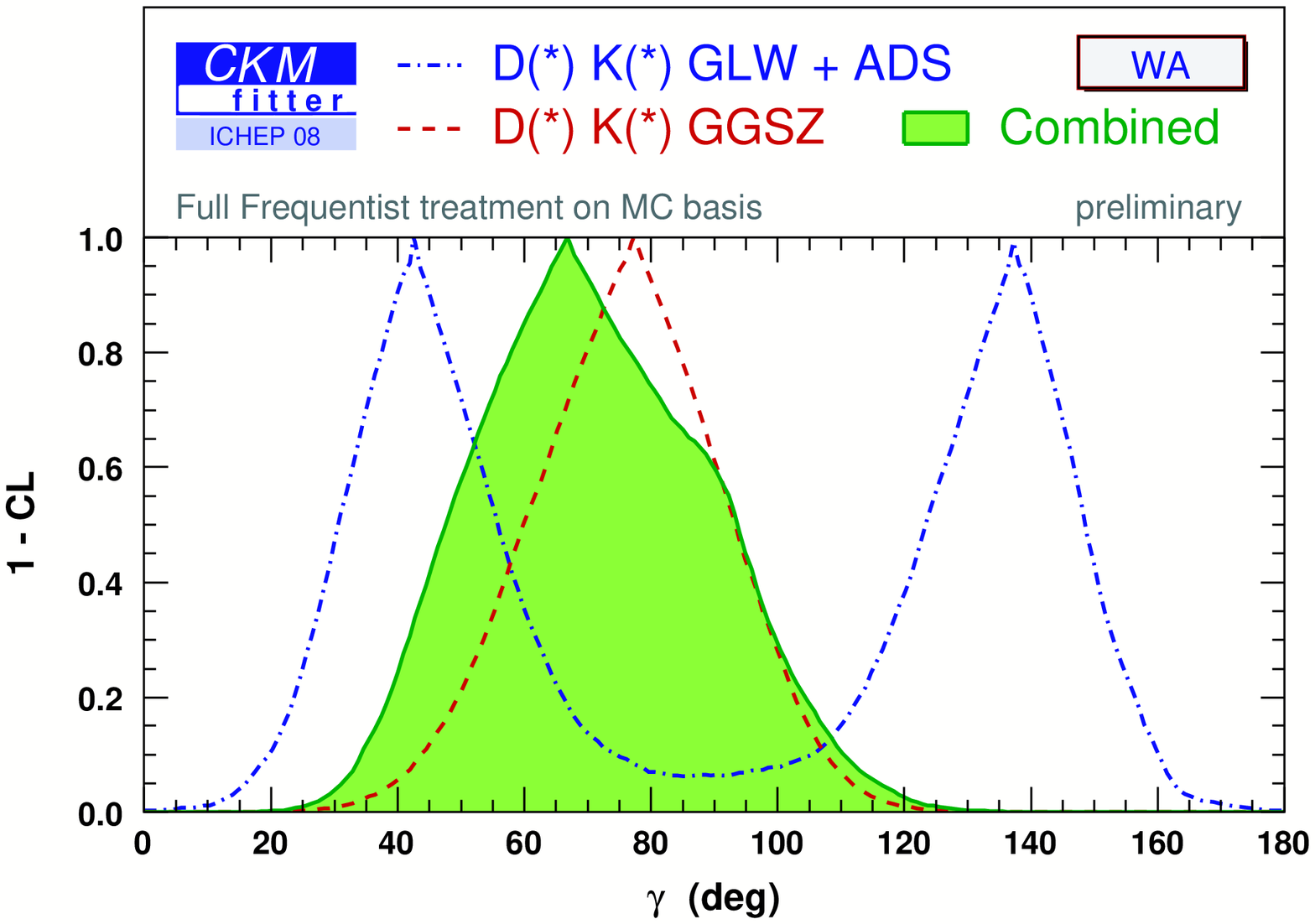,height=2.5in}\hspace
{5mm}\vspace{-5mm} \epsfig{figure=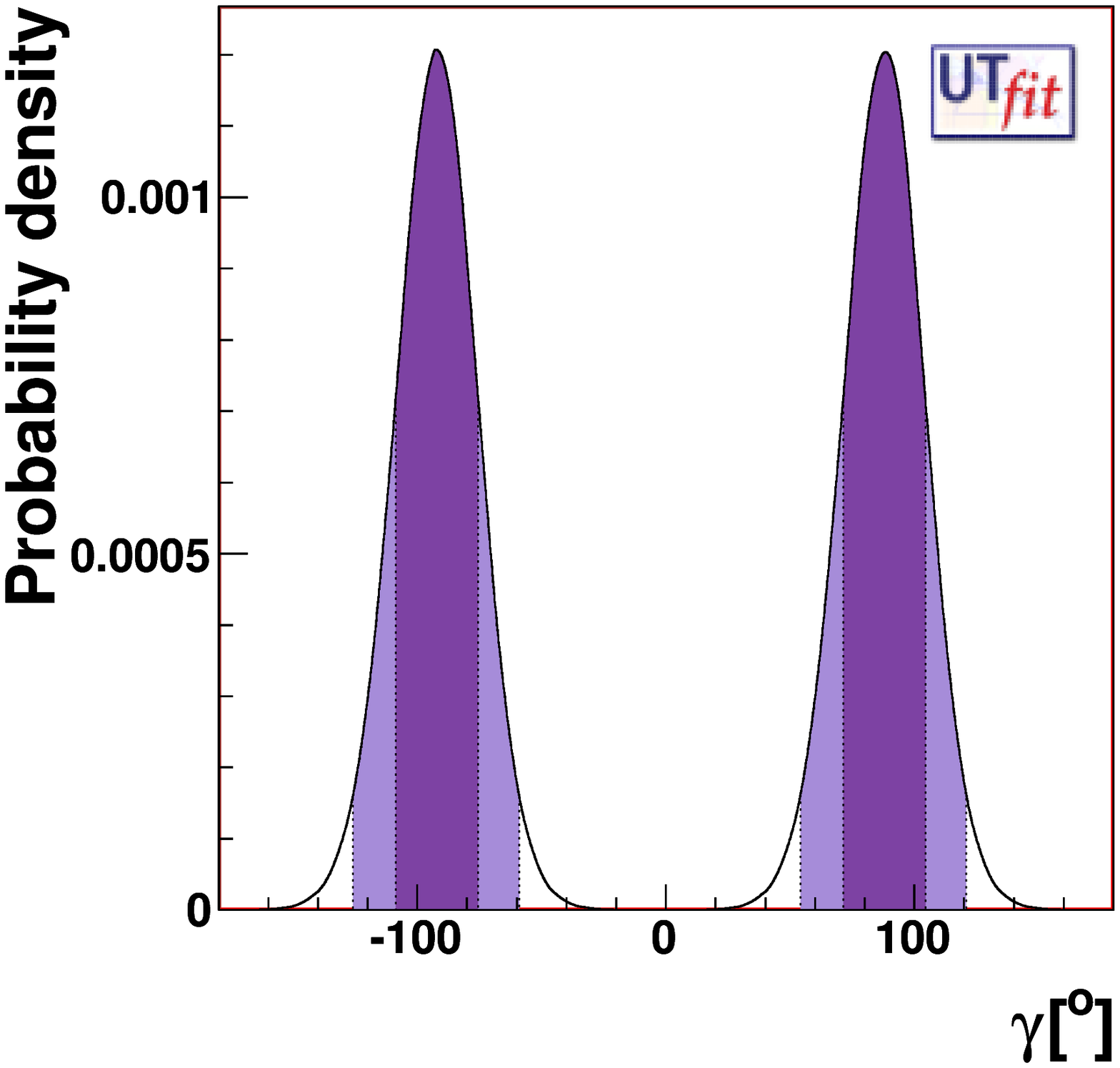,height=2.5in} } 
\caption{The experimental confidence levels for $\gamma$ as determined separately by
the CKM fitter and UT fit groups.} \label{gamma-avg}
\end{figure}

\subsubsection{The Angle {$\chi$}}
 \label{sec:transversity}
The angle $\chi$ shown in Figure~\ref{six_tri} is the phase that appears in the box
diagram for $B_s$ mixing, similar to the diagram for $B^0$ mixing shown in
Figure~\ref{bmix}, but with the $d$ quark replaced by an $s$ quark. The analogous mode
to $B^0\to J/\psi K_s$ in the $B_s$ system is $B_s\to J/\psi\eta$. The Feynman
diagrams are shown in Figure~\ref{psi_ssbar}. This is very similar to measuring $\beta$
so $\chi$ is often called $\beta_s$.\footnote{Note that $\phi_s \ne -2\chi$, since
$-2\chi={\rm arg}(V_{tb}V_{ts}^*)^2/(V_{cb}V_{cs}^*)^2$, whereas
$\phi_s$ is ${\rm arg}(M_{12}/\Gamma_{12})$, with ${\rm arg}(M_{12})= V_{tb}V_{ts}^*)^2/(V_{cb}V_{cs}^*)^2$,
and ${\rm arg}(\Gamma_{12})$ is a linear combination of $(V_{cb}V_{cs}^*)^2$,
$V_{cb}V_{cs}^*V_{ub}V_{us}^*$, and $(V_{ub}V_{us}^*)^2.$}

  \begin{figure}[ht]
  \begin{center}
  \includegraphics[width=3in]{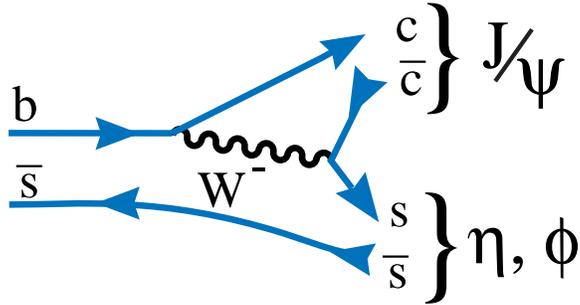}
  \caption[] {
   The Feynman diagram for the decay $B_s\to J/\psi\eta{\rm ~or~}\phi$.
  }
  \label{psi_ssbar}
  \end{center}
  \end{figure}

Since there are usually two photons present in the $\eta$ decay, experiments at hadron
colliders, which can perform time-dependent studies of $B_s$ mesons, preferentially use the $J/\psi\phi$
final state.  This, unfortunately, introduces another complexity into the problem; as
the $B_s$ is spinless the total angular momentum of the final state particles must be
zero. For a decay into a vector and scalar, such as $J/\psi\eta$, this forces the
vector $J/\psi$ to be fully longitudinally polarized with respect to the decay axis.
For a vector-vector final state both angular momentum state vectors are either longitudinal (L), both
are transverse with linear polarization vectors parallel ($\|$) or they are perpendicular
($\bot$) to one another \cite{Rosner1}. Another way of viewing this is that a spin-0
$B$ decay into two massive vector mesons can form $CP$ even states with L=0 or 2, and a
$CP$ odd state with L=1. The relative populations in the two $CP$ states are determined by
strong interactions dynamics, but to study the weak phase here we are not particularly
interested in the actual amount, unless of course one state dominated. We do not
expect this to be the case however, since the SU(3) related decay $B^0\to J/\psi
K^{*0}$, $K^{*0}\to K^+\pi^-$ has a substantial components of both $CP$ states; the PDG
quotes gives the longitudinal fraction as $(80\pm8\pm 5)$\% \cite{PDG}.

The even and odd $CP$ components can be disentangled by measuring the appropriate
angular quantities of each event.  Following Dighe \etal~\cite{Rosner2}, we can
decompose the decay amplitude for a $B_s$ as
\begin{equation}
A(B_s\to J/\psi\phi)=A_0(m_{\phi})/E_{\phi}{\bep}^{*L}_{J/\psi}
-A_{\|}\bep^{*T}_{J/\psi}/\sqrt{2} -iA_{\bot}{\bep}^*_{\phi}\cdot\hat{\bf
p}/\sqrt{2},
\end{equation}
where ${\bep}_{J/\psi}$ and ${\bep}_{\phi}$ are polarization 3-vectors in the $J/\psi$
rest frame, $\hat{\bf p}$  is a unit vector giving the direction of the $\phi$
momentum in the $J/\psi$ rest frame, and $E_{\phi}$ is the energy of the $\phi$ in the
$J/\psi$ rest frame. We note that the corresponding amplitude for the $\bar B_s$ decay
are $\bar A_0 = A_0$, $\bar A_\| = A_\|$, and $\bar A_\bot = - A_\bot$. The amplitudes
are normalized so that
\begin{equation}
d \Gamma(B_s \to J/\psi \phi)/dt = |A_0|^2 + |A_\||^2 + |A_\bot|^2~~~.
\end{equation}

The $\phi$ meson direction in the $J/\psi$ rest frame defines the $\hat x$ direction. The $\hat z$ direction is perpendicular to the decay plane of the $K^+ K^-$ system , where $p_y(K^+)\ge 0.$ The decay direction of the $\ell^+$ in the $J/\psi$ rest frame is described by the angles $(\theta,~\phi)$. The angle $\psi$ is that formed by the $K^+$ direction with the $\hat x$-axis in the $\phi$ rest frame. Figure~\ref{transversityf} shows the angles.

\begin{figure}[ht]
 \begin{center}
   \includegraphics[width=4.5in]{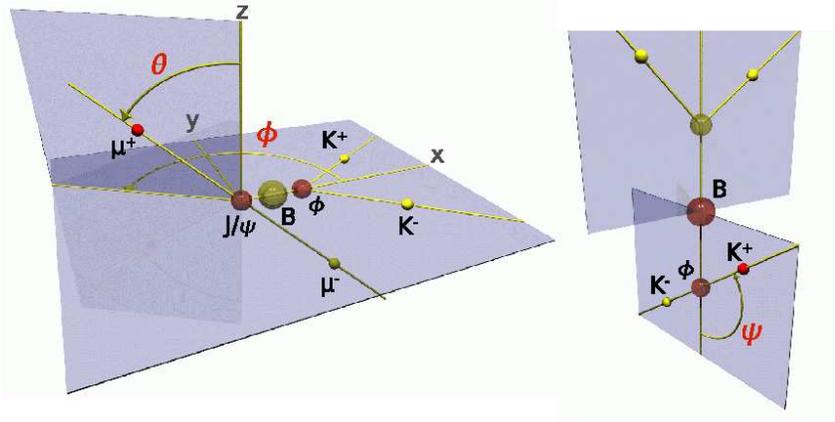}
   \caption[] {
    Pictoral description of the decay angles. On the left $\theta$ and $\phi$ defined
    in the $J/\psi$ rest frame and on the right $\psi$ defined in the
    $\phi$ rest frame. (From T. Kuhr \cite{trans-fig}.).
   }
   \label{transversityf}
   \end{center}
   \end{figure}

The decay width can be written as $$ \frac{d^4 \Gamma [B_s \to (\ell^+\ell^-)_{J/\psi}
(K^+K^-)_{\phi}]} {d \cos \theta~d \phi~d \cos \psi~dt} = \frac{9}{32 \pi} [2 |A_0|^2
\cos^2 \psi (1 - \sin^2 \theta \cos^2 \phi)
$$
$$
+ \sin^2 \psi \{ |A_\parallel|^2  (1 - \sin^2 \theta \sin^2 \phi) + |A_\perp|^2 \sin^2
\theta - {\rm Im}(A_\parallel^* A_\perp) \sin 2 \theta \sin \phi \}~~~
$$
\begin{equation} \label{eqn:threeangle}
+\frac{1}{\sqrt{2}}\sin 2 \psi \{{\rm Re}(A_0^* A_\parallel) \sin^2 \theta \sin 2 \phi
+ {\rm Im}(A_0^* A_\perp) \sin 2 \theta \cos \phi \} ]~~~.
\end{equation}
The decay rate for $\overline{B}_s$ can be found by replacing $A_{\bot}$ in the above
expression with $-A_{\bot}$.

Another complexity arises from the expectation that the width difference
$\Delta\Gamma_s/\Gamma_s\approx 15$\%. This complicates the time dependent rate
equations. For convenience, setting $\overrightarrow{\rho}\equiv(\cos\theta,\phi,\cos\psi)$, we have for the decay width for $B_s$:
\begin{eqnarray}
\frac{d^4 P(t,\vec\rho)}{dtd\vec\rho}
&\propto& |A_0|^2 {\cal T}_+ f_1(\vec\rho) + |A_\parallel|^2 {\cal T}_+ f_2(\vec\rho) \nonumber \\
&+& |A_\perp|^2 {\cal T}_- f_3(\vec\rho) + |A_\parallel||A_\perp| {\cal U}_+ f_4(\vec\rho) \nonumber \\
&+& |A_0||A_\parallel| \cos(\delta_{\parallel}) {\cal T}_+ f_5(\vec\rho) \nonumber \\
&+& |A_0||A_\perp| {\cal V}_+ f_6(\vec\rho), \label{eqn:Bs} 
\label{eq:timedepangles}
\end{eqnarray}
where
\begin{eqnarray*}
{\cal T}_\pm &=& \left[(1\pm\cos(2\chi))e^{-\Gamma_Lt}+(1\mp\cos(2\chi))e^{-\Gamma_Ht}\right]/2, \\
{\cal U}_{\pm} &=&\pm e^{-\Gamma t}\times\left[\sin(\delta_{\perp}-\delta_{\parallel})\cos(\Delta m_st) \right.\\
               &-&\cos(\delta_{\perp}-\delta_{\parallel})\cos(2\chi)\sin(\Delta m_st)                \\
               &\pm&\left.\cos(\delta_{\perp}-\delta_{\parallel})\sin(2\chi)\sinh(\Delta\Gamma t/2)
                  \right], \nonumber\\
{\cal V}_{\pm} &=&\pm e^{-\Gamma t}\times\left[\sin(\delta_{\perp})\cos(\Delta m_st) \right.\\
               &-&\cos(\delta_{\perp})\cos(2\chi)\sin(\Delta m_st)  \\
               &\pm&\left.\cos(\delta_{\perp})\sin(2\chi)\sinh(\Delta\Gamma t/2)\right].\\
 f_1(\vec{\rho})&=& 2\cos^2\psi(1-\sin^2\theta \cos^2\phi), \\
 f_2(\vec{\rho})&=& \sin^2\psi(1-\sin^2\theta \sin^2\phi), \\
 f_3(\vec{\rho})&=& \sin^2\psi \sin^2\theta, \\
 f_4(\vec{\rho})&=& -\sin^2\psi \sin(2\theta) \sin\phi,
\\
 f_5(\vec{\rho})&=& \sin(2\psi) \sin^2\theta \sin(2\phi) / \sqrt{2}, \\
 f_6(\vec{\rho})&=& \sin(2\psi) \sin(2\theta) \cos\phi / \sqrt{2}.
\end{eqnarray*}

The quantities $\delta_{\bot}$ and $\delta_{\|}$ are the strong phases of $A_\bot$ and
$A_{\|}$ relative to $A_0$, respectively \cite{DDF}. The expression for $\overline{B}_s$ mesons can be found by substituting ${\cal U_+}\ra{\cal U_-}$ and ${\cal
V_+}\ra{\cal V_-}$.

The most interesting quantities to be extracted from the data are $\chi$ and $\Delta\Gamma$. There are many experimental challenges: the angular and lifetime distributions must be corrected for experimental acceptances; flavor tagging efficiencies and dilutions must be evaluated; backgrounds must be measured.
Both the CDF \cite{CDF-chi} and D0 \cite{D0-chi} experiments have done this complicated analysis. Updated results as of this writing are summarized by the CKM fitter derived limits shown in Figure~\ref{Charles-phi_s}.

\begin{figure}[htb]
\begin{center}
\centerline{\epsfig{figure=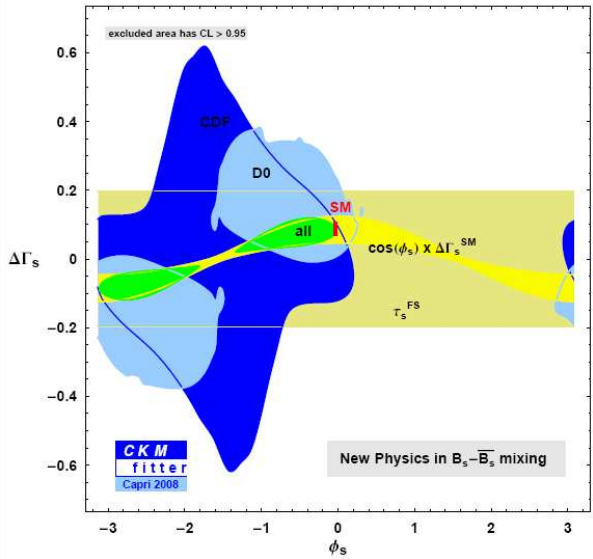,height=3in}}
   \caption{
    Constraints at 68\% confidence level in the ($\phi_s$, $\Delta\Gamma_s$)
plane. Overlaid are the constraints from the CDF and D0 measurements, the constraint from $\Delta\Gamma_s=\cos\phi_s\Delta\Gamma_s^{SM}$, the constraint
from the flavor specific $B_s$ lifetime \cite{HFAG}, and the
overall combination. The SM prediction is also given. From \cite{Charles}.
   }
   \label{Charles-phi_s}
   \end{center}
   \end{figure}

The Standard Model allowed region is a very thin vertical band centered near zero at $\phi_s$
of -0.036$\pm$0.002 (shown in red).  The region labeled ``all" (green) shows the
allowed region at 68\% confidence level. Although the fit uses several input components besides the $CP$ asymmetry measurements in $B_s\to J/\psi\phi$, including
use of measured total widths introduced via the constraint equation $\Delta\Gamma_s=\cos\phi_s\Delta\Gamma_s^{SM}$, it is the measurement of $\phi_s$ that dominates the fit result. We see that there is a discrepancy that may be as
large as 2.7 standard deviations \cite{Charles}. While this is as not yet significant, it is
very tantalizing. The LHCb experiment plans to vastly improve this measurement \cite{Blouw}.

It has been pointed out, however, that there is likely an S-wave $K^+K^-$ contribution in the region of the $\phi$ that contributes 5-10\% of the event rate as estimated using $D_s^+$ decays \cite{Stone-Zhang}. For example, analysis of the analogous channel $B^0\to J/\psi K^{*0}$ reveals about 8\% S-wave in the $K\pi$ system under the $K^{*0}$, and BaBar has used this to extract a value for $\cos 2\beta$ thus removing an ambiguity in $\beta$ \cite{BaBar-psiKstar}. The S-wave amplitude and phase needs to be added to Eq.~\ref{eq:timedepangles}. Note that the errors will increase due to the addition of another amplitude and phase. The S-wave can manifest itself as a $\pi^+\pi^-$, so it is suggested that the decay $B_s\to J/\psi f_0(980)$ be used to measure $\chi$; here the $f_0\to\pi^+\pi^-$ \cite{Stone-Zhang}; the estimate is that the useful $f_0$ rate would be about 20\% of the $\phi$ rate, but the $J/\psi f_0$ is a $CP$ eigenstate, so an an angular analysis is unnecessary, and these events may provide a determination of $\chi$ with an error comparable to that using $J/\psi\phi$.

\subsubsection{Measurements of Direct {\it\textbf CP} Violation in {$B\to K\pi$} Decays}
Time integrated asymmetries of $B$ mesons produced at the $\Upsilon$(4S) resonance can only be due to direct $CP$ asymmetry, as the mixing generated asymmetry must integrate to zero due to the fact that the initial state has $J^{PC}=1^{--}$. The first evidence for
such direct $CP$ violation at the greater than four standard deviation level in the $K^{\mp}\pi^{\pm}$ final state was given by BaBar \cite{BaBar-Kpi}.
The latest BaBar result is \cite{BaBarlate-Kpi}
\begin{equation}
{\cal{A}}_{K^{-}\pi^{+}}=\frac{\Gamma(\overline{B}^0\to K^-\pi^+)-\Gamma(B^0\to K^+\pi^-)}
{\Gamma(\overline{B}^0\to K^-\pi^+)+\Gamma(B^0\to K^+\pi^-)}=-0.107\pm0.016^{+0.006}_{-0.004}~,
\end{equation}
showing a large statistical significance.

This result was confirmed by the Belle collaboration, but Belle also measured the isospin conjugate mode.
Consider the two-body decays of $B$ mesons into a kaon and a pion, shown in Figure~\ref{KpidiagSS}. For netural and charged decays, it can proceed via a tree level diagram (a) or a Penguin diagram (b). There are two additional decay diagrams allowed for the $B^-$, the color-suppressed tree level diagram (c) and the elusive ``Electroweak" Penguin diagram in (d). (So named because of the intermediate $\gamma$ or $Z$ boson.)
\begin{figure}[htb]
\begin{center}
\centerline{\epsfig{figure=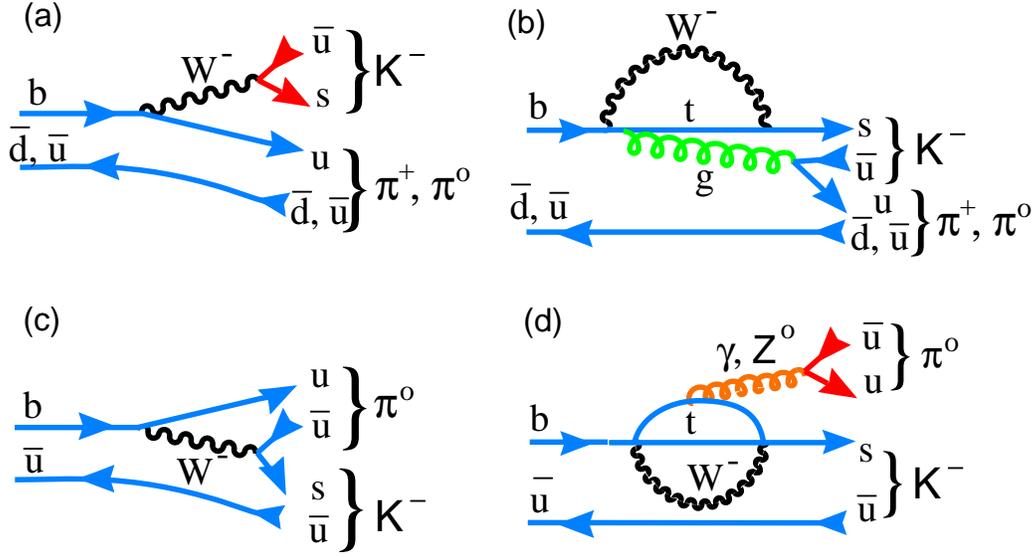,height=3in}}
   \caption{
    Processes for $\overline{B}^0\to K^-\pi^+$ and $B^-\to K^-\pi^0$, (a) via tree and (b) Penguin diagrams, and $B^-\to K^-\pi^0$ (c) via color-suppressed tree and (d)``Electroweak" Penguin diagrams.   }
   \label{KpidiagSS}
   \end{center}
   \end{figure}
Since it is expected that diagrams (c) and (d) are small, the direct $CP$ violating asymmetries in both charged and neutral modes should be the same. Yet Belle observed \cite{Belle-Kpi}
\begin{equation}
{\cal{A}}_{K^{\mp}\pi^{0}}=\frac{\Gamma({B}^-\to K^-\pi^0)-\Gamma(B^+\to K^+\pi^0)}
{\Gamma({B}^-\to K^-\pi^0)+\Gamma(B^+\to K^+\pi^0)}=0.07\pm0.016\pm0.01~.
\end{equation}
The Belle data are shown in Figure~\ref{Kpidata}.
\begin{figure}[htb]
\begin{center}
\centerline{\epsfig{figure=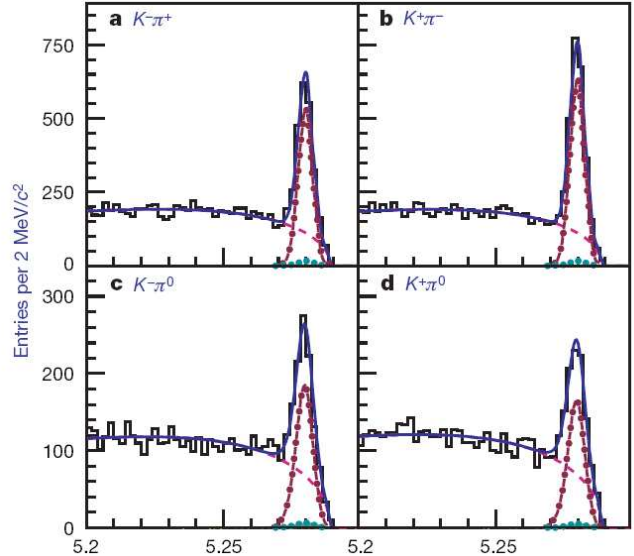,height=3in}}
   \caption{
   The beam constrained mass distributions from Belle in units of GeV for the four different indicated final states. The dotted curves indicate peaking backgrounds, the dot-dashed curves the signal and the solid curves the sum. The differences in event numbers between the charge conjugate modes are apparent. }
   \label{Kpidata}
   \end{center}
   \end{figure}

The difference between ${\cal{A}}_{K^{-}\pi^{+}}$ and ${\cal{A}}_{K^{\mp}\pi^{0}}$ is not naively expected in the Standard Model and Belle suggests that this may be a sign of New Physics. Peskin commented on this possibility \cite{Peskin}.

\subsection{Conclusions from {\it\textbf CP} Violation Measurements}

All $CP$ violation in the quark sector is proportional to the parameter $\eta$ in
Eq.~\ref{eq:CKM}. In fact all $CP$ asymmetries are proportional to the ``Jarlskog
Invariant", $J=A^2\lambda^6\eta$, which represents the equal area of all the CKM
triangles \cite{Jarlskog}. Since we know the value of these three numbers, we do know the amount of $CP$
violation we can expect, even without making the measurements. We also can estimate
the amount of $CP$ violation necessary using cosmology. To reproduce the observed baryon
to entropy ratio requires many orders of magnitude more $CP$ violation than thus far found
in heavy quark decays \cite{BAU}. Thus we believe there are new sources of $CP$ violation that
have not yet been found.

\section{The CKM Parameter {$|V_{cb}|$}}

There are two experimental methods to determine \vcb:  the {\bf exclusive}  method, where \vcb\ is extracted by studying the exclusive
  $\overline{B} \to D^{(*)}\ell^-\bar{\nu}$ decay process; and  the {\bf inclusive}   method, which uses the semileptonic decay width of $b$-hadron decays. In both methods, the extraction of \vcb\ is systematics limited and the dominant errors are from theory.
The inclusive and exclusive determinations of $|V_{cb}|$ rely on different theoretical calculations of the hadronic matrix element needed to extract it from measured quantities,
and  make use of different techniques which, to a large extent, have uncorrelated experimental uncertainties. Thus, the comparison between inclusive and exclusive decays allows us to test our understanding of hadronic effects in semileptonic decays. The latest determinations differ by more than 2$\sigma$, with the inclusive method having a stated error half of the size of the exclusive one.

\subsubsection{Beauty Quark Mass Definitions}\label{sec:quark_masses}
Due to confinement and the non-perturbative aspect of the strong interaction, the concept of the quark masses cannot be tied to an intuitive picture of the rest mass of a particle, as for leptons.  Rather, quark masses must be considered as couplings of the SM Lagrangian that have to be determined from processes that depend on them.  As  such the $b$-quark mass ($m_b$) is a scheme-dependent, renormalised quantity.

In principle, any renormalisation scheme or definition of quark masses is possible.  In the framework of QCD perturbation theory the difference between two mass schemes can be determined as a series in powers of $\alpha_s$.  Therefore, higher-order terms in the perturbative expansion of a quantity that depends on quark masses are affected by the particular scheme employed.  There are schemes that are more appropriate and more convenient for some purposes than others.   Here we examine the main quark mass definitions commonly used in the description of $B$ decays.

\begin{itemize}
\item {\bf Pole mass: }
The pole mass definition is gauge-invariant and infrared-safe~\cite{pole_mass} to all orders in perturbation theory and has been used as the standard mass definition of many perturbative computations in the past.  By construction, it is directly related to the concept of the mass of a free quark.
The presence of a renormalon ambiguity~\cite{infrared} makes the numerical value of the pole mass an order-dependent quantity,  leading to large perturbative corrections for Heavy Quark Effective Theory (HQET) parameters (see below for a discussion of HQET).  These shortcomings  are avoided by using quark mass definitions that reduce the infrared sensitivity by removing the $\Lambda_{\rm QCD}$ renormalon of the pole mass. Such quark mass definitions are generically called ``short-distance" masses.

\item{ \bf $\overline{\rm MS}$ mass: }
The most common short-distance mass definition is the $\overline{\rm MS}$ mass $\overline{m}_b(\mu)$ \cite{msbar}, where
the scale $\mu$  is typically chosen to be the order of the characteristic energy scale of the process.
 In the $\overline{\rm MS}$  scheme the subtracted divergencies do not contain any infrared sensitive terms and the  $\overline{\rm MS}$ mass is only sensitive to scales larger than $m_b$.  The $\overline{\rm MS}$ mass arises naturally in processes where the $b$-quark is far off-shell, but  it  is less adequate  when the $b$-quark has non-relativistic energies.

\item{ \bf Kinetic mass: }
The shortcomings of the pole and the $\overline{\rm MS}$ mass in describing non-relativistic $b$-quarks can be resolved by so-called threshold masses \cite{threshold}, that are free of an ambiguity of order $\Lambda_{\rm QCD}$ and  are defined through subtractions that contain universal contributions for the dynamics of non-relativistic quarks.  Since the subtractions are not unique, an arbitrary number of threshold masses can be constructed. The kinetic mass is defined as~\cite{bigi,ref:2}:
\begin{equation}
m_{b,{\rm kin}}(\mu_{\rm kin}) = m_{b,{\rm pole}} - [\bar \Lambda(\mu_{\rm kin})]_{\rm pert} - \left[ \frac{\mu_{\pi}^2(\mu_{\rm kin})}{2m_{b,{\rm kin}}(\mu_{\rm kin})}\right]_{\rm pert} + \ldots ,
\end{equation}
where $\mu_{\rm kin}$ is the nominal kinetic mass renormalisation scale. 
For $\mu_{\rm kin}\to 0$ the kinetic mass reduces to the pole mass.

\item{\bf {$1S$} mass: }
The kinetic mass depends on an explicit subtraction scale to remove the universal infrared sensitive contributions associated with the non-relativistic $b$-quark dynamics.  The $1S$ mass~\cite{1s_scheme} achieves the same task without a factorisation scale, since it is directly related to a physical quantity.  The $b$-quark $1S$ mass is defined as half of the perturbative contribution to the mass of the $\Upsilon(S_1)$ in the limit $m_b \gg m_b v \gg m_b v^2 \gg \Lambda_{\rm QCD}$.
\end{itemize}

A list of $b$-quark mass determinations, converted into the $\overline{\rm MS}$ mass scheme is shown in Table~\ref{t:publ1mass}.
\begin{table}
\begin{center}
\caption{List of $m_b$ determinations converted into the $\overline{\rm MS}$ mass scheme. \label{t:publ1mass}}
\begin{tabular}{cc} \hline
$m_b(m_b)$ GeV  &Method\\ \hline
$4.243 \pm 0.042$& From $B\to X_s\gamma$ and $B\to X_c\ell\nu$ fit \cite{gardi} \\
$4.25\pm0.02\pm0.11$ & Lattice UKQCD \cite{Lattice_mb} \\
$4.346\pm0.07$ & $\Upsilon(1S)$ NNNLO \cite{penin}\\
$4.164\pm0.025$ &low-moment sum rules NNNLO \cite{khun}\\
$4.17\pm0.05$ & $\Upsilon$ sum rules NNLO \cite{Hoang} \\
 \hline
\end{tabular}
\end{center}
\end{table}

\subsection{Determination Based on Exclusive Semileptonic {\it\textbf B} Decays}

The exclusive \vcb\ determination is obtained by studying the decays
$B\to D^{*} \ell \nu$ and $B\to D \ell \nu$, where $\ell$ denotes either an electron or a muon. The exclusive measurements of a single hadronic final state, e.g. the ground state $D$ or $D^*$, restrict the dynamics of the process. The remaining degrees of freedom,
usually connected to different helicity states of the charmed hadron, can be expressed in terms of form factors, depending on the invariant mass of the lepton-$\nu$ pair, $\qsq$. The shapes of those form factors are unknown but can be measured. However, the overall normalization of these functions needs to be determined from theoretical calculations.

Isgur and Wise formulated a theoretical breakthrough in the late 1980's. They found that in the limit of an
infinitely heavy quark masses QCD possess additional flavor and spin symmetries. They showed that since the heavier $b$ and $c$ quarks have masses much heavier than the scale of the QCD coupling constant, that they are heavy enough to posses this symmetry but that corrections for the fact that the quark mass was not infinite had to be made. They showed there was a systematic method of making these corrections by expanding a series in terms of the inverse quark mass \cite{Isgur-Wise-BBook}. This theory is known as Heavy Quark Effective Theory (HQET).

When studying $b$ quark decays into $c$ quarks, it is convenient to view the process in four-velocity transfer ($w$) space as opposed to four-momentum transfer space, because at maximum four-velocity transfer, where $w$ equals one, the form-factor in lowest order is unity, i.e. the $b$ transforms directly into a $c$ quark without any velocity change.
The value of \vcb\ can be extracted by studying the decay rate for the process \btods\ as a function of the recoil kinematics
of the \dsp\ meson. Specifically, HQET predicts that:
\begin{eqnarray}
\frac{d\Gamma(B\to D^{*} \ell \nu)}{dw} = \frac{G_F^2| V_{cb}| ^2}{48\pi ^3}{\cal K}(w){\cal F}(w)^2,
\end{eqnarray}
where $w$ is the product of the four-velocities of the
\dsp\ and the \bbar\ mesons, related to $q^2$.
${\cal K}(w)$ is a known phase space
factor and the form factor ${\cal F}(w)$ is generally expressed as the product of a normalization
factor  and a shape function described by three form factors,
constrained by dispersion relations \cite{grinstein}:
\begin{eqnarray}
{\cal F}(w)^2 &= &\frac{|h_{A_1}(w)|^2 }{(1-r)^2}\left[ (w-r)^2\left(1-R_2(w)\right)^2 +2(1-2wr-r^2)\left(1-\frac{w-1}{w+1}R^2_2(w)\right)\right], \\
\frac{h_{A_1}(w)}{h_{A_1}(1)} &= &  1 - 8 \rho^2 z +(53 \rho^2-15)z^2 -(231\rho^2-91) z^3,
\nonumber \\
R_1(w)&= &  R_1(1) -0.12(w-1)+0.05(w-1)^2, \nonumber \\
R_2(w)&= &  R_2(1) -0.11(w-1)-0.06(w-1)^2, \nonumber
\end{eqnarray}
where  $r=m_{D^*}/m_B$, $z = \frac{\sqrt{w+1}-\sqrt{2}}{\sqrt{w+1}+\sqrt{2}}$; The linear slope of the form-factor is given by the
parameter $\rho^2$, and must be determined from the data. In the infinite mass limit, ${\cal F}(w=1)= h_{A_1}(w=1)=1$;  for finite quark masses, non-perturbative effects can be expressed in powers of $1/m_Q$.
There are several different corrections to the infinite mass value
${\cal F}(1)=1$:
\begin{equation}
{\cal F}(1) =\eta _{\rm QED}\eta _A \left[ 1 + \delta _{1/m_Q^2} + ...\right]
\end{equation}
Note that the first term
in the non-perturbative expansion in powers of $1/m_Q$ vanishes \cite{Chay:1990da}.
QED corrections up to leading logarithmic order  give $\eta _A = 0.960\pm 0.007$. Different estimates of the
$1/m_Q^2$ corrections, involving terms proportional to $1/m_b^2$,
$1/m_c^2$ and $1/(m_bm_c)$, have been performed in a quark model
with QCD sum rules, and, more recently, with an HQET based lattice
gauge calculation. The best estimate  comes
from lattice QCD, $h_{A_1}(1)={\cal F}(1)=0.921\pm 0.013 \pm 0.020$ \cite{Laiho:2008pn}. This result does not include a 0.7\% QED correction.

 Since the phase-space factor ${\cal K}(w)$ tends to zero
as $w\rightarrow 1$, the decay rate vanishes and the
accuracy of the \vcb\ value extracted with this method depends
upon experimental and theoretical uncertainties in the extrapolation.
Experiments determine the product $|{\cal F}(1)\cdot V_{cb}| ^2$
by fitting the measured ${d\Gamma}/{dw}$ distribution.

 This decay has been analyzed by CLEO, BaBar and Belle using $B$ mesons from the $\Upsilon \rm(4S)$ decay, and by ALEPH, DELPHI, and OPAL at the $\rm Z^0$ center of mass energy.
 Experiments that exploit the $\Upsilon \rm(4S)$ have the advantage that $w$
resolution is quite good. However, they suffer from lower
statistics near $w=1$ in the decay $B\rightarrow  D^{*+} \ell
\nu$ due to the lower reconstruction efficiency of the slow
$\pi^{\pm}$. On the other hand, the decay $B\rightarrow D^{*0}
\ell \nu$ is not affected by this problem \cite{Adam:2002uw}.  In addition, kinematic constraints enable these  experiments to identify the final state including
$D^{*}$ without large contamination from the poorly known portion of
semileptonic decays with a hadronic system recoiling against the
lepton-$\nu$ pair with masses higher than the $D$ and $D^{*}$,
commonly identified as `$D^{**}$'. $B$
-factories and CLEO fit for the signal and background components in the distribution of the
cosine of the angle  between the direction of the
$B$ and the direction of the $D^*\ell$ system.
At LEP,
 $B$ mesons are produced with a large variable momentum  (about 30 GeV on average), giving a relatively poor
$w$ resolution and limited physics background rejection capabilities.
By contrast, LEP experiments benefit from an efficiency that is only
mildly dependent upon $w$.

LEP experiments extracted $|V_{cb}|$ by performing a two-parameter fit, for \fvcb\ and the slope $\rho^2$.
The first measurements of both ratio $R_1$ and $R_2$, and  $\rho^2$, were made by the CLEO collaboration \cite{CLEO-Vcb}. Belle \cite{bellen} and BaBar  \cite{hep-ex/0602023} improved upon these measurements using the  $\overline{B}^0 \to D^{*+}e^-\bar{\nu}_e$ decay.  They determined $R_1$, $R_2$, and $\rho^2$ using an unbinned maximum likelihood fit to the full decay
distribution. BaBar and CLEO results are combined to give:
$R_1=1.396\pm 0.060\pm 0.035\pm 0.027$,
$R_2=0.885\pm 0.040\pm 0.022\pm 0.013$, and $\rho^2=1.145\pm 0.059\pm
0.030\pm 0.035$.  The stated uncertainties are the statistical from
the data and systematic uncertainty, respectively.

Table~\ref{t:publ}
summarizes the available data.
Values of \fvcb\  from different experiments can be combined if
they are extracted using the same $\cal{F} (\it{ w})$
parametrization.
All measurements included in the exclusive $|V_{cb}|$ world average relying on the above form factor ratios $R_1$ and $R_2$.
\begin{table}
\caption{Experimental results based on $B\to D^*\ell \nu$ after the correction to common inputs and the world average as obtained by HFAG \cite{hfag08}. All numbers are corrected to use R$_1$ and R$_2$ from a global fit to CLEO and BaBar data. The newest BELLE result is not included in the world average.
\label{t:publ}}
\begin{center}
\begin{tabular}{lcccc} \hline
experiment & \fvcb\ $(\times 10^{-3})$ & $\rho^2$ & $\rm Corr_{stat}$ & Branching fraction(\%)  \\ \hline
 ALEPH \cite{LEP}  &  31.6$\pm$1.8$\pm$1.3 & 0.50$\pm$0.20$\pm$0.09 & 94\% &5.44$\pm$0.25$\pm$0.20	 \\
BaBar \cite{BaBar(excl)}      &  33.9$\pm$0.3$\pm$1.1 & 1.18$\pm$0.05$\pm$0.03 & 27\% &4.53$\pm$0.04$\pm$0.13\\
BaBar \cite{babar07}      &  34.9$\pm$0.8$\pm$1.4 & 1.11$\pm$0.06$\pm$0.08 & 90\% &5.40$\pm$0.16$\pm$0.25\\
BaBar  \cite{babar08}  &  35.7$\pm$0.2$\pm$ 1.2 & 1.20$\pm$0.02$\pm$0.07 & 38\% &-\\
BELLE\cite{bellen}    &  34.7$\pm$  0.2$\pm$ 1.0 & 1.16$\pm$ 0.04$\pm$ 0.03 & 91\% & ${\dagger}$ \\
BELLE  \cite{belle1}         &  34.7$\pm$ 0.2$\pm$1.0 & 1.16$\pm$0.04$\pm$0.03 & 91\% &4.75$\pm$0.25$\pm$ 0.19\\
 CLEO \cite{cleods}            &  41.3$\pm$1.3$\pm$1.8 & 1.37$\pm$0.08$\pm$0.18 & 91\% &6.02$\pm$ 0.19$\pm$0.20\\
 DELPHI(excl)\cite{LEP}          &  36.3$\pm$1.8$\pm$1.9 & 1.04$\pm$0.14$\pm$0.15 & 89\% &5.53$\pm$0.19$\pm$0.34\\
 DELPHI(part rec)\cite{LEP}      &  35.8$\pm$1.4$\pm$2.3 & 1.18$\pm$0.13$\pm$ 0.25 & 94\% &5.00$\pm$0.15$\pm$0.18	 \\
 OPAL(excl)\cite{LEP}   &  36.9$\pm$1.6$\pm$1.5 & 1.24$\pm$0.20$\pm$0.14 & 77\% &5.17$\pm$ 0.20$\pm$0.38\\
 OPAL(par rec)\cite{LEP}  &  37.6$\pm$1.2$\pm$2.4 & 1.14$\pm$0.13$\pm$0.27 & 94\% & 5.63$\pm$0.27$\pm$ 0.43\\
  \hline
  world average &  $35.4 \pm 0.5$ & $1.16\pm 0.05$  & 20\%& 5.05 $\pm$ 0.10 \\
    \hline
\end{tabular}
${\dagger}$ - Not used in the World Average
\end{center}
\end{table}
Using the  \fvcb\ world average in Table \ref{t:publ} and $ h_{A_1}(1) = 0.921 \pm 0.013\pm0.020$ \cite{Laiho:2008pn}, we find
$$ |V_{cb}| = (38.2 \pm 0.5_{exp} \pm 1.0_{theo}) \times 10^{-3},$$
where the dominant error is theoretical, and it will be difficult to improve upon.

The study of the decay $ B \to D \ell \nu$ poses new challenges
from the experimental point of view.  The differential decay rate for $B\to D \ell \nu$ can be
expressed as \cite{CLN}
\begin{equation}
\frac{d\Gamma_D}{dw} (B\to D\ell\nu)= \frac{G_F^2|
V_{cb}|^2}{48\pi^3}{\cal K_D}(w){\cal G}(w)^2,
\end{equation}
where
${\cal K_D}(w)$ is the phase space factor and the form factor ${\cal G}(w)$
is generally expressed as the product of a normalization factor
${\cal G}(1)$ and a function, $g_D(w)$, constrained by dispersion
relations \cite{grinstein}.
The strategy to extract $\rm {\cal G}(1)|V_{cb}|$ is identical to
that used for the $B\to D^{\star} \ell \nu$ decay. However,
${d\Gamma_D}/{dw}$ is more heavily suppressed near $w=1$ than
${d\Gamma_{D^*}}/{dw}$ due to the helicity mismatch between
initial and final states. Moreover, this channel is much more
challenging to isolate from the dominant background $B\to
D^\star \ell \nu$ as well as from fake $D$-$\ell$
combinations.
Table \ref{t:publ1} shows the results of two-dimensional fits to $|V_{cb}| {\cal G}(1)$ and $\rho^2$ for different experiments and the world average.
\begin{table}
\caption{Experimental results from $B\to D\ell\nu$ using corrections to the common inputs and world averages. \label{t:publ1}}
\begin{center}
\begin{tabular}{lcccc} \hline
experiment &$\rm {\cal G}(1)|V_{cb}| (\times 10^{-3})$ & $\rho_D^2$ & $\rm Corr_{stat}$ & Branching fraction(\%) \\ \hline
 ALEPH\cite{LEP}    &  38.1$\pm$11.8$\pm$6.1 & 0.91$\pm$0.98$\pm$ 0.36 & 98\% & $2.25\pm0.18\pm0.36$\\
BaBar \cite{bard}        &  42.3$\pm$1.9$\pm$1.0 & 1.20$\pm$0.09$\pm$0.04 & 95\% &-\\
BaBar \cite{bard1}  &  43.8$\pm$0.8$\pm$2.1 & 1.22$\pm$0.04$\pm$0.06 & 63\% &-\\
BaBar \cite{bard2} &  - & -& - & $2.20\pm0.11\pm0.12$\\
BELLE \cite{bd}         &  40.7$\pm$4.4$\pm$5.1 & 1.12$\pm$0.22$\pm$0.14 & 96\% &$2.09\pm0.12\pm0.39$\\
 CLEO \cite{cd}        &  44.5$\pm$5.9$\pm$3.4 & 1.27$\pm$0.25$\pm$0.14 & 95\% &$2.10\pm 0.13\pm0.15$\\
  \hline
  World average &  $42.4 \pm 0.7 \pm 1.4$ & $1.19\pm 0.04 \pm0.04$  & 96\% &$2.16\pm0.12$ \\
    \hline
\end{tabular}
\end{center}
\end{table}

In the limit of infinite quark masses, ${\cal G}(w=1)$ coincides with the
Isgur-Wise function~\cite{IWS}. In this case there is no suppression of
$1/m_Q$ corrections and QCD effects on ${\cal G}(1)$ are
calculated with less accuracy than ${\cal F}(1)$.
Corrections to this prediction have recently been calculated with improved precision, based on unquenched lattice QCD~\cite{Okamoto:2005hr}, specifically
${\cal G}(1) = 1.074\pm 0.018\pm 0.016$.  Using this result  we get
$$ |V_{cb}|=(39.5\pm1.4_{exp}\pm0.9_{theo})\times10^{-3}$$consistent with the value extracted from $B\to D^{\star} \ell \nu$ decay,
but with an experimental uncertainty about twice as large.

BaBar recently has also studied the differential decay widths for the decays $B^-\to \Dz\ell\nu$
 and $B^-\to D^{\star 0}\ell\nu$ to extract the ratio ${\cal G}(1)/ {\cal F}(1)= 1.23\pm 0.09$ \cite{Aubert:2008yv}, compatible with the lattice theory prediction of 1.16 $\pm$ 0.04 .

\subsection{ $B \to D^{**} \ell\nu$ Decays}
It is important to understand the composition of the inclusive $B$ semileptonic decay rate in terms of exclusive final states for use in semileptonic $B$ decay analyses.
The  $B \to D^{(*)} \ell \nu$  decays  are well measured, but a sizeable fraction of semileptonic  $B$ decay are to
$D^{**} \ell \nu$. The $D^{**}$ resonances have larger masses than the $D^*$ and not well studied.\footnote{
$D^{**}$ refers to the resonant states, $D^*_0$, $D^*_1$, $D^*_2$ and $D^{}_1$.} The measurements of these branching fractions require a good understanding  of the different $D^{(*)}n\pi$ systems, which can be either resonant or non-resonant, each with characteristic decay properties.

 There are four orbitally excited states with L=1.  They can be grouped in two pairs according to the value of the spin on the light system, $ j = $L$ \pm 1/2$~ (L=1).
States with $j = 3/2$ can have $\rm{J}^{\rm{P}}= 1^+ $and $2^+$ state.  The $1^+$ state decays only through $D^*\pi$, and the $2^+$ through $D\pi$ or $D^*\pi$.  Parity and angular momentum conservation imply that in the $2^+$ the $D^*$ and $\pi$ are in a D wave but allow both S and D waves in the $1^+$ state.
However, if the heavy quark spin is assumed to decouple, conservation of $j = 3/2$ forbids S waves even in the $1^+$ state. A large D-wave component and the fact that the masses of these states are not far from threshold imply that the $j = 3/2$ states are narrow.  These states have been observed with a typical width of 20 MeV/$c^2$.   On the contrary, $j = 1/2$  states can have $\rm{J}^{\rm{P}}= 0^+ $and $1^+$, so they are expected to decay mainly through an S wave and manifest as broad resonances, with typical widths of several hundred MeV/$c^2$.

  The ALEPH \cite{LEP}, CLEO\cite{cleod1d2}, DELPHI \cite{LEP}, and D0\cite{D12D0}  experiments have reported evidence of the narrow resonant states ($D_1$ and $D_2^*$) in semileptonic decays, whereas more recent measurements by the BaBar  \cite{b22} and Belle \cite{matsumoto} experiments provide semi-inclusive measurements to $D^{(*)} \pi \ell \nu$ final states \cite{kuzmin}.
  \begin{table}
\begin{center}
\caption{Inclusive versus sum of exclusive measured $B$ semileptonic branching fractions (\%).
\label{dss}}
\footnotesize{
\begin{tabular}{l|cc|cc|cc} \hline
$\cal{B}$(\%) & \multicolumn{2}{c|}{BaBar} & \multicolumn{2}{c|}{Belle}
& \multicolumn{2}{c}{World Average} \\
& $B^0$ & $B^-$ &$B^0$ & $B^-$ &$B^0$ & $B^-$ \\
\hline
$B \to D \ell \nu$                                      & 2.20 $\pm$ 0.16       & 2.30 $\pm$ 0.10       &2.09 $\pm$
0.16    &$-$                            &2.16 $\pm$ 0.12        &2.32 $\pm$ 0.09\\
$B \to D^* \ell \nu$                                    & 4.53 $\pm$ 0.14       & 5.37 $\pm$ 0.21       &4.42 $\pm
$ 0.25  &$-$                            &5.05 $\pm$ 0.10        &5.66 $\pm$ 0.18\\
$B \to D \pi \ell \nu$                                  & 0.42 $\pm$ 0.09       & 0.42 $\pm$ 0.07
&0.43 $
\pm$ 0.09       &0.42 $\pm$ 0.06        &0.43 $\pm$ 0.06        &0.42 $\pm$ 0.05\\
$B \to D^* \pi \ell \nu$                                & 0.48 $\pm$ 0.09       & 0.59 $\pm$ 0.06       &0.57 $
\pm$ 0.22       &0.68 $\pm$ 0.11        &0.49 $\pm$ 0.08        &0.61 $\pm$ 0.05\\
\hline
$\Sigma$(Exc.)                          & 7.63 $\pm$ 0.25       & 8.68 $\pm$ 0.25       &7.51 $\pm$ 0.73
&$-$                            &8.13 $\pm$ 0.19        &9.01 $\pm$ 0.21\\
Inc.                                            & 10.14$\pm$ 0.43       & 10.90 $\pm$ 0.47      &10.46 $\pm$ 0.38
&11.17 $\pm$ 0.38       &10.33 $\pm$ 0.28       &10.99 $\pm$ 0.28\\
\hline
Inc. - $\Sigma$(Exc.)   & 2.51 $\pm$ 0.50       & 2.22 $\pm$ 0.53       &2.95 $\pm$
0.82    &$-$                            &2.20 $\pm$ 0.34        &1.98 $\pm$ 0.35\\ \hline
\end{tabular}
}
\end{center}
\end{table}

The differences between the measured inclusive semileptonic branching fraction and the sum of all exclusive $B$ semileptonic measurements  for Belle, BaBar and World averages are given in Table \ref{dss} for $B^0$ and $B^-$ decays. In the case where multiple measurements exist, only the most precise measurements have been used in the BaBar and Belle columns, i.e. no attempt at an average is made. In all cases the sum of the exclusive  components does  not saturate the $B$ semileptonic rate.

All measured rates for the $D^{**}$ narrow states are in good agreement. Experimental results seem to point towards a larger rate for broader states. If it is due mainly to $D'_1$ and $D^*_0$ decay channels, these results disagree with the prediction of QCD sum rules.
However, Belle set an upper limit for the $D'_1$ channel below the rate measured by BaBar and DELPHI. More measurements need to be performed to elucidate this puzzle.

\subsection{Determination Based on Inclusive Semileptonic {\it\textbf B} Decays}

Inclusive determinations of $|V_{cb}|$ are obtained using combined fits to inclusive $B$~decay distributions~\cite{Bauer:2004ve,Buchmuller:2005zv}. These determinations are based on calculations of the semileptonic decay rate in the frameworks of the Operator Product Expansion (OPE)~\cite{wilson} and HQET~\cite{Bauer:2004ve,Benson:2003kp}.
They predict the semileptonic decay rate
in terms of  $|V_{cb}|$, the $b$-quark mass $m_b$, and non-perturbative matrix elements.
The spectator model decay rate is the leading term in a well-defined
expansion controlled by  the parameter
$\Lambda _{\rm QCD}/m_b$~\cite{Benson:2003kp,gremm-kap,falk,Gambino:2004qm} with
non-perturbative corrections arising to order
 $1/m_b^2$. The key issue in this approach is the ability to separate perturbative and non-perturbative
 corrections (expressed in powers of  $\alpha _s$).
Thus the accuracy of an inclusive determination of $|V_{cb}|$ is limited by our knowledge of the heavy quark nonperturbative matrix elements and the $b$ quark mass.  It also relies upon the assumption of local quark hadron duality. This is the statement that a hadronic matrix element can be related pointwise to a theoretical expression formulated in terms of quark and gluon variables.

  Perturbative and non-perturbative corrections depend on the $m_b$ definition, i.e. the non-perturbative expansion scheme, as well as the non-perturbative matrix elements that enter the expansion.  In order to determine these parameters, Heavy Quark Expansions (HQE) \cite{Bauer:2004ve,Gambino:2004qm,Benson:2004sg} express the semileptonic decay width  $\Gamma_{\rm SL}$, moments of the lepton energy and hadron mass spectra in $B\to X_c\ell\nu$~decays
in terms of the running kinetic quark masses $m_b^{\rm kin}$ and $m_c^{\rm kin}$ as well as the $b$-quark mass $m_b^{\rm 1S}$ in the 1S expansion scheme. These schemes should ultimately yield consistent results for $|V_{cb}|$.  The precision of the $b$-quark mass is also important for $|V_{ub}|$.

The shape of the lepton spectrum and of the hadronic mass spectrum provide constraints on the heavy quark expansion, which allows for the calculation of the properties of $B \to X_c \ell \nu$ transitions.  So far, measurements of the hadronic mass distribution and the leptonic spectrum have been made by BaBar~\cite{bb}, Belle~\cite{mx}, CLEO~\cite{cl,chen2001}, DELPHI~\cite{battaglia}. CDF\cite{cdfmom}  provides only the measurement of the hadronic mass spectrum with a lepton momentum cut of  0.6 GeV in the $B$ rest frame.

  The inclusive semileptonic width can be expressed as
\begin{eqnarray}\label{eq:b2c}
\Gamma (\overline{B} \to X_c \ell \bar \nu) & = &\frac{G_F^2 m_b^5 |V_{cb}|^2}{192 \pi^3} ( f(\rho) + k(\rho) \frac{\mu_\pi^2}{2 m_b^2} + g(\rho)\frac{\mu_G^2}{2m_b^2} \\ \nonumber
~ & ~ &  +
d(\rho)\frac{\rho_D^3}{m_b^3}+l(\rho)\frac{\rho_{LS}^3}{m_b^3}  +{\cal O}(m_b^{-4})),
\end{eqnarray}
where   $\rho = m_c^2/m_b^2$,  and  $\mu_\pi^2$,  $\mu_G^2$, $\rho_D$ and $\rho_{LS}$ are non-perturbative matrix elements of local operators in HQET.
To make use of equations such as Eq.~\ref{eq:b2c}, the values of the non-perturbative expansion parameters must be determined experimentally.   Although some of these parameters are related to the mass splitting of pseudoscalar and vector meson states, most non-perturbative parameters are not so easily obtained.  Measurements of the moments of different kinematic distributions in inclusive $B$-decays are used to  gain access to these parameters. The first moment of a distribution is given by:
\begin{equation}
\langle M_1 \rangle = \int M_1(\vec{x}) d \vec{x} \frac{d \Gamma}{d \vec{x}},
\end{equation}
corresponding to the mean.  Subsequent (central) moments are calculated around the first moment,
\begin{equation}
\langle M_n \rangle = \langle ( \vec{x} - M_1) ^n \rangle,
\end{equation}
corresponding to a distribution's width, kurtosis and so on.

A $\overline{B} \to X_c \ell \bar \nu$ decay observable calculated with the OPE is a double expansion in terms of the strong coupling $\alpha_s(m_b)$ and the ratio $\Lambda_{\rm QCD}/m_b$.  Observables are typically calculated with the cuts used in the experimental determination for background suppression, and enhanced sensitivity such as the lepton energy cut.
The spectral moments are defined as:
\begin{eqnarray}
<E_{\ell}^n E_x^m({M^2}_{x})^\ell > & = &\frac{1}{\Gamma_0} \int_{\hat{E}_{\ell}^{\rm min}} dE_\ell \int dE_x \int dM_x\frac{d \Gamma}{dM_XdE_xdE_{\ell}}E_{\ell}^n E_x^m({M^2}_x)^\ell \\ \nonumber
 & = & f_0 [ n, \hat{E}_{\ell}] + f_1 [ n, \hat{E}_{\ell}] \frac{\bar \Lambda}{m_b} \\\nonumber
~ & ~ & +\sum_{i=1}^{2} f_{i+1} [n, \hat{E}_{\ell}] \frac{\lambda_i}{m_b^2} + {\cal O }\left( \alpha_s, \frac{\Lambda_{\rm QCD}^3}{m_b^3} \right),
\end{eqnarray}
where
\begin{equation}
\Gamma_0 =\frac{G_F^2 m_b^5 |V_{cb}|^2}{192 \pi^3}.
\end{equation}

The measurement of $|V_{cb}|$ from inclusive decays requires that these decays be adequately described by the OPE formalism.  The motivation of the moment approach is to exploit the degree of experimental and theoretical understanding of each moment and for different $\hat{E}_{\ell}^{\rm min}$, directly examining the dependence of the various coefficient functions $f_i$ on these terms.
The possibility of deviations from the OPE predictions due to quark hadron duality violations have been raised~\cite{isgur}.  To compare the OPE predictions with data, one also has to define how uncertainties from $1/m_b^3$ corrections are estimated.  These uncertainties are hard to quantify reliably, as the only information on the size of the matrix elements of the dimension six operators comes from dimensional analysis.  The only method to check the reliability  of $|V_{cb}|$ extraction from inclusive semileptonic decays is how well the OPE fits the data.

To compare with theoretical predictions, the moments are measured with a well defined cut on
  the lepton momentum in the $B$ rest frame. The measured hadronic mass distribution and lepton energy
  spectrum are affected by a variety of experimental factors such as detector resolution, accessible phase space, radiative effects. It is particularly
  important  to measure the largest fraction of the accessible
phase space in order to reduce both theoretical and experimental uncertainties. Each experiment has
focused on lowering the lepton energy cut.

The hadronic mass spectrum in $B \to X_c \ell  \nu$ decays can be split into three contributions
corresponding to $D$, $D^*$, and $D^{**}$, where  $D^{**}$ here
stands for any neutral charmed state, resonant or not, other than $D$ and $D^*$.
Belle \cite{mx}, BaBar \cite{bb} and CLEO \cite{cl}  explored the moments of the hadronic mass spectrum $M_X^2$ as a function of the lepton momentum cuts.   CLEO performs
a fit for the contributions of signal and background to the full three-dimensional
 differential decay rate
distribution as a function of the reconstructed quantities $q^2$, $M_X^2$, $\cos{\theta_{\ell}}$.
Belle \cite{belle} and BaBar \cite{bb}  use a sample where one of the $B$ mesons, produced in pairs from
$\Upsilon (4S)$ decays, is fully
reconstructed and the signal side is tagged by a well identified lepton.
The 4-momentum $p_X$ of the hadronic system~$X$, recoiling against the lepton and neutrino, is determined by summing the 4-momenta of the remaining charged tracks and unmatched clusters.
Belle reconstructs the full hadronic mass spectrum and measures the first, second central and second non-central
moments of the unfolded $M^2_X$~spectrum in $B \to X_c\ell\nu$,  for lepton
energy thresholds, $E_\mathrm{min}$, varying from 0.7 to 1.9 GeV~\cite{mx}.
BaBar extracts the moments from the measured distributions
 using a calibration curve derived from Monte Carlo data, with a minimum momentum for the electron
in the $B$ meson rest frame of 0.9 GeV.
The latest  BaBar analysis \cite{mx}  measures the first, second central and second non-central
moments of the $M^2_X$~spectrum for  $E_\mathrm{min}$ from 0.9 to 1.9 GeV.
The main systematic errors originate from background estimation, unfolding and signal model dependence.

DELPHI follows a different approach in extracting the moments, measuring  the invariant
 mass distribution of the $D^{**}$ component only and fixing
the $D$ and $D^*$ components.  DELPHI measures the first moment with respect to the spin averaged mass
of $D$ and $D^*$.  At LEP $b$-quarks were created with an energy of approximately
30 GeV allowing  the measurement of the hadronic mass moments without a cut on
 the lepton energy \cite{battaglia}.

The shape of the lepton spectrum provides further constraints on the OPE. These measurements are sensitive to higher order OPE parameters and  are considerably more precise experimentally.
Moments of the lepton momentum with a cut $p_{\ell}\ge 1.0$ GeV/c
 have been measured by the CLEO collaboration \cite{prd2003}. BaBar \cite{bb}
 extract up to the third moment of this distribution,
using a low momentum cut of  $p_{\ell} \ge 0.6$ GeV/c. Both BaBar and CLEO use dilepton samples.
The most recent measurement of the  electron energy spectrum is from Belle  \cite{belle}. Events are selected by  fully reconstructing one of the $B$ mesons, produced in pairs from
$\Upsilon (4S)$ decays and it determines the true electron energy spectrum by unfolding~\cite{ref:13} the measured spectrum in the $B$ meson rest frame. Belle measures $B^0$ and $B^+$ weighted average partial branching fractions~$\mathcal{B}(B \to X_c\ell\nu)_{E_\ell>E_\mathrm{min}}$ and the first four moments of the electron energy spectrum in $B \to X_ce\nu$, for $E_\mathrm{min}$ from 0.4 to 2.0~GeV~\cite{el}.
All the lepton moment measurements are consistent with theory and with the moment of the
 hadronic and $b \to s \gamma$ photon energy spectrum. Hadronic and lepton energy measurements are
consistent within their errors. When compared with theory there is no sign of inconsistencies.
 The B-factories provide the most precise HQE parameter estimates.

\subsubsection{HQE Parameters}
Using the moment measurements described above, it is possible to determine the CKM matrix element $|V_{cb}|$ and HQE parameters by performing global fits  in the kinetic and 1S $b$-quark mass schemes~\cite{bellefit}.   The photon energy moments in $B\to X_s\gamma$ decays~\cite{Abe:2005cv}  are also  included in order to constrain the $b$-quark mass more precisely.
Measurements that are not matched by theoretical
predictions and those with high cutoff energies are excluded (i.e. semileptonic moments with
$E_\mathrm{min}>1.5$~GeV and  photon energy moments with
$E_\mathrm{min}>2$~GeV).  The results are preliminary.

The inclusive spectral moments of $B\to X_c \ell \nu$ decays have been
derived in the 1S scheme up to
$\mathcal{O}(1/m_b^3)$~\cite{Bauer:2004ve}. The theoretical
expressions for the truncated moments are given in terms of
HQE parameters with coefficients determined by theory, as functions of
$E_\mathrm{min}$. The non-perturbative corrections are parametrized in terms of the following non-perturbative parameters:
$\Lambda$ at $\mathcal{O}(m_b)$, $\lambda_1$ and $\lambda_2$ at $\mathcal{O}(1/m_b^2)$, and
 $\tau_1$, $\tau_2$, $\tau_3$, $\tau_4$, $\rho_1$ and $\rho_2$  at $\mathcal{O}(1/m_b^3)$.  In Table  \ref{1smom} one finds the following results for the fit parameters \cite{bellefit}. The first error is from the fit including experimental and
theory errors, and the second error (on $|V_{cb}|$ only) is due to the
uncertainty on the average $B$~lifetime.  If the same fit is performed to all measured moments of inclusive distributions in $B \to X_c \ell \nu$ and $B \to s \gamma$ decays, the $|V_{cb}|$ and $m_b$  values obtained are in Table \ref{1smom}. The  fit results for
        $|V_{cb}|$ and $m_b^\mathrm{1S}$ to
    $B\to X_c\ell\nu$~data only and $B\to X_c\ell\nu$ and $B\to
    X_s\gamma$~data combined are displayed in Fig. \ref{fig:3}.
\begin{table}
\caption{Experimental results using corrections to the common inputs and world averages for the $1S$ scheme. \label{1smom}}
\begin{center}
\begin{small}
\begin{tabular}{lcccc} \hline
 &$|V_{cb}| (\times 10^{-3})$ & $m_b^\mathrm{1S}$ (GeV)&  $\lambda_1 (\rm GeV^2)$ &$\chi^2/$n.d.f.\\ \hline
$ X_c\ell\nu+X_s\gamma$   & $41. 81\pm 0.34_\mathrm{fit}\pm 0.08_{\tau_B}$ & $4.700\pm 0.030$ & $-0.315 \pm 0.026$ & 24.7/56\\
$ X_c\ell\nu$   & $42.03\pm 0.42_\mathrm{fit}\pm 0.08_{\tau_B}$ & $4.656\pm 0.060$ &  $-0.343  \pm 0.046$ & 19.0/45\\ \hline
\end{tabular}
\end{small}
\end{center}
\end{table}

\begin{figure}
\begin{center}
  \includegraphics[width=0.45\textwidth]{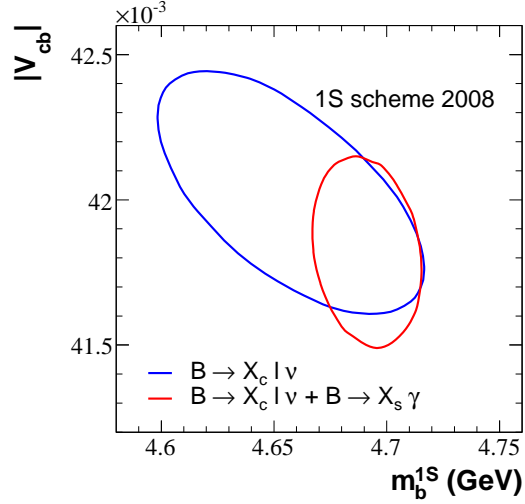}
  \caption{Fit results for
        $|V_{cb}|$ and $m_b^\mathrm{1S}$ to
    $B\to X_c\ell\nu$~data only (larger blue ellipse) and $B\to X_c\ell\nu$ and $B\to
    X_s\gamma$~data combined (smaller red ellipse)  for the $1S$ scheme. The ellipses are for $\Delta\chi^2=1$ \cite{CKM}.}
    \label{fig:3}
\end{center}
\end{figure}

Spectral moments of $B\to X_c\ell\nu$~decays have been derived up to
$\mathcal{O}(1/m^3_b)$ in the kinetic scheme \cite{Gambino:2004qm}. The
theoretical expressions used in the fit contain improved
calculations of the perturbative corrections to the lepton energy
moments~\cite{ref:2} and account for the $E_\mathrm{min}$~dependence
of the perturbative corrections to the hadronic mass
moments~\cite{Uraltsev:2004in}. For the $B\to X_s\gamma$~moments, the
(biased) OPE prediction and the bias correction have been
calculated~\cite{Benson:2004sg}.  All these expressions depend on the $b$- and $c$-quark masses
$m_b(\mu)$ and $m_c(\mu)$, the non-perturbative parameters,
defined at the scale $\mu=1$~GeV: $\mu^2_\pi(\mu)$ and $\mu^2_G(\mu)$ ($\mathcal{O}(1/m^2_b)$),
$\tilde\rho^3_D(\mu)$ and $\rho^3_{LS}(\mu)$ ($\mathcal{O}(1/m^3_b)$),
and $\alpha_s$. The CKM element $|V_{cb}|$ is a
free parameter in the fit, related to the semileptonic
width $\Gamma(B\to X_c\ell\nu)$~\cite{Benson:2003kp}.
The fit results for various inputs and correlations are shown in Table \ref{mom} .
\begin{table}
\caption{Experimental results using corrections to the common inputs and world averages for the kinetic scheme. The first error is from the fit (experimental error, non-perturbative and bias corrections), and the second error is a 1.5\% uncertainty from the theoretical expression for the semileptonic width~\cite{Benson:2003kp}. \label{mom}}
\begin{center}
\begin{tabular}{lcccc} \hline
 &$|V_{cb}| (\times 10^{-3})$ & $m_b$ (GeV)&  $\mu^2_\pi (\rm GeV^2)$ &$\chi^2/$n.d.f.\\ \hline
$ X_c\ell\nu+X_s\gamma$   & $41.67 \pm 0.47_{fit}\pm0.08_{\tau_B}\pm0.58_{th}$ & $4.601  \pm 0.034$ & $0.440 \pm 0.040$ & 29.7/57\\
$ X_c\ell\nu+X_s\gamma$${\dagger}$   & $40.85 \pm 0.68_{fit}\pm0.08_{\tau_B}\pm0.57_{th}$ & $4.605  \pm 0.031$ &$0.312  \pm 0.060$ & 54.2/46\\
$ X_c\ell\nu$    & $41.48\pm0.47_{fit}\pm0.08_{\tau_B}\pm0.58_{th}$ & $4.659  \pm 0.049$ &  $0.428  \pm 0.044$ & 24.1/46\\
\hline
\end{tabular}
\end{center}
${\dagger}$ - New theoretical correlations are applied \cite{christoph}.
\end{table}

 All the measured moments of inclusive distributions in $B \to X_c \ell \nu$ and $B \to s \gamma$ decays are used in a fit to extract $|V_{cb}|$ and the $b$ and $c$ quark masses. The $|V_{cb}|$ and $m_b$ values obtained are listed in the Table \ref{mom}.  The default fit also gives $m_c = 1.16 \pm 0.05$ GeV, and $\mu^2_G = 0.27 \pm 0.04$ GeV$^2$.
The errors are experimental and theoretical (HQE and $\Gamma_{\rm SL}$) respectively.
 In this fit the following variations were considered when  in the extraction of the HQ parameters, $\pm$20 MeV for the $b$ and $c$ quark masses, $\pm$20\% for $\mu^2_\pi$ and $\mu^2_G$,  $\pm$30\% for the  3$^{rd}$ order non perturbative terms and $\alpha_2=
0.22\pm0.04$ for the perturbative corrections. The bias corrections uncertainties for $B\to s\gamma$ were varied by the full amount of their value.

There are open issues relating to the global fits. First of all, the $\chi^2/$ n.d.f. are very small, pointing to an underestimate in the theoretical correlations. In a recent study  \cite{christoph}, the theoretical correlations used in the fit were scrutinized, and new correlation coefficients were derived from the theory expressions using a ``toy Monte Carlo" approach, showing that the theoretical correlations were largely underestimated. The result of this new fit is shown in Table \ref{mom}.
The second issue is related the size of the theoretical error. Recently the NNLO full two-loop calculations become available\cite{NNLO_full}. In the Kinetic scheme NNLO calculations include an estimate of the non-BLM terms and lead to a roughly 0.6 \% reduction of the $|V_{cb}|$ value $-0.25 \times 10^{-3}$. In the 1S scheme the shift on
$|V_{cb}|$ is of about $-0.14 \times 10^{-3}$ \cite{mel}. From the new power corrections at NLO  we expect the chromo-magnetic corrections to be more important as the tree level corrections are more important, and a change of about 20-30\% in the extracted value of $\mu_\pi^2$ in the pole expansion (may be less with the other schemes) \cite{beicherlange}. HQE has been carried out up to  $1/m_b^4$ and the effects are expected to be of the order $\delta^{(4)}\Gamma/\Gamma \approx 0.25\%$ \cite{dassinfer}.
All these newly calculated results can be used to scrutinize the earlier error estimates. In the kinetic scheme, the full NNLO value for $A_{pert} =0.919$, which is in good agreement with $A_{pert} =\Gamma(B\to X_c\ell\nu)/\Gamma(B\to X_c\ell\nu)_{tree} =0.908\pm0.009$.
In the 1S scheme the estimated uncertainty from the non-BLM two-loop is half of the BLM part, equivalent to  1.5\%
of the tree-level rate, more than 3 times the actual correction.
The new results are not implemented in the global fits as they are available mainly in the pole-mass scheme.

\subsection{Outlook}

The error on the inclusive and exclusive determination of $V_{cb}$ is limited by theory. The two method have a 2$\sigma$ disagreement, even when results in the same experiment are compared.

For exclusive determinations, the experimental  determinations in $B \to D^* \ell \nu$ from different experiments are not in agreement.  Improving the statistical error  for the determination from $B \to D \ell \nu$ will help in elucidating the origin of this discrepancy.

In the inclusive method,  errors of less than 2\% are quoted. However, the latest theoretical results and the introduction of better correlation between theoretical error show shifts in the central value larger than the quoted fit error. The situation needs to be reevaluated when all the new calculations and corrections are implemented in the fit. Another puzzling result of the global fit to  $B\to X_c\ell\nu$ moments only is the value of $m^{kin}_b$, which seems to be in disagreement with other determinations, such as the ones  in Table~\ref{t:publ1mass}.\footnote{For a detailed comparison between $m^{kin}_b$ and different mass determinations see A. Hoang, talk at Joint Workshop on $|V_{ub}|$ and  $ |V_{cb}|$ at the B-Factories Heidelberg, December 14-16, 2007.}

\section{ The CKM parameter $\vub$}
The parameter $\vub$ determines one of the sides of the unitarity triangle, and thus affects one of the crucial tests of the Yukawa sector of the Standard Model. Also in this case, there are two
 general methods to determine this parameter, using $B$ meson semileptonic decays. The first approach relies on the determination of branching fractions and form factor
determinations of exclusive semileptonic decays, such as $B\to \pi \ell \bar{\nu}_\ell$. The relationship between experimental measurements and $\vub$ requires a
theoretical prediction of the hadronic form factors governing these decays. The complementary approach relies on measurements of inclusive properties of $B$ meson semileptonic decays.
In this case, the hadronic matrix element is evaluated in the context of the OPE.

Both methods pose challenges to both experimenters and theorists. The branching fractions are small, and a substantial
background induced by the dominant $b\to c\ell \bar{\nu}_\ell$ needs to be suppressed or accounted for. The large data samples accumulated at the two b-factories, Belle and BaBar, have
made possible the development of new experimental techniques that have reduced the experimental errors substantially. In parallel, theorists devoted considerable efforts
 to devise measurable quantities for which reliable predictions could be produced, and to improve the accuracy of the predictions through refinements in the calculation.
 Although the precision of the stated errors improved,
so did the difference in central values between the inclusive and exclusive estimates of $\vub$, at least in most determinations. Possible interpretations of this discrepancy will be discussed.

\subsection{Determinations Based on Exclusive {\it\textbf B} Semileptonic Decays}

The decay $B\to \pi \ell \bar{\nu}_\ell$ is the simplest to interpret, as it is affected by a single form factor. The differential decay width
is given by
\begin{equation}
\frac{d\Gamma}{d\qsq}=\frac{G_F^2 \vub^2}{192 \pi ^3 m_B^3}\lambda (\qsq)^{3/2}\mid f_+(\qsq ) \mid ^2
\end{equation}
where $G_F$ is the Fermi constant, $\lambda (\qsq)=(\qsq +m_B^2-m_\pi^2)^2-4m_B^2m_\pi^2$, and $f_+(\qsq)$ is the relevant form factor.

The first experimental challenge is to reconstruct this exclusive channel without significant background from the dominant charm
semileptonic decays, and the additional background
component from
other $b\to u\ell \bar{\nu}$ transitions.
The advantage of $\epm$ B-factories is that the $B$ decaying semileptonically originates from the decay $\epm\to B\bar{B}$. Thus, if the companion $B$ is fully reconstructed,
the $\bar{\nu}_\ell$ 4-momentum can be inferred from the rest of the event.

CLEO pioneered this approach by reconstructing the $\nu$ from the missing energy ($E_{miss}\equiv 2E_B-\Sigma _i E_i$), and momentum ($\vec{p}_{miss}\equiv \Sigma _i \vec{p}_i$)
 in the event; in these definitions the index $i$ runs over the well reconstructed tracks and photons, and cuts are applied to enhance the probability that the only missing particle in the event is the $\bar{\nu}$. This approach allows the application of a low momentum cut of 1.5 GeV on the lepton for $B\to \pi \ell \nub$ and 2 GeV for $B\to \rho \ell \nub$.
Averaging the results from these two exclusive channels, they obtain $\vub = (3.3\pm 0.2 ^{+0.3}_{-0.4}\pm 0.7)\times 10^{-3}$. Using their relatively modest full data set (16 fb$^{-1}$) at the center-of-mass energy of the $\Ufs$ and considering only the $B\to \pi \ell \nub$ channel, they get  the branching fraction ${\cal B}(B\to \pi^+\ell \nub)= (1.37\pm 0.15\pm 0.11)\times 10^{-4}$, and $\vub = (3.6\pm 0.4 \pm 0.2 \pm ^{+0.6}_{-0.4})\times 10^{-3}$ \cite{Adam:2007pv}.

BaBar  \cite{Aubert:2006px} uses a sample of 206 fb$^{-1}$ to obtain
${\cal B}(B\to \pi^+\ell \nub) = (1.46\pm 0.07\pm 0.08)\times 10^{-4}$.  Recently, the availability of very large data sets at Belle and BaBar have made possible tagged analyses, where semileptonic decays are studied in samples where the other $B$ is fully reconstructed, thus defining the event kinematics even more precisely. The first implementation relies on the the partial reconstruction of exclusive $B\to D^{(\star)} \ell ^+\nu _\ell$ decay to tag the presence of a $B\bar{B}$ event \cite{Hokuue:2006nr,Aubert:2006ry}.

 Belle uses fully reconstructed hadronic tags, achieving the best kinematic constraints and thus the highest background suppression capabilities and the most precise determination of  $\qsq$. Branching fractions obtained with this technique have a bigger statistical error because of the penalty introduced by the tag requirement, but the overall error is already comparable with the other methods. Table~\ref{pilnu:tab} summarizes the present status of the experimental information on this decay. Both the total branching fraction for $B\to \pi \ell \nub$ and the partial branching fraction for $\qsq \ge 16 $ GeV$^2$ are shown. The latter partial branching fraction is useful to extract $\vub$ using unquenched lattice calculations, as this is the only $\qsq$ interval where their calculation is reliable.

\begin{table}[hbt]
\caption{\label{pilnu:tab} Partial and total branching fractions, in units of $10^{-4}$, for the decay $B\to \pi \ell^+\nu_{\ell}$. Whenever possible, $\Bz$ and $\Bp$ data are combined.}
\begin{center}
{\small
\begin{tabular}{llll}\hline\hline
Experiment & total & $\qsq >16$ GeV$^2$ & Method \\ \hline
CLEO \cite{Adam:2007pv} & $1.37\pm 0.15 \pm 0.11$ & $0.41\pm 0.08 \pm 0.04$ & Untagged Analysis \\
BaBar  \cite{Aubert:2006px} & $1.46 \pm 0.07 \pm 0.08$ & $0.38 \pm 0.04 \pm 0.04$ & Untagged Analysis\\
Belle  \cite{Hokuue:2006nr}& $1.38 \pm 0.19\pm 0.14 \pm 0.03$ & $0.36 \pm 0.10 \pm 0.04 \pm 0.01$ & $B\to D^{(\star)} \ell \nu$ Tag\\
Belle$^{\dagger}$ \cite{Abe:2006gb} & $1.49 \pm 0.26\pm 0.06$ & $0.31\pm 0.12 \pm 0.01$ & $\Bz\to\pim\ell ^+\nu _\ell$ (hadron tags) \\
Belle$^{\dagger}$ \cite{Abe:2006gb} & $1.53 \pm 0.20 \pm 0.06$ & $ 0.39\pm 0.12\pm 0.02$ & Combined $\Bz$ and $\Bp$ tags\\
BaBar \cite{Aubert:2006ry} & $1.33 \pm 0.17 \pm 0.11$ & $0.46 \pm 0.10 \pm 0.06$ & Combined $\Bz$ and $\Bp$ tags\\
\hline
combined\cite{hfag08} & $1.34 \pm 0.06 \pm 0.05$ & $0.37 \pm 0.03 \pm 0.02$ & HFAG ICHEP08 \\\hline
\end{tabular}}
${\dagger}$ - Preliminary results.
\end{center}
\end{table}

In order to interpret these results, we need theoretical predictions for the form factor $f_+(\qsq)$. This problem can be split into two parts: the determination of the form factor
normalization, $f_+(0)$, and the functional form of the $\qsq$ dependence. Form factor predictions have been produced with quark models \cite{ISGW2} and QCD sum rule calculations \cite{Ball:2004ye}.
Lattice calculations provide evaluations of $f_+(\qsq)$ at specific values of $q^2$ or, equivalently, pion momenta ($p_\pi$). Authors then fit these data points with a variety of shapes. Typically a dominant pole
shape has been used in the literature. Nowadays more complex functional forms are preferred.  Becirevic and Kaidalov (BK) \cite{Becirevic:1999kt} suggest using
\begin{equation}
f_+(\qsq)|_{BK}= \frac{c_B(1-\alpha)}{(1-\qsq/\mbstar^2)(1-\alpha \qsq/\mbstar^2)},
\end{equation}
where $c_B\mbstar^2$ is the residue of the form factor at $\qsq=\mbstar^2$, and $\mbstar ^2/\alpha$  is the squared mass of an effective 1$^-$ $\bstarpr$ excited state.
Ball and Zwicky \cite{Ball:2004ye} propose
\begin{equation}
f_+(\qsq)|_{BZ}=\frac{r_1}{1-\qsq/\mbstar^2}+\frac{r_2}{1-\alpha\qsq/\mbstar ^2},
\end{equation}
where the parameters $r_1$, $r_2$, and $\alpha$ are fitted from available data.
Lastly, parameterizations that allow the application of constraints derived from  soft-collinear effective theory (SCET), and dispersion relations have been proposed by Boyd, Grinstein, and Lebedev \cite{grinstein}, and
later pursued also by Hill \cite{Hill:2005}. They define
\begin{equation}
f_+(t)=\frac{1}{P(t)\Phi(t,t_0)}\Sigma _0^\infty a_k(t_0)z(t,t_0)^k
\end{equation}
where $t$ is $(p_B-p_\pi)^2$, defined beyond the physical region, $t_\pm =(m_B\pm m_\pi)^2$, and $t_0$ is an expansion point. BGD use $t_0=0.65 t_-$. The parameter $\alpha _K$ allows the modeling of different functional forms, and the variable
\begin{equation}
z(t,t_0)=\frac{\sqrt{t_+-t}-\sqrt{t_+-t_0}}{\sqrt{t_+-t}+\sqrt{t_+-t_0}}
\end{equation}
maps $t_+<t<\infty$ onto $|z|=1$ and $-\infty<t<t_+$ onto the $z$ interval [-1,1].

 All are
refinements of the old ansatz of a simple pole shape, now rarely used. The first lattice calculations were carried out with the quenched approximation that ignores vacuum polarization effects
\cite{Bowler:1999xn,Shigemitsu:2002wh}. In 2004, preliminary unquenched results were presented by the Fermilab/MILC \cite{Okamoto:2004xg}  and HPQCD \cite{Dalgic:2006dt} collaborations. These calculations employed the MILC collaboration
$N_f= 2+1$ unquenched configurations, which attain the most realistic values of the quark masses so far. Using these calculations and the most recent value of the partial branching fraction ${\cal B}(B\to \pi \ell \nu)$ for $\qsq >16$ GeV$^2$, shown in Table \ref{pilnu:tab}, we obtain $\vub=(3.51\pm 0.096_{exp} \pm 0.49_{th})\times 10^{-3}$ with the HPQCD normalization, and $\vub=(3.70\pm 0.10_{exp} \pm 0.37_{th})\times 10^{-3}$ with the Fermilab/MILC normalization. Fits to experimental data, combining lattice predictions, and dispersion relations reduce the theoretical errors. For example, \cite{Arnesen:2005ez} obtains $\vub = (3.5\pm 0.17 \pm 0.44)\times 10^{-3}$, and, more recently, ~\cite{Flynn:2007rs} obtains
$\vub = (3.47 \pm 0.29)\times 10^{-3}$.
The most recent lattice calculation \cite{Bailey:2008wp} performs a simultaneous fit of improved lattice numerical Monte Carlo results and the 12-bin BaBar experimental data on $\vub f_+(\qsq )$ \cite{BaBarff} and derives $\vub =(3.38 \pm 0.35)\times 10^{-3}$. The $\sim$ 10\% error includes theoretical and experimental errors, not easily separable.

\subsection{Determinations Based on Inclusive {\it\textbf B} Semileptonic Decays}
Inclusive determinations of $\vub$ rely on the heavy quark expansion (HQE), which combines perturbative QCD with an expansion in terms of $1/m_b$, which accounts for non-perturbative
effects. Although the possible breaking of the assumption of local quark hadron duality may produce unquantified errors, other non-perturbative uncertainties can be evaluated with systematic
improvements, and their uncertainties are easier to assess than the ones of unquenched lattice QCD or QCD sum-rules. This statement applies to the
total charmless semileptonic width $\Gamma _u ^{SL}$. In the OPE approach, discussed in the $\vcb$ section, the observation by Chay, Georgi and Grinstein that there are no non-perturbative corrections of order $\Lambda _{\rm QCD}/m_b$ \cite{Chay:1990da}
has inspired the hope that this approach would lead to a more precise determination of this important parameter. Before discussing the methods used to relate data with theory, it is useful to
observe one of the possible expressions available in the literature for the total semileptonic width \cite{Lange:2005yw}, including an exact two-loop expression for the perturbative expansion and second-order power corrections. They obtain
\begin{eqnarray}
\Gamma _u &  =  & \frac{G_F^2\vub ^2 m_b^5}{192\pi ^3}\{1+\alpha _S(\mu)\left[-0.768+2.122\frac{\mu_\star}{m_b}\right] \\ \nonumber
& & +\alpha _S^2(\mu)\left[-2.158+1.019\ln{\frac{m_b}{\mu}}+(1.249+2.184\ln{\frac{\mu}{\mu _\star}}+0.386\frac{\mu_\pi ^2}{\mu_\star^2}\mu_\star/m_b)+0.811\frac{\mu _\star^2}{m_b^2}\right]
\\ \nonumber
 & & -\frac{3(\mu _\pi^2-\lambda _2)}{m_b^2}+...\}
\end{eqnarray}
where  $\mu \sim m_b$ is the scale at which $\alpha _S$ needs to be evaluated, while  the scale $\mu _\star$ applies to
the non-perturbative expansion parameters, namely the $b$ quark mass $m_b$, the chromo-magnetic operator,  $\lambda _2$, and the kinetic operator, $\mu _\pi$. It is clear that, even restricting our attention to the total width, a precise knowledge of the $b$ quark mass is critical, and considerable theoretical effort has been devoted to a reliable extraction of this parameter from experimental observables.  Other uncertainties, such as the effects of weak annihilation or violations of quark-hadron duality, will be discussed later.

Charmless $B$ meson semileptonic decays constitute only about 1\%
of the total semileptonic width. Thus the big challenge for
experimentalists is to identify techniques to suppress this large background.
For example,
the first evidence for  $B$ meson charmless semileptonic decays came from the study of the end point of the lepton spectrum, where
leptons from $b\to c\ell\nu$ processes are forbidden due to the larger mass of the hadronic system formed by the $c$ quark
\cite{Bartelt:1993xh}. While this was very important first evidence that $\vub \ne 0$, very quickly several authors pointed out that this region of phase space is ill suited to a precise determination of $\vub$ because near the end point the OPE does not converge and an infinite series of contributions needs to be taken into account \cite{Isgur:1992iv, Bigi:1993ex, Neubert:94QBI}.
Thus, a large effort has gone into developing experimental techniques that would feature a low lepton energy $E_\ell$ cut and an acceptable signal to background ratio. Table \ref{tab:endpoint} summarizes the present status of the $\vub$ determination with this approach.

\begin{table}[hbt]
\begin{center}

\caption{\label{tab:endpoint} Summary of the experimental $\vub$ determinations using the lepton endpoint; $\vub$ is extracted with the ``shape function'' method \cite{Lange:2005yw}.}
\begin{tabular}{llllll}
\hline\hline
Experiment & Lepton energy range  & \multicolumn{2}{c}{${\cal L}_{int}$ (fb$^{-1}$)}& $\Delta {\cal B}(X_u\ell \bar{\nu}_\ell)$ ($\times 10^{4}$) & $\vub$ ($\times 10^3$) \\
~~& ~~ & $\Ufs$ &  cont. & ~~ & ~~ \\
\hline
CLEO \cite{cleo:2002} & 2.6 GeV $>E_\ell>$  2.1 GeV & 9.1 & 4.3 & $3.28 \pm 0.23\pm 0.75 $ & $3.94\pm 0.46 ^{+0.37}_{-0.33}$\\
Belle \cite{Belle:2005} & 2.6 GeV $>E_\ell>$  1.9 GeV & 27 &8.8 & $5.72 \pm 0.41 \pm 0.65 $ & $4.74 \pm 0.44^{+0.35}_{-0.30}$\\
BaBar \cite{BaBar:2006} & 2.6 GeV $>E_\ell>$  2.0 GeV & 80 & 9.5 & $5.72 \pm 0.41 \pm 0.65 $ & $4.29\pm 024 ^{+0.35}_{-0.30}$\\\hline
\end{tabular}
\end{center}
\end{table}

Next a whole host of papers proposed alternative ``model independent'' approaches to measure $\vub$ from inclusive decays \cite{Bigi:1997dn,Bauer:2001yb,Lange:2005yw}, with the common goal of identifying a region of phase space where experimentalists can suppress the $b\to c$ background, and where the OPE works.  The first proposal by Bigi, Dikeman, and Uraltsev proposed considering semileptonic events where $M_X\le 1.5$ GeV \cite{Bigi:1997dn}. However, Bauer, Ligeti, and Luke pointed out \cite{Bauer:2000xf} that the kinematic limit $m_X^2\sim m_D^2$ has the same properties of the lepton end point, and spoils the convergence of the OPE; the same authors proposed using $d\Gamma/dq^2$ up to $q^2=m_D^2$ and argue that this distribution is better behaved in the kinematic region of interest.
 This is the theoretical foundation of the so called ``improved end point'' method, which encompasses the simultaneous study of $E_\ell$ and $\qsq$. BaBar \cite{BaBar:improved} used this technique, inferring the $\nu$ 4-momentum from the missing momentum ($\vec{p}_{miss}$) in the event. There results give $|V_{ub}|$ central values between (3.88-4.93)$\times 10 ^{-3}$, depending on the explicit model and with total errors of $\approx$10\% on each value.

An alternative approach \cite{Lange:2005yw}, incorporates hadronic structure functions to model the region of large hadronic energy and small invariant mass, not well modeled by the OPE, and applies the  OPE to the kinematic region where the hadronic kinematic variables scale with $M_B$, and smoothly interpolates between them. This approach is commonly referred to as  ``shape function'' method, and uses the $\gamma$ spectrum in inclusive $B\to X_S \gamma$  to reduce the theoretical
uncertainties. It combines the experimental data on high momentum leptons from $B$ decays with the constraints from inclusive radiative decays, to produce a precise value of $\vub$ inclusive.

 The study of charmless inclusive semileptonic decays benefits from the use of hadronic tags. Belle used the information of the tag momentum to boost the electron into the $B$ meson rest frame and to select a sample of high purity. They then reconstruct the $\nu$ energy and momentum from the measured 4-momentum vectors of the $\Ufs$, $B$ tag, lepton, and the additional tracks not used to form the tag or the lepton. They then evaluate the invariant mass $M_X$ and the quantity $P^+=|E_X-\vec{P}_X|$, where $\vec{P}_X=\vec{P}_{beam}-\vec{p}_{tag}-\vec{p}_\ell-\vec{p}_\nu$. Both $M_X$ and $P^+$ are smaller for $b\to u$ transitions. They define different signal region for $p_\ell>1$ GeV: $P_+<0.66$ GeV/c, $M_X<1.7$ GeV/c$^2$, and $M_X<1.7$ GeV/c$^2$ combined with $\qsq >8$ GeV$^2$/c$^2$. They evaluate the partial branching fractions in each of them, and extract $\vub$  directly from the partial branching fractions, normalized by corresponding theoretical scale factors  $R(\Delta\Phi)$, evaluated with the shape function method \cite{blnp}. The authors observe that different kinematic cuts give different values of $\vub$ and speculate that this may be due to additional theoretical uncertainties not completely accounted for.
Recently, BaBar has also used the same technique in slightly different kinematic regions:
$P_+<0.66$ GeV/c, $M_X<1.55$ GeV/c$^2$, and $M_X<1.7$ GeV/c$^2$ combined with $\qsq >8$ GeV$^2$/c$^2$. They use both the shape function method \cite{blnp} and the dressed gluon exponentation method \cite{dge}. The results obtained with this approach are summarized in Table ~\ref{tab:vubtagged}.

\begin{table}[hbt]
\begin{center}
\caption{\label{tab:vubtagged} Inclusive $\vub$ determinations with tagged samples.}
\begin{tabular}{llll}
\hline\hline
Experiment &  \multicolumn{2}{c}{${\cal L}_{int}$(fb$^{-1}$)} &   $\vub \times 10^3$ \\
~~&  $\Ufs$ &  cont. & ~~ \\
\hline
Belle \cite{Belle:2005} & 253 & 18 &  $4.09 \pm 0.44^{+0.35}_{-0.30}$\\
BaBar \cite{BaBar:2006} &347.4 & ~~ &  $4.21\pm 0.20 ^{+0.32}_{-0.27}$\\

\hline
\end{tabular}
\end{center}
\end{table}


In order to provide predictions that are most suited to different experimental cuts, theorists
have made available
the triple differential width $\frac{d^3\Gamma}{dE_\ell d\qsq dq_0}$ given by
\begin{equation}
\frac{G_F^2\vub ^2}{8\pi^3}\{\qsq W_1-[2E_\ell^2-2q_0E_\ell+\frac{\qsq}{2}]W_2+\qsq (2E_\ell-q_0)W_3\}\times\theta(q_0-E_\ell)\frac{\qsq}{4E_\ell})\theta(E_\ell)
\theta(\qsq)\theta(q_0-\sqrt{\qsq}),
\end{equation}
where $q_0$ is the energy of the lepton-$\nu$ pair and $E_\ell$ is the energy of the charged lepton in the $B$ meson rest frame, and $W_{1-3}$ are the three structure functions relevant if we assume massless leptons. Reference \cite{Gambino:2007rp} computes the functions $W_i(q_0,\qsq)$ as a convolution at fixed $\qsq$ between non-perturbative distributions $F_i(K_+,\qsq;\mu)$
and the perturbative functions $W_i^{pert}(q_0,\qsq)$
\begin{equation}
W_i(q_0,\qsq)=\int dk_+ F_i(k_+,\qsq;\mu)W_i^{pert}\left(q_o-\frac{k_+}{2}\left(1-\frac{\qsq}{m_bM_B}\right),\qsq;m_b\right).
\end{equation}

Perturbative corrections to the structure functions $W_{1-3}$ are now known up to order ${\cal O}(\alpha _s^2\beta_0)$ \cite{Aquila:2005hq} and power corrections are included through ${\cal O}(m_b^2)$. The separation between perturbative and non-perturbative physics is set by a cut-off scale $\mu= 1$ GeV, in the ``kinetic scheme'' \cite{Uraltsev:2004in}, which takes input parameters from fits to the $b\to c$ moments. Alternative choices of kinematical variable have been used, for example,  in the SCET approach, the variables $P^+\equiv |E_X-\vec{P}_X|$ and $P^-\equiv |E_X+\vec{P}_X|$ are used. In particular,
the ``shape function'' method \cite{Lange:2005yw} provides theoretical expression for the triple differential ${d^3\Gamma}/{dE_\ell dP^+dP^-}$ and relates it to moments of the shape function extracted from $B\to X_S\gamma$. Finally, the approach originally proposed by Gardi \cite{Gardi:2004ia} uses resummed perturbation theory in momentum space to provide a perturbative calculation of the on-shell
decay spectrum in the entire phase space. The method used, dressed gluon exponentiation, is a general resummation formalism for inclusive distributions near a kinematic threshold. Finally, another model based on soft-gluon resummation and an analytical time-like QCD coupling has been proposed \cite{Aglietti:2006yb}. Clearly a lot of work has gone into bringing to fruition the original promise that the inclusive $\vub$ determination is a more precise method to determine this important CKM parameter. However, Figure~\ref{vubincl:fig} shows that different methods provide central values of $\vub$ that often differ beyond the stated errors. Moreover, these estimates are generally significantly higher than the value of $\vub$ extracted from $B\to \pi \ell \bar{\nu}_\ell$ and the one that is obtained from global unitarity triangle fits.

\begin{figure}
\begin{center}
\includegraphics[width=0.45\textwidth]{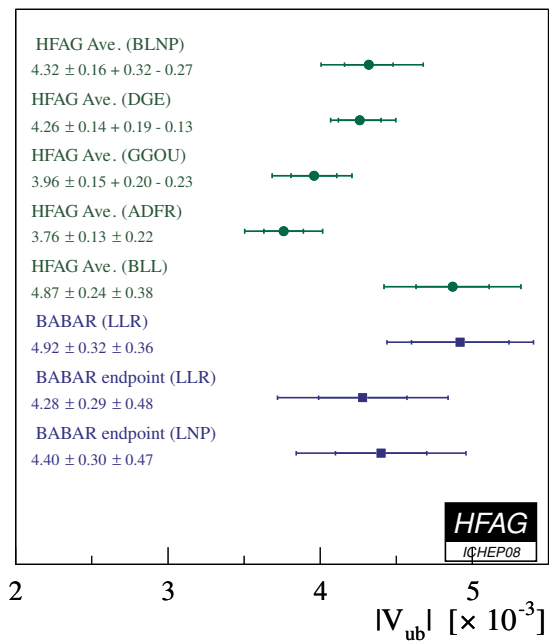}
\caption{Summary of Inclusive $\vub$ determinations. \label{vubincl:fig}}
\end{center}
\end{figure}

An effect that can influence the inclusive $\vub$ is the contribution due to topologies where the incoming $b\bar{q}$ pair annihilates into a $W$ boson, with a concomitant soft gluon emission.
These contributions are known as ``weak annihilation'' and they appear as a delta function at the end point of the lepton spectrum. This effect introduces a difference between $\Bz$ and $\Bp$ semileptonic decays.

A first attempt to quantify the annihilation effect was performed by CLEO \cite{Rosner:2006zz}, studying the high $\qsq$ contributions to charmless semileptonic decays. They used inclusive data on a sample including both charged and neutral $B$ mesons. They used a variety of models to set the limit on the fractional contribution of annihilation diagrams, $\Gamma _{WA}/\Gamma _{b\to u}< 7.4$ \% at 90 \% confidence level. BaBar has presented a measurement of ${\cal B}(\Bz\to X_u\ell \bar{\nu}_\ell)$ at the end point region which, combined with previous inclusive measurement, sets
the limit
\begin{equation}
\Gamma _{WA}/\Gamma _{b\to u}< \frac{3.8 \%}{f_{WA}(2.3-2.6)}
\end{equation}
 at 90 \% confidence level, where  $f_{WA}(2.3-2.6)$ represents the fraction of weak annihilation rate contributing in the momentum interval $\Delta p_\ell$ comprised between 2.3 and 2.6 GeV used in this analysis, believed to be close to unity.

 Finally, Voloshin \cite{Voloshin:2001} has suggested that the difference in semileptonic widths of the $D^0$ and $\Ds$ mesons can assess non-factorizable terms through the relationship
 \begin{equation}
 \frac{(\Gamma_{sl}(D^0)-\Gamma_{sl}(\Ds))}{\Gamma _0(c\to s \ell \nu)}= 3.4 \left(\frac{f_D}{0.22\ {\rm GeV}}\right)^2(B_1^{ns}-B_2^{ns}),
 \end{equation}
 where $\Gamma _0(c\to s \ell \nu)$ is the bare parton semileptonic rate $c\to s \ell\nu$, $f_D$ is the $D$ decay constant, and $B_1^{ns}-B_2^{ns}$ is the non-factorizable term that affects charmless semileptonic decays as well. Using this formula with the CLEO-c results for inclusive $\Dz$ semileptonic decay \cite{Adam:2006nu}, the sum of $\Ds$ exclusive semileptonic decays \cite{ds-semil}, and $m_c(m_c)=1.268$ GeV \cite{Steinhauser:2008pm}, we obtain an estimate of non-factorizable effects equal to (9$\pm$5)\%.
 Note that annihilation effect induced uncertainties are estimated to contribute 1.6\% in the BLNP approach, 1.5\% in the DGE approach, and (+0.0,-3.1)\% in the GGOU approach.

 In summary, the values if $\vub$ extracted with the inclusive method have quoted errors between 5.5\% and 10\%, with central values that change well outside these uncertainties and are all
 considerably larger than the $\vub$ value obtained from the exclusive branching fraction ${\cal B}(B\to \pi \ell \nu)$.

\section{Rare {\it\textbf B} Decays}
In general, we define as ``rare'' $B$ decays processes that are suppressed at tree level.  They are interesting because they are typically
mediated by loop diagrams which may be characterized by a matrix element whose strength is comparable to components with similar Feynman diagrams, where new particles appear in loops. Thus evidence for new physics may appear either through enhancements in branching fractions relative to the Standard Model expectation, or through interference effects.

\subsection{$B\to \tau\nu$}
\label{sec:tau_nu}
The decay $B\to\tau\nu$ is affected by two quantities of great interest, the quark mixing parameter $\vub$ and the pseudoscalar decay constant
$f_B$. In fact, the leptonic branching fraction is given by
\begin{equation}
{\cal B}(B^+\to \tau^+ \nu)=\frac{G_F^2m_B m_\tau ^2}{8\pi}\left(1-\frac{m_\tau^2}{m_B^2}\right)^2f_B^2\vub ^2\tau_B,
\end{equation}
where $m_\tau$ is the mass of the $\tau$ lepton, and $\tau_B$ is the charged $B$ lifetime. Theoretical predictions for $f_B$ are summarized in Table~\ref{tab:fb}.
The most recent value exploits the full machinery of unquenched Lattice QCD, and has a precision of 7\%.
\begin{table}
\begin{center}
\caption{\label{tab:fb} Summary of recent theoretical evaluations of the decay constant $f_B$.}
\begin{tabular}{lc}
\hline
Method & $f_B$ (MeV) \\ \hline
HPQCD (unquenched) \cite{Gray:2005ad} & $216\pm 9 \pm 19 \pm 4 \pm 6 $\\
LQCD (chiral extrapolation) \cite{Guo:2006nt} & $209.4\pm 9.7 \pm 1.0$\\
FNAL-MILC-HPQCD \cite{Bernard:2007zz} & $197\pm 6 \pm 12$ \\ \hline
\end{tabular}
\end{center}
\end{table}

Belle performed the first measurement of the branching fraction ${\cal B}(\Bp\to \taup\nu)$ using a tagged sample of fully reconstructed $B$ decays \cite{belle:taunuhad}. They later used a tagged sample of semileptonic decays \cite{belle:taunulep}.  BaBar also published studies performed using fully reconstructed $B$ decays \cite{babar:taunuhad} and semileptonic tags \cite{:2008gx}. The results are listed in Table~\ref{tab:btaunu}. The resulting average experimental value ${\cal B}(B\to \tau \nu)$ is $(1.73\pm 0.35)\times 10^{-4}$.

\begin{table}
\begin{center}
\caption{\label{tab:btaunu} Summary of experimental determinations of ${\cal B}(B\to \tau \nu)$.}
\begin{tabular}{llc}
\hline
Experiment & Method & ${\cal B}(B\to \tau \nu)\times 10^4$ \\
\hline
Belle\cite{belle:taunuhad} & Hadronic tag sample & $1.79 ^{+0.56+0.46}_{-0.49-0.51}$ \\
Belle\cite{belle:taunulep} & Semileptonic tag sample & $1.65 ^{+0.38+0.35}_{-0.37-0.37}$ \\
BaBar\cite{babar:taunuhad} & Hadronic tag sample  & $1.8 ^{+0.9}_{-0.8}\pm 0.4 \pm 0.2$ \\
BaBar\cite{:2008gx}& Semileptonic tag sample & $1.8 \pm 0.8 \pm 0.1$ \\ \hline
Average & ~~ & $ 1.73 \pm 0.35$ \\ \hline
\end{tabular}
\end{center}
\end{table}

\begin{figure}
\begin{center}
\includegraphics[width=0.45\textwidth]{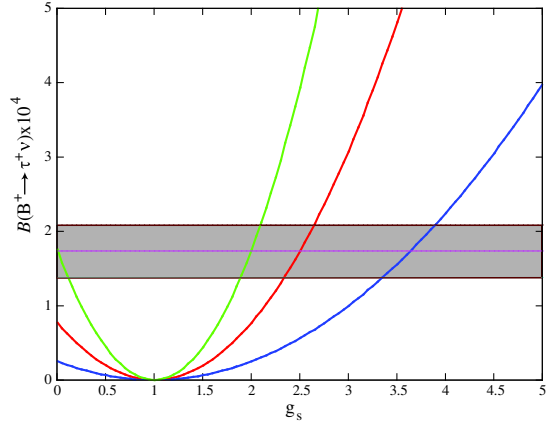}
\caption{${\cal B}(B\to \tau\nu)$ as a function of the effective scalar coupling $g_s$. The shaded region is the measure data $\pm$1$ \sigma$ The range of theoretical predictions is obtained by assuming $\vub=(3.38\pm 0.35)\times 10^{-3}$ \cite{Bailey:2008wp}, and $f_B=(0.197\pm 0.06 \pm 0.012$) MeV \cite{Bernard:2007zz}. (The curves are theoretical and explained in the text. \label{fig:taunu}}
\end{center}
\end{figure}

The Standard Model prediction for ${\cal B}(B\to \tau\nu)$ is $(0.77^{+0.98}_{-0.52})\times 10^{-4}$, obtained using the exclusive value of $\vub$ \cite{Bailey:2008wp} and the most recent Lattice QCD value of $f_B$ \cite{Bernard:2007zz}. The range has been obtained by calculating the Standard Model branching fraction using the  values $\vub=\vub _{excl}\pm \sigma(\vub )$
and $f_B=f_B \pm \sigma(f_B)$. An intriguing possibility is the enhancement of this branching fraction due to new physics.
For example in a 2-Higgs-doublet model (2HDM) \cite{Trine:2008qv} based on supersymmetric extensions of the Standard Model, the ratio
\begin{equation}
\frac{{\cal B}(B\to \tau \nu)}{{\cal B}(B\to \tau \nu)|_{SM}}=|1-g_s|^2
\end{equation}
where $g_s$ is the effective scalar coupling. Fig.~\ref{fig:taunu} shows the predicted value of ${\cal B}(B\to\tau \nu)$ as a function of $g_s$ and the band representing the measured value, which constrains $g_s$ to be less than 0.1 or between 1.8 and 3.9. A better understanding of the $\vub$ and $f_B$ inputs is necessary to improve on this estimate.

Note that a quantity that is sensitive to the same coupling is the ratio $\frac{{\cal B}(B\to D\tau \nu)}{{\cal B}(B\to D \ell \nu)}$. The present experimental value for this parameter is
$(41.6\pm 11.7\pm 5.2)$\% \cite{PDG}, which constrains $g_s$ to be $\le 1.79$. Thus these two measurements provide complementary constraints, which have very different systematic uncertainties. Thus an improvement in the knowledge of leptonic and semileptonic branching fractions involving a $\tau$ lepton in the final state is very important.

\subsection{Radiative  {$B$} Decays}

Radiative $B$ decays involving $b \to s(d) \gamma$ transitions are very sensitive to new physics processes. These processes are ideals for indirect searches
for physics beyond the Standard Model,  such as two-Higgs doublet models
, supersymmetric theories  and extended technicolor scenarios \cite{new_physi_bsg}. Hence, comparison of
results from these theories with experimental measurements places constraints upon new
physics. Moreover, $b\to X_s\gamma$ is an ideal laboratory for studying the dynamics of the $b$-quark inside
the $B$ meson: since the motion of the $b$-quark inside the $B$ meson is universal, information gained
from a measurement of the energy spectrum of the emitted photon in  this transition is applicable to
other processes, for instance semileptonic decays.

In general, in the OPE, the amplitude $A$ for a process can be expressed as sum
\begin{equation}
A=\langle {\cal{H}}_{\rm eff}\rangle=\sum_i C_i(\mu,M_W)\langle Q_i(\mu)\rangle,
\end{equation}
where the $Q_i$ are local operators, the $C_i$ are Wilson coefficients, and $\mu$ is the QCD renormalization scale \cite{BBL}.
The $B\to X_s\gamma$  branching fraction directly probes the Wilson coefficient $C_7$. However, some new physics contributions may leave the
$B\to X_s\gamma$ rate unaltered, with changes to the $C_7$ amplitude. In this case the direct $CP$ asymmetry is sensitive to new
phases  that may appear in the decay  loop \cite{cpasy} and the $B\to X_s\ell^+\ell^-$ transition may provide information on the sign of the amplitude, since it also probes $C_9$ and $C_{10}$.

The $B\to X_s\gamma$ branching fraction, as a function of a photon energy cut-off $E_0$,  is related to the $B\to X_c \nu \ell$ transition by
\begin{eqnarray}
{\cal{B}}(B\to X_s\gamma ~;~E_\gamma>E_0\ {\rm GeV})  & = & {\cal{B}}(B\to X_c e \bar{\nu})_{exp} \left|\frac{V^*_{ts}V_{td}}{V_{cb}}\right|^2\frac{6\alpha_{em}}{\pi C}  \times \left[ P(E_0)-N(E_0)\right] \\\nonumber
 &= & \left|\frac{V_{cb}}{V_{ub}}\right|^2\left[\frac{\Gamma(\bar{B}\to X_c e\bar{\nu})}{\Gamma(\bar{B}\to X_u e\bar{\nu})}\right]
\end{eqnarray}
where the perturbative corrections $P(E_0)$ are defined as
\begin{eqnarray}
 \left|\frac{V^*_{ts}V_{td}}{V_{cb}}\right|^2
 \frac{6\alpha_{em}}{\pi}  P(E_0) = \left|\frac{V_{cb}}{V_{ub}}\right|^2&&\left[\frac{\Gamma(b\to s \gamma)_{E>E_0}}{\Gamma(b\to c e \bar{\nu})}\right]_{LO}\times
  \\\nonumber
&& \left(  1+ {\cal O}(\alpha_s)_{NLO}+ {\cal O}(\alpha)_{NLO}+ {\cal O}(\alpha^2_s)_{NNLO} \right),\\\nonumber
\end{eqnarray}
and the non-perturbative $N(E_0)$ terms are  ${\cal O}(\frac{\Lambda^2}{m_b^2})_{LO+m^{NLO }_b}$$+{\cal O}(\frac{\Lambda^2}{m_c^2})_{LO+ m^{NLO}_c}$ $+
{\cal O}(\alpha_s \frac{\Lambda}{m_b})_{NLO+ m^{LO}_b} $.
Note that the minimum value of $E_0$ for which this relation is valid is  1.6 GeV.
The theoretical error from NLO perturbative calculations is about 10\%, dominated by  the renormalization scale
dependence and a charm quark mass uncertainty.  This uncertainty is mainly due to the change from the ratio in the pole scheme  $m^{pole}_c/m^{pole}_b=0.29\pm0.02$ to the ratio $m^{\overline{\rm MS}}_c/m^{\overline{\rm MS}}_b=0.22\pm0.04$ in the $\overline{\rm MS}$  scheme (see section \ref{sec:quark_masses}). The large $m_c$ dependence is due to the fact that $m_c$ first enters at NLO matrix elements. Hence, the natural scale at which $m_c$ should be renormalized  can only be determined by
dedicated calculations at NNLO.
Two predictions for the branching fraction at NNLO  have been given:
\begin{eqnarray}
{\cal {B}}(B\to X_s\gamma ~;~{E_{\gamma>1.6~{\rm GeV}}})&=&(3.15\pm 0.23)\times 10^{-4},\\\nonumber
                                                        &=&(2.98\pm 0.26)\times10^{-4},
\end{eqnarray}
from \cite{Misiak} and \cite{matt}, respectively.
Both agree with the world average in Table \ref{t:br1}.
The theoretical error of  7\% is obtained by quadratically adding the uncertainty due to non-perturbative corrections (5\%) \cite{LNP},
the uncertainty due to missing higher-order corrections (3\%), the  $m_c$ interpolation ambiguity (3\%) and the parametric uncertainty (3\%). Not included are some known NNLO and non-perturbative corrections. The size of these neglected contributions is about
1.6\%, which is smaller than the
present theoretical error. While progress in the calculation of the perturbative corrections is expected in the future, the uncertainty on the non-perturbative effects will not be easily reduced as they are very difficult to estimate.

Experimentally, two methods are used to extract the  $b\to X_s\gamma$
signal: the  fully inclusive method and  the semi-inclusive method. In the fully inclusive method, events containing a hard photon consistent with
$B\to X_s\gamma$ are selected. In this method
the  subtraction of a very large background, primarily from $q\bar{q}$ continuum events, is the main issue.
 There are several options to reduce this large background.   Particularly effective in the suppression of continuum backgrounds is
the requirement of a high momentum lepton, tagging the semileptonic decay of the accompanying $B$
meson. Alternatively, a fully reconstructed  $B$ can be used to identify $B$ decays.
In this measurement the large background increases towards the lower photon energy, making it impossible to measure the full photon spectrum.  Hence, all measurements  require a minimum photon energy where the signal to noise ratio is still acceptable. The total branching fraction is then  obtained by extrapolating the signal to the full phase space.  This extrapolation is based on theoretical models and is an irreducible source of systematic error.
It has been argued that the energy scale $\Delta = m_b -2E_\gamma^{\rm min}$ is significantly smaller than $m_b$ and therefore
the above extrapolation error is underestimated.  Therefore it is  important to measure the $b\to s\gamma$ photon energy spectrum as precisely as possible.
So far we have been able to lower the minimum photon energy to 1.7 GeV.

In the semi-inclusive method, the $B\to X_s\gamma$ branching fraction is determined by summing up  exclusive modes, with an extrapolation procedure to
account for the unobserved modes.
The semi-inclusive analysis provides a more precise photon energy in the $B$-meson rest frame and usually provides the flavor of
the decay (i.e. $b$ or $\overline{b}$). The extrapolation to account for the unobserved modes is is the key issue in this approach; it is based on isospin symmetry and
the  $B$-meson Monte Carlo hadronization model, which is hardly precise and reliable.
All the available measurements are summarized in Table \ref{t:br1}. Note that the CLEO measurement on the branching fraction includes  $b \to d \gamma$ events, which is expected to be quite small $\sim |V_{td}/V_{ts}|^2 \approx 5$\%.

At the parton level in the two-body decay $b \to s \gamma$, the energy of the photon is $E_\gamma\approx m_b/2$ in the $b$-quark rest frame.
However, the $b \to X_s \gamma$ is not mono-chromatic, due to several effects including the width of the $X_s$ mass distribution, gluon emission and the {\it Fermi motion} of the $b$-quark in the $B$ meson.
The non-perturbative Fermi motion effects will raise the photon energy above $m_b/2$, while the gluon emission will give a long low-energy tail.
The low-energy tail can be described by HQE a part  non-perturbative effects that can be modeled by process-independent shape functions described by a few
parameters, e.g. the $b$ quark mass and the Fermi momentum
($\mu_\pi^2$) that are considered to be universal in the following inclusive
$B$ decays: $B\to X_s\gamma$, $B\to X_s \ell\ell$, $B\to X_c\ell\nu$  and  $B\to X_u\ell\nu$.
Recently a significant effort was made to combine all available data for
$B\to X_s\gamma$ and $B\to X_c\ell\nu$  to determine the HQE parameters.
Using these  parameters, $B\to X_s\gamma$ branching fraction results
are combined together to provide a rather precise branching fraction. The method to combine these measurement has been provided by Limosani \cite{tony} and it is different from the method used by HFAG, as explained below.

The extrapolated branching fractions do not include the published model uncertainties and the uncertainty on the extrapolation factor. These errors are included in the average by recalculation in the framework of a
particular ansatz. Correlations between different measurements have been ignored.
The parametric error on $m_b$ is evaluated by varying $m_b$ within its uncertainty.
The world average has been calculated taking into account the correlations, when available,\footnote{For the time being only Belle provides these correlations.} between partial branching fraction measured at different photon energy thresholds for each single analysis. In the branching fraction average we use the energy threshold of each photon energy spectrum that corresponds to the optimal overall uncertainty on the full rate after extrapolation. This differs from the HFAG method, which uses the lowest energy thresholds for each spectrum measurement. The HFAG method penalizes analyses that quote measurements at low $E_{min}$, which suffer from larger systematic uncertainties.

Two different calculations were used to extrapolate the measured partial branching fractions down to a photon energy lower threshold of 1.6 GeV.
The extrapolation factors were determined using  $m_b^{\overline{\rm MS}}=4.243 \pm 0.042$ GeV as input  \cite{dge}, or  with $m_b^{SF}=4.63 \pm 0.04~{\rm GeV}$ and  $\mu_\pi^2=0.272 ^{+0.056}_{-0.076} {\rm GeV}^2$ as input \cite{matt}.
The two world averages are listed in Table \ref{t:br1} and they agree within their theoretical errors.

The agreement between the measured $B\to X_s\gamma$ branching
fraction and the theoretical prediction constrains various new
physics scenarios.  One of the most popular examples is the lower bound
on the type-II charged Higgs mass, since it always constructively
interferes with the SM amplitude.  The current limit is around 200 GeV for any $\tan \beta$
if no other destructive new physics contribution exists \cite{misiaktype2}.  This limit is
significantly higher than the direct search limit.

\begin{table}[hbt]
\caption{Measured branching fractions, minimum photon energy, and branching fractions for $E_{min}=1.6$ GeV photon energy for $b \to s \gamma$. The  third error is the model uncertainty quoted by the experiment. Two world averages are  calculated extrapolating all the branching fractions
using two different theoretical calculations \cite{dge} and \cite{matt}.
The Belle semi-inclusive measurement sums up 16 modes, BaBar 38 modes. The Belle inclusive analysis calculates the branching fraction for different photon energy cut-offs.
\label{t:br1}}
\begin{center}
    \begin{tabular}{llccc}\hline
Experiment & $E_{\mathrm{min}}$ & $\mathcal{B}$ $(10^{-6})$
&  $\mathcal{B}_{\rm Modif.}(10^{-6})$\cite{matt}
 & $\mathcal{B}_{\rm Modif.}(10^{-6})$\cite{dge}\\
  & & at $E_{\mathrm{min}}$ & $E_{\mathrm{min}}=1.6$ GeV & $E_{\mathrm{min}}=1.6$ GeV \\\hline
  CLEO(Incl.) \cite{cleoi}        & $2.0$  & $306 \pm 41 \pm 26$   & $337$ & $323$ \\
  Belle(Semi-inc) \cite{belles}   & $2.24$ & $336 \pm 53 \pm 42$  & $496$ & $434$ \\
BaBar(Semi-inc) \cite{babars}  & $1.9$  & $327 \pm 18^{+55}_{-40}$ & $354$  & $337$ \\
 BaBar(Incl.) \cite{babari}      & $1.9$  & $367 \pm 29 \pm 34$  &  $397$ & $378$ \\
 BaBar(recoil) \cite{babarf}        & $1.9$  & $366 \pm 85 \pm 60$    & $396$ & $377$ \\
Belle(Incl.) \cite{bellei}       & $1.7$  & $331 \pm 19 \pm 37$ &   $337$ & $333$ \\
Belle(Incl.) \cite{bellei}       & $1.8$  & $324 \pm 17 \pm 24$  &  $339$ & $329$ \\
Belle(Incl.) \cite{bellei}     & $1.9$  & $312 \pm 15 \pm 16$   &  $338$    & $321$ \\
 Belle(Incl.) \cite{bellei}       & $2.0$  & $294 \pm 14 \pm 12$  &  $334$ & $310$
  \\ \hline
  Average              & 1.6  & -- & $350 \pm 14_{exp} \pm 5_{m_b} \pm 8_{\mu_\pi^2}$ & $322 \pm 14_{exp} \pm 3_{m_b}$  \\
  & & &  $\chi^2/ndf=7.3/8$  &  $\chi^2/ndf=8.3/8$ \\
\hline
\end{tabular}
%
\end{center}
\end{table}

New physics contributions may leave the
$B\to X_s\gamma$ rate unaltered, with new  phases  appearing in the decay  loop \cite{cpasy}.
Since the Standard Model prediction of the $CP$ asymmetry ($A_{CPs\gamma}$) is zero in the limit of U-spin symmetry \cite{thurt},
significant non-zero values would be evidence for new phenomena.
$A_{CPs\gamma}$ has been measured by Belle, BaBar and CLEO. The values of all such measurements are listed in Table~\ref{cpsg}.
All measurements are in agreement with the Standard Model predictions. BaBar also report a $CP$-asymmetry for $b \to (d+s) \gamma$ of $A_{CPs\gamma}=\err{-0.11}{0.12}{0.02}$ \cite{babari},
measured using an inclusive analysis with a lepton tag.

\begin{table}[hbt]
\caption{$CP$ Asymmetries for exclusive and inclusive $b\to s\gamma$ transitions for $B^0$, $B^\pm$ and $B^0/B^\pm$ admixture.
\label{cpsg}}
\begin{center}
\begin{tabular}{lcc}
\hline
Experiment & $K^*\gamma$   & $A_{CP}$  \\ \hline
BaBar \cite{311} &   & {$-0.013\pm 0.036\pm 0.010$}    \\
Belle \cite{46} &  & {$-0.015\pm 0.044\pm 0.012$} \\
CLEO \cite{72} &    &  $0.08\pm 0.13\pm 0.03$ \\
Average \cite{hfag08}  &  & $-0.010 \pm 0.028$   \\\hline
Experiment & $X_s\gamma$   & $A_{CP}$  \\ \hline
BaBar \cite{2} &   &$\err{-0.011}{0.030}{0.014}$\\
Belle \cite{51}&   & $\err{0.002}{0.050}{0.030}$\\
CLEO \cite{74} &    & $\err{-0.079}{0.108}{0.022}$  \\
Average \cite{hfag08}  &  &$-0.012 \pm 0.028$                                \\
\hline
\end{tabular}
\end{center}
\end{table}

The B-factories also attempted to measure the polarization of the photon in the $b \to s \gamma$ transition, which can provide a test of the Standard Model, which
predicts the photon to be mainly left-handed \cite{cpasy}. The  measurements rely on either the exploitation of the $B^0-\bar{B^0}$ interference  \cite{cpasy} or the interference between higher-Kaon resonances decaying into $K\pi\pi^0$ \cite{pol}. However, both Belle and BaBar \cite{exppol} measurements are not precise enough to pin down the photon polarization.
The list of established decay modes of the type $B \to K^{(*)} X\gamma$, where $X$ is one or more flavorless mesons, is increasing. These channels are interesting as they can provide a measurement of the polarization of the photon. Many decay involving kaon resonances modes (e.g. $K^*(892), K_1(1270), K^*_2(1430)$) are now well established and their branching ratios  agree with the Standard Model predictions.

\subsection{{$B\to \ell\ell$}}

Flavor changing neutral current (FCNC) decays proceed only through loop diagrams in the Standard Model, and are further reduced by helicity and Glashow-Iliopoulos-Maiani (GIM) suppression. Thus even the largest branching fraction predicted by the Standard Model, ${\cal B}_{SM}(\Bs\to \mu^+\mu^-)\sim 4\times 10^{-9}$, is very tiny. New physics can enter either at tree level, for example through the presence of an additional $Z^\prime$ boson, or through new particles appearing in loops, and can increase the actual value of these branching fractions.
Decay modes involving two muons in the final state are particularly interesting because they are amenable to measurements in experiments operating at hadronic machines. In fact the best upper limits come from CDF and D0, which are reaching a sensitivity comparable to the SM prediction. Table \ref{mumu:tab} shows a summary of the present experimental values.

\begin{table}
\begin{center}
\caption{\label{mumu:tab} Summary of upper limits for $B\to \ell\ell$ decays.}
\begin{tabular}{llc}
\hline
Mode & Experiment & UL 90\% CL ($\times 10^8$)\\
\hline
$\Bz\to\mup\mum$ &BaBar\cite{babarmm} & 5.2\\
$\Bz\to\mup\mum$ &CDF\cite{cdfmm} & 1.5\\
$\Bs\to\mup\mum$ &CDF\cite{cdfmm} & 4.7\\
$\Bs\to\mup\mum$ &D0\cite{d0mms} & 7.5 \\
$\Bz\to e^+e^-$ & BaBar\cite{babaree} & 11.3\\
$\Bz\to e^+e^-$  & Belle\cite{belleee} & 19\\
$\Bz\to e^+e^-$ & CDF\cite{cdee} & 8.3 \\
$\Bs\to e^+e^-$ & CDF\cite{cdee} & 28 \\
\hline
\end{tabular}
\end{center}
\end{table}

\subsection{{$B\to X_s \ell\ell$}}

The physics of $B\to X_s\ell\ell$ is governed by the Wilson coefficients $C_7$, $C_9$ and $C_{10}$, which describe
the strengths of the corresponding short-distance operators in the effective Hamiltonian, i.e. the
electromagnetic operator $O_7$ and the semileptonic vector and axial-vector operators $O_9$ and $O_{10}$,
respectively \cite{BBL}. The Wilson coefficients are experimental observables. Contributions from new
physics appear in the experiment as deviations from the SM values, which have been calculated
to next-to-next-to-leading order (NNLO).
The experimental knowledge on the Wilson coefficient $C_7$ comes from the inclusive $b\to X_s\gamma$
branching fraction, which determines its absolute value to about 20\% accuracy,
but not its sign. The partial $b\to X_s\ell\ell$
decay rate in the lepton invariant mass range below the $J/\psi$ resonance is sensitive to the sign of
$C_7$.
Measurements of the inclusive $b\to X_s\ell\ell$
decay rate have been published by  Belle \cite{bellefisrt} and BaBar \cite{second} (see
Table~\ref{xll}, who also report a direct $CP$ asymmetry compatible with zero, see Table \ref{cpll}.

The inclusive $B\to X_s\ell\ell$ branching fraction, which constrains $C_9$  and
$C_{10}$  \cite{alilungji},  gives no information on the individual signs
and magnitudes of these coefficients. To further pin down the values of these coefficients, it is
necessary to exploit interference effects between the contributions from different operators. This
is possible in  $B\to X_s\ell\ell$ decays by evaluating the differential inclusive decay rate as a function of
the lepton invariant mass, $q^2=m(\ell\ell)^2$, or by measuring the forward-backward asymmetry
in the exclusive decay $B \to K^*\ell\ell$.
\begin{table}
\caption{Summary of inclusive and exclusive $b\to X_s\ell\ell$ branching fractions. \label{xll}}
\begin{center}
\begin{tabular}{lccccc}
\hline
Mode & BaBar\cite{3}  & Belle\cite{31} & CDF\cite{65} & Average\cite{hfag08}  \\ \hline
 $K^+\ell^+\ell^-$                                 & 
$\aerr{0.38}{0.09}{0.08}{0.02}$                & 
$\aerr{0.53}{0.06}{0.05}{0.03}$            & 
--                                                             & 
$0.49 \pm 0.05$                                   \\
$K^+e^+e^-$                                       & 
{$\aerr{0.42}{0.12}{0.11}{0.02}$}                 & 
$\aerr{0.57}{0.09}{0.08}{0.03}$           & 
--                                                                            & 
$0.52 \pm 0.07$                                   \\

$K^+\mu^+\mu^-$                                   & 
{ $\aerr{0.31}{0.15}{0.12}{0.03}$}                & 
$\berr{0.53}{0.08}{0.07}{0.03}$            & 
$0.59\pm0.15\pm0.04$                     & 
$\cerr{0.50}{0.07}{0.06}$                         \\

$K^{*+}\ell^+\ell^-$                              & 
{$\aerr{0.73}{0.50}{0.42}{0.21}$}                 & 
 $\aerr{1.24}{0.23}{0.20}{0.12}$           & 
--                                                                      & 
$\cerr{1.15}{0.23}{0.21}$                         \\

$K^{*+}e^+e^-$                                    & 
{$\aerr{0.75}{0.76}{0.65}{0.38}$}                 & 
$\aerr{1.64}{0.50}{0.42}{0.18}$            & 
--                                                                           & 
$\cerr{1.42}{0.43}{0.39}$                         \\

$K^{*+}\mu^+\mu^-$                                & 
{$\aerr{0.97}{0.94}{0.69}{0.14}$}                 & 
$\aerr{1.14}{0.32}{0.27}{0.10}$            & 
--                                                                         & 
$\cerr{1.12}{0.32}{0.27}$                         \\ \hline
\end{tabular}
 \end{center}
\end{table}

\begin{table}
\caption{$CP$ Asymmetries for exclusive and inclusive  $b\to X_s\ell\ell$ transitions for $B^0$, $B^\pm$ and $B^0/B^\pm$ admixture. \label{cpll}}
\begin{center}
\begin{tabular}{lcccc}
\hline
  Mode &  BaBar\cite{12,14}  & Belle \cite{31}     &  Average\cite{hfag08}   \\ \hline

$K^+\ell\ell$                                     & 
$\err{-0.07}{0.22}{0.02}$                 & 
$\err{-0.04}{0.10}{0.02}$                & 
$-0.05 \pm 0.09$                                  \\
$K^{*+}\ell\ell$                                  & 
--                                       & 
 $\err{-0.13}{0.17}{0.02}$                  & 
$-0.13 \pm 0.17$                                  \\
$K^0\ell\ell$                                     & 
--                                                   & 
 $\err{-0.08}{0.12}{0.02}$              & 
$-0.08 \pm 0.12$                                  \\
$K^{*0}\ell\ell$                                  & 
--                                                       & 
$\err{-0.10}{0.10}{0.02}$            & 
$-0.10 \pm 0.10$                                  \\
$K^*\ell\ell$                                     & 
$\err{0.03}{0.23}{0.03}$             & 
--                                             & 
$0.03 \pm 0.23$                                   \\
$s\ell\ell$                                       & 
$\err{-0.22}{0.26}{0.02}$                & 
--                                                      & 
$-0.22 \pm 0.26$                                  \\
$K^*\ell\ell$                                     & 
$\err{0.03}{0.23}{0.03}$                      & 
--                                                      & 
$0.03 \pm 0.23$                                   \\
\hline
\end{tabular}
\end{center}
\end{table}

The forward-backward asymmetry in $B \to K^*\ell\ell$, defined as
\begin{equation}
A_{FB}(q^2) = {N(q^2;\,\theta_{B\ell^+}>\theta_{B\ell^-})
           - N(q^2;\,\theta_{B\ell^+}<\theta_{B\ell^-}) \over
             N(q^2;\,\theta_{B\ell^+}>\theta_{B\ell^-})
           + N(q^2;\,\theta_{B\ell^+}<\theta_{B\ell^-})},
\label{eq:afb}
\end{equation}
is a function of $q^2$ and of $\theta_{B\ell^-}$, the angle between the negative lepton and the $B$ meson. It
is an ideal quantity to disentangle the Wilson coefficients $C_i$ since
the numerator of Eq.~\ref{eq:afb} can be expressed as
\begin{equation}
-C_{10}\xi(q^2) \times \left[{\rm Re}(C_9) F_1 + {1\over q^2} C_7
 F_2\right]
\end{equation}
where $\xi$ is a function of $q^2$, and $F_{1,2}$ are functions of form
factors.
It is straightforward to determine $C_{10}$, ${\rm Re}(C_9)$ and the
sign of $C_7$ from the $A_{FB}$ distribution as a function of $q^2$, using
the value of $|C_7|$ from  $B\to X_s\gamma$ and a few more assumptions:
phases of $C_{10}$ and $C_7$ are neglected, and higher order corrections
are known.  These assumptions should be examined by comparing the
results with the inclusive $B\to X_s\ell\ell$  differential branching fraction as
a function of $q^2$, since it is also sensitive to $C_9$ and $C_{10}$ in
a different way.
Most of the eight individual $B\to K (K^*)\ell\ell$ modes have been established.
Both experiments have searched for asymmetries with respect to lepton flavour. Currently available
data clearly favour a negative sign for $C_7$, as predicted by the Standard Model.

In some SUSY scenarios
the sign of the $b\to s\gamma$ amplitude ($C_7$) can be opposite to the SM
prediction, while the transition rate may be the same.
Within the SM there is a zero crossing point of the forward-backward
asymmetry in the low $q^2$ region, while it disappears with the opposite
sign $C_7$ if the sign of ${\rm Re}(C_9)$ is the same as in the SM.  In
another model with $SU(2)$ singlet down-type quarks, tree-level $Z$
flavor-changing-neutral-currents are induced.  In this case, the larger
effect is expected on the axial-vector coupling ($C_{10}$) to the
dilepton than on the vector coupling ($C_9$).  Because the
forward-backward asymmetry is proportional to the axial-vector coupling,
the sign of the asymmetry can be opposite to the SM.  The same new
physics effect is also a possibility for $B\to\phi K_S$ where anomalous
mixing-induced $CP$ violation can occur.

Belle attempted to  extract $C_9$ and $C_{10}$ from $A_{FB}$ in
$B\to K^*\ell \ell$ with 357~fb$^{-1}$ data \cite{a910}. The Belle analysis constrains the sign of the product $C_9
C_{10}$ to be negative as in the SM.  In this study higher order QCD
correction terms are assumed to be the same as in the Standard Model.  Only the
leading order terms, $C_9$ and $C_{10}$, are allowed to float in the fit to the
data.

Since $B^0\to K^{0*}\ell^+\ell^-$ is an all charged particle final state, LHCb may be
able to measure the zero crossing point with a better precision than a $B$ factory.  However, a model independent analysis requires measurement of the forward-backward
asymmetry in  $B\to X_s\ell\ell$, which is only possible at a super-$B$ factory.

The precision of the $B\to K (K^*)\ell\ell$ branching fractions is dominated by the theoretical errors, which have large model dependent irreducible
uncertainties  in the form-factors,  which can be as large as a factor two.
These uncertainties are much smaller if  the ratio $R_{K^{(*)}}={\cal{B}}(B\to K (K^*) \mu^+\mu^-)/{\cal{B}}(B\to K (K^*) e^+e^-)$ is measured.
In the SM, the  $B\to K e^+e^-$ and $B\to K \mu^+\mu^-$ branching fraction are almost equal, a part from a small phase space difference due to
the lepton masses.  However, the  $B\to K^*e^+e^-$ branching fraction is expected to be
 larger than the $B\to K^*\mu^+\mu^-$ branching fraction, due to a larger interference contribution from $B\to K^*\gamma$ in $B\to K^*e^+e^-$.
In new physics models with a different Higgs sector than that of the SM,
scalar and pseudo-scalar types of interactions may arise in $b \to s \ell\ell$.
Therefore, $R_K={\cal{B}}(B\to K \mu^+\mu^-)/{\cal{B}}(B\to K e^+e^-)$ is an observable that is sensitive to new physics. For example,  a neutral SUSY Higgs contribution can
significantly enhance only the $B\to K (K^*) \mu^+\mu^-$ channel at large $\tan\beta$.
The current world average for the branching fractions gives:
$R_K=0.96\pm0.26$,
dominated by the statistical error.

Finally, the transition $b\to s \nu\bar{\nu}$  is theoretically much cleaner than its charged counterpart, thanks to the absence of the photon penguin
diagram and hadronic long-distance effects (charmonium resonances). From the experimental
point of view these modes are very challenging due to the presence of the
two neutrinos. Searches for such modes at the B factories use a fully reconstructed $B$ sample or semileptonic tagged mode. Both Belle and BaBar report limits for these modes.

\section{Current Status of Overall CKM fits}
\label{sec:CKMfits}

We have seen that there are no startling departures from the Standard Model. Yet, we need to formulate quantitatively how restrictive the measurements are. First of all, we can demonstrate that the data are consistent among the different measurements by plotting the extracted values of $\overline{\rho}$ versus $\overline{\eta}$ along, with their experimental and theoretical errors, shown in Figure~\ref{CKMfits}, from many of the measurements described above, also including the measurement of $CP$ violation in the neutral kaon system, designed as ``$\epsilon_K$." Physics beyond the Standard Model could be revealed by seeing a particular measurement that is not in agreement with the others. The width of the bands are the one-standard deviation combined overall errors. For many of the measurements, e.g. $|V_{ub}|$, $\epsilon_K$, $\Delta m_d$ and $\Delta m_s$, the  theoretical errors dominate and thus how they are treated is important. The CKM fitter \cite{CKMfitter} group generally treats the errors more conservatively than the UT fit group \cite{UTfit}. The former uses a frequentist statistical approach, while the later a Bayesian approach. The Bayesian errors are smaller, but if the question is one of seeing new physics or not, a conservative approach may be better.
\begin{figure}[htb]
\begin{center}
\vspace{7mm}
\includegraphics[width=4in]{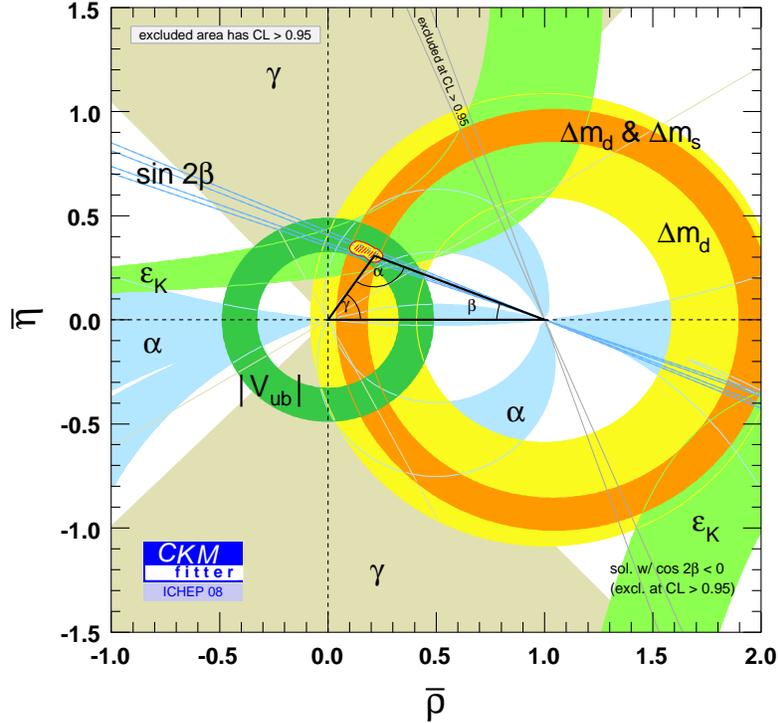}
\vspace{-3mm}
   \caption[] {
    The projections in the $\overline{\rho}$-$\overline{\eta}$ plane from measurements of the angles and sides of the unitarity triangle. The colored area at the apex shows the 95\% confidence level point.  (From the CKM fitter group \cite{Charles}).
   }
   \label{CKMfits}
   \end{center}
   \end{figure}
The results for the Standard Model CKM parameters from both groups are listed in Table~\ref{tab:CKM}.
\begin{table}[htb]
\begin{center}
\caption{Values of Wolfenstein parameters from CKM fits assuming the Standard Model}
\label{tab:CKM}
\begin{tabular}{lcccc}\hline\hline
Group & $\overline{\rho}$ &$\overline{\eta}$ & $A$ & $\lambda$ \\\hline
CKM fitter & $0.135^{+0.033}_{-0.016}$ & $0.345^{+0.015}_{-0.018}$ & $0.795^{+0.025}_{-0.015}$ & $0.2252\pm 0.0008$ \\
UT fit & 0.155$\pm$0.022 & 0.342$\pm$0.014 &  &\\
\hline
\end{tabular}
\end{center}
\end{table}
The central values of $\overline{\rho}$ and $\overline{\eta}$ are in agreement. The value of $\lambda$ is in agreement with that found by Blucher and Marciano \cite{MB}. The fact that there is an overlap region common to all the measurements shows no obvious deviation from the Standard Model, which does not predict values for $\overline{\rho}$ and $\overline{\eta}$, but merely that the various measurements must find the same values.

To search for deviations from the SM it is useful to compare measurements involving $CP$ violating loop diagrams, where imaginary parts of amplitudes come into play, with measurements of the sides of triangle. The UT fit group shows in Figure~\ref{UT-comp} a comparison in the $\overline{\rho}$-$\overline{\eta}$ plane of angle measurements with those of the sides of the unitarity triangle. The overlap regions in both cases are consistent.
\begin{figure}[htb]
\begin{center}
\centerline{\epsfig{figure=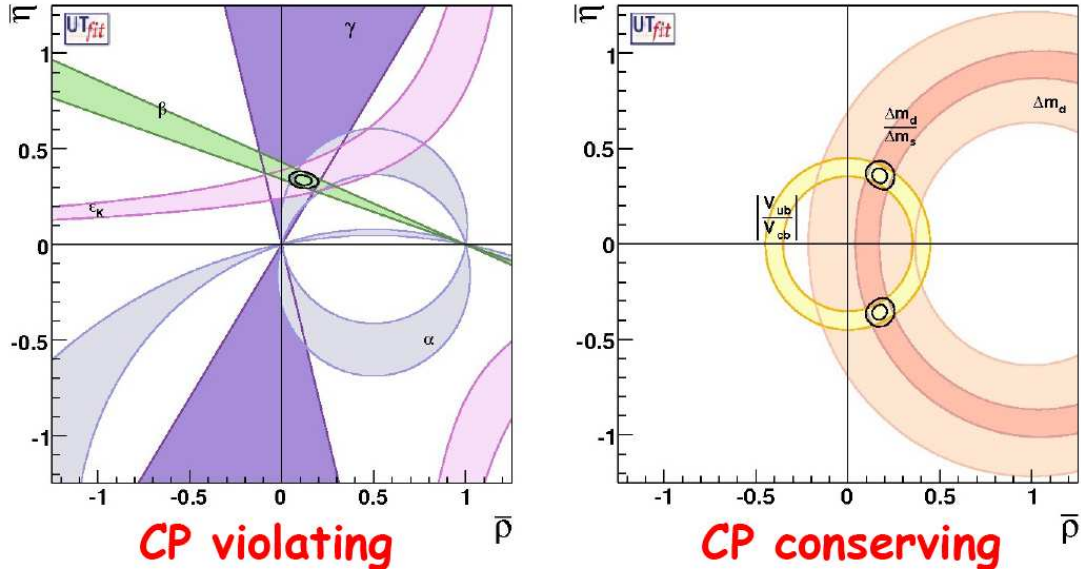,height=3in}}
   \caption{
   Constraints at 68\% confidence level in the $\overline{\rho}$-$\overline{\eta}$
plane. On the left side are measurements of the $CP$ violating angles $\alpha$, $\beta$
and $\gamma$ and $\epsilon_K$ in the neutral kaon system. On the right side are measurements of $|V_{ub}|/V_{cb}|$ and the ratio of $B_s$ to $B_d$ mixing. (From the UT fit group \cite{UTfit-CKM}).
   }
   \label{UT-comp}
   \end{center}
   \end{figure}

It remains to set limits on New Physics. We allow for NP to contribute to the mixing amplitude of either $B_d$ or $B_s$ mesons, which we consider separately. Then the amplitude of this second order ($\Delta F=2$) flavor changing interaction can be expressed as
\begin{equation}
\langle B_q \left| {\cal H}^{SM+NP}_{\Delta F=2} \right|\overline{B}_q \rangle=
\left[{\rm Re} (\Delta_q)+i {\rm Im}(\Delta_q)\right]\cdot
\langle B_q \left| {\cal H}^{SM+NP}_{\Delta F=2} \right|\overline{B}_q \rangle~~,
\end{equation}
where $q$ specifies the type of neutral $B$ meson, and $\Delta_q$ gives the relative size of NP to SM physics. (The SM point has Re$(\Delta_q)$=1 and Im$(\Delta_q)$=0.)

 To set limits it is necessary to make some simplifying, but very reasonable, assumptions \cite{NMFV}.  We assume that there is no NP in the tree level observables starting with the magnitudes $|V_{ud}|$, $|V_{us}|$, $|V_{cb}|$, and $|V_{ub}|$. Since since $\gamma$ is measured using tree level $B^{\mp}$ decays we include it in this category.
 Now we incorporate the measurement of $\beta$ by noting that $\alpha=\pi-\gamma-\beta^{\rm meas}$, allowing that the measured $\beta^{\rm meas}=\beta^{\rm SM}-{\rm Arg}(\Delta_d)$. Also used are measurements of the semileptonic asymmetries (see Eq.~\ref{eq:asl}) \cite{ASL}, and the constraint equation $\Delta\Gamma_s=\cos\phi_s\Delta\Gamma_s^{SM}$ \cite{LenzN}.
 The allowed range in the $\overline{\rho}$-$\overline{\eta}$ plane for NP via $B$ mixing is shown in Figure~\ref{NP}.

 \begin{figure}[htb]
\begin{center}
\centerline{\epsfig{figure=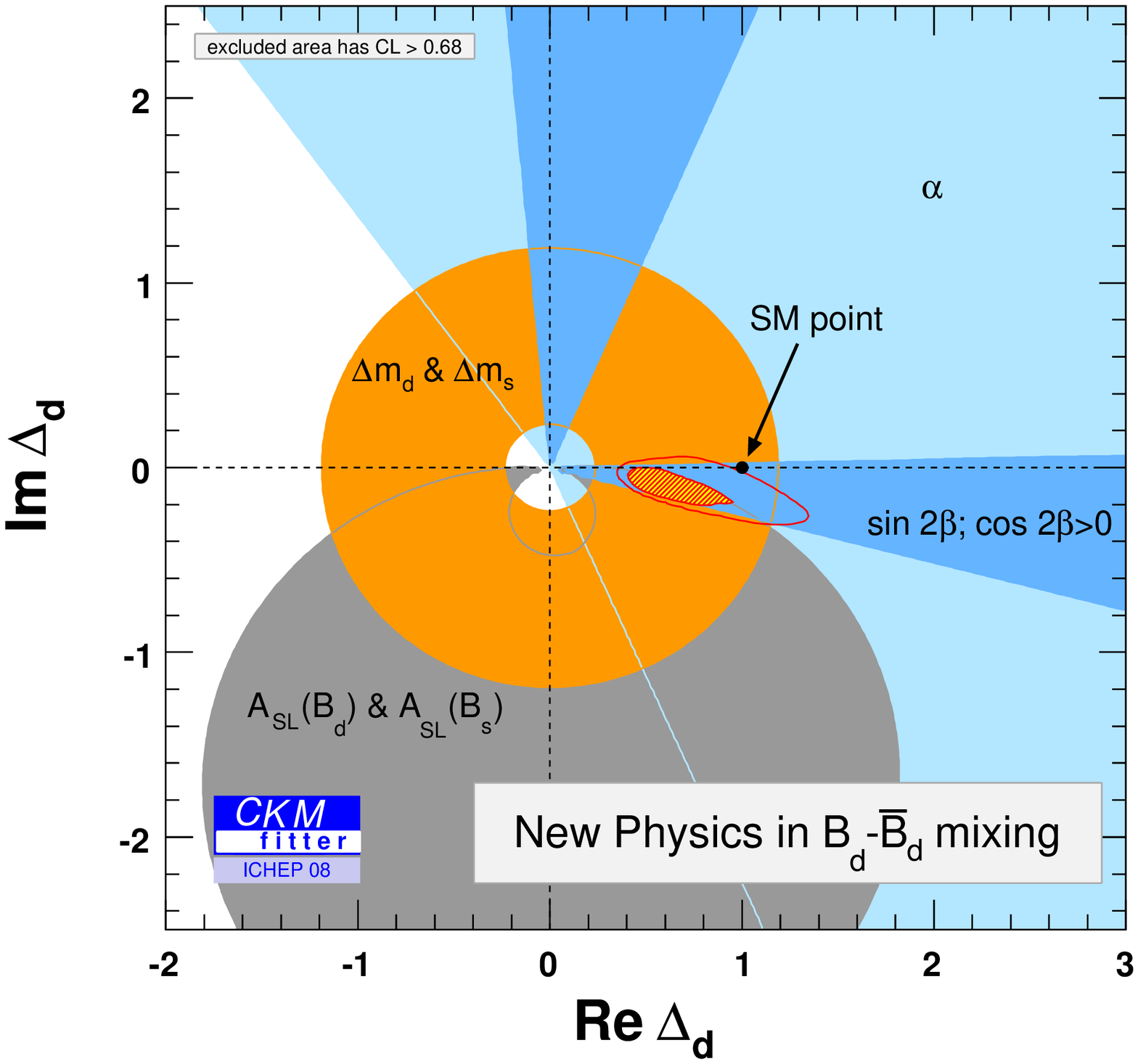,height=3in}\hspace{5mm}
\epsfig{figure=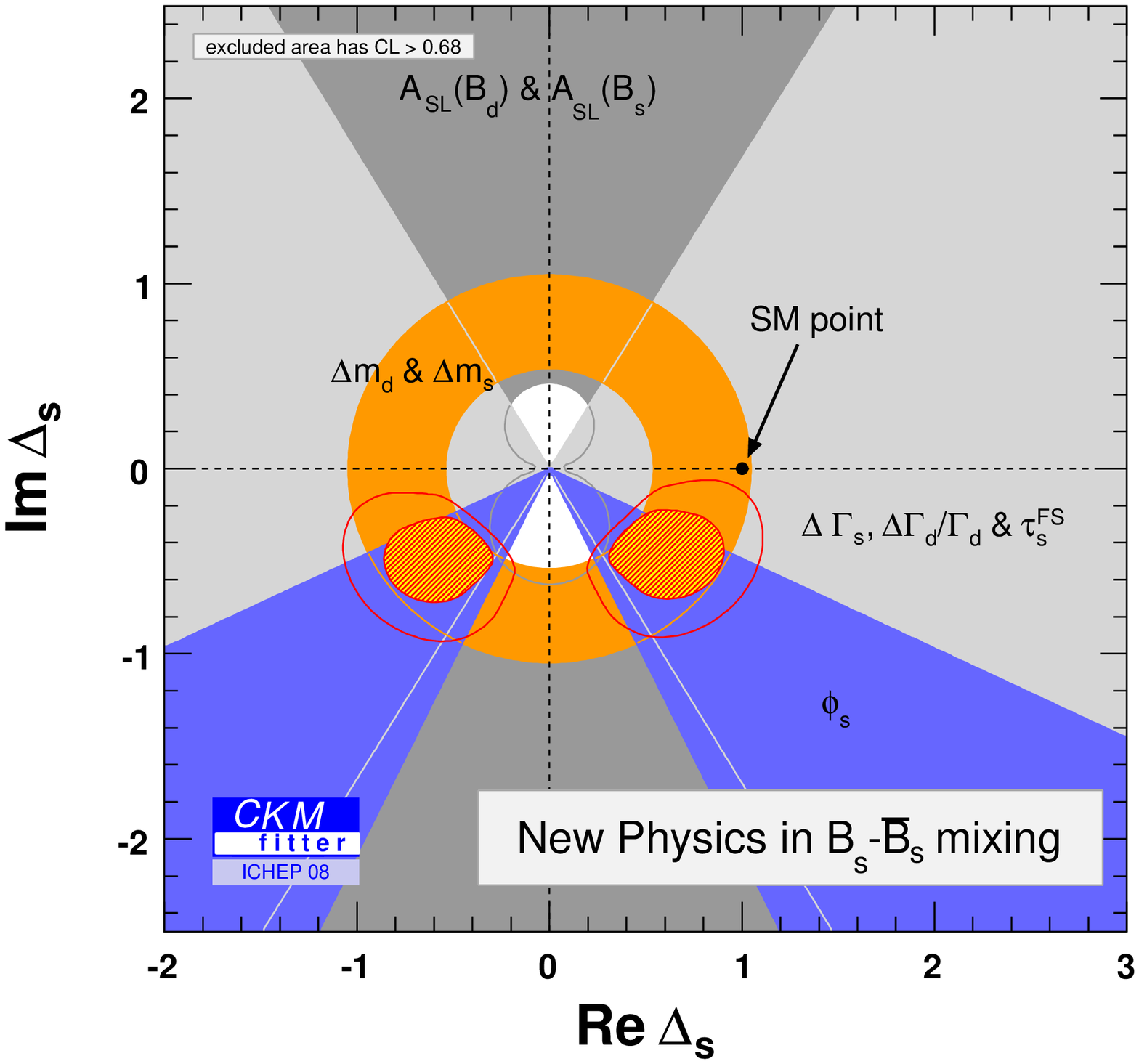,height=3in}}
   \caption{
Constraints on the imaginary part versus the real part of a New Physics amplitude proceedings via $B$ mixing for (left) $B_d$ mesons and (right) $B_s$ mesons. The SM point is at (1,0). The contours shown are all at 68\% confidence level. (From the CM fitter group \cite{Charles}).
   }
   \label{NP}
   \end{center}
   \end{figure}

In both case quite large values of the real part of NP amplitudes are
allowed. In the $B_s$ case a large negative value of the imaginary part
is also preferred, but again the effect is only 2.7$\sigma$.
We can conclude that we need to do a lot more work to constrain NP
in these transitions.

\section{Conclusions}
The goals of our studies are to elucidate the features of the electroweak interactions and the fundamental constituents of nature. The $b$ was discovered at fixed target experiment in a proton beam, by the first observation of narrow $\Upsilon$ states in 1977. Experiments at several $e^+e^-$ colliders have increased our knowledge enormously. CLEO provided the first fully reconstructed $B$ mesons and measurements of their mass. The Mark II and MAC experiments at PEP gave the first measurements of the $b$ lifetime, which were confirmed by PETRA experiments. ARGUS first measured $B^0$ mixing. CLEO, ARGUS and the LEP experiments competed on the first precision measurements of the CKM parameters $V_{cb}$ and $V_{ub}$. The two most recent, BaBar and Belle, have provided measurements of $CP$ violation.

The hadron collider experiments have measured $B_s$ mixing and are starting to probe $CP$ violation in the $B_s$ system. The torch will soon be passed to first hadron collider experiment designed to measure $b$ decays, LHCb. The ATLAS and CMS experiments will also contribute.

In the 31 years since the discovery of the $b$ quark we have learned a great deal about the physics of $b$-flavored hadrons. We have seen that the Standard Model describes most decays well, yet studies continue as any new particles that we believe must exist have influence, especially on rare and $CP$ violating $B$ decays. We expect that future experiments will learn a great deal about these new particles either by observing their effects on $B$ decays, or by seeing that they do not have observable effects. The latter case is one that experimentalists do not like, but it can be just as important in firming up our understanding of the basic constituents of matter and the forces that operate between them.

\section*{Acknowledgements}
We thank Phillip Urquijo
and Antonio Limosani for useful discussions concerning the CKM elements.
M. Artuso and S. Stone thank the U. S. National Science Foundation for support.

\end{physmath}

\begin{thebibliography}{999}
\bibitem{Cabibbo} N. Cabibbo, Phys. Rev. Lett. {\bf 10}, 531 (1963).

\bibitem{CP} J. H. Christenson, J. W. Cronin, V. L. Fitch and R. Turlay, Phys. Rev.
    Lett., {\bf 13}, 138 (1964). In $CP$, $C$ stands for charge conservation invariance
    and $P$ for parity invariance. The operation of $C$ takes a particle to
    anti-particle state and the operation of $P$ changes left to right. See the article by
    John Ellis in the CERN courier \url{http://cerncourier.com/cws/article/cern/28092}.
    Also see Yosi Nir in \url{http://www.symmetrymagazine.org/cms/?pid=1000194}. Streaming
    video of several talks is available at
    \url{http://www-project.slac.stanford.edu/streaming-media/CPViolation/CPViolation.htm}.

\bibitem{Bigi-Sanda} An excellent resource is the book by I. I. Bigi and A. I. Sanda,
    ``CP Violation," Cambridge Monographs on Particle Physics, Nuclear Physics and
    Cosmology (No. 9), (1999).

\bibitem{TR} J. J. Aubert \etal, Phys. Rev. Lett. {\bf 33}, 1404 (1974); J. E.
    Augustin \etal, Phys. Rev. Lett. {\bf 33}, 1406 (1974).

\bibitem{Glashow} J. D. Bjorken and S. L. Glashow, Phys. Lett. {\bf 11}, 255 (1964);
    S. L. Glashow, J. Iliopoulos and L. Maiani, Phys. Rev. {\bf D2}, 1412 (1972); S.
    L. Glashow, ``Charm: An Invention Awaits Discovery," IV Int. Conf. on Experimental
    Meson Spectroscopy, Apr, 1974, Boston, Mass. Published in AIP Conf. Proc. {\bf
    21}, 387 (1974).

\bibitem{KM} M. Kobayashi and T. Maskawa, Progress in Theoretical Physics {\bf 49},
    652 (1973).

\bibitem{PDG} Particle Data Group, C. Amsler et al., Phys. Lett. B {\bf 667}, 1
    (2008).

\bibitem{SM} S. L. Glashow, Nucl. Phys. {\bf 22} 579, (1961); S. Weinberg, Phys. Rev.
    Lett. {\bf 19} 1264, (1967); A. Salam in Elementary Particle Physics (Nobel Symp.
    N.8), Ed. N.Svartholm, Almquist and Wiksells, Stockholm (1968), p.367.

\bibitem{Higgs} To view a cartoon of the mechanism of Higgs mass generation see
    \url{http://www.exploratorium.edu/origins/cern/ideas/higgs.html}.

\bibitem{Wolf} L. Wolfenstein, Phys. Rev. Lett. {\bf 51}, 1945 (1983).

\bibitem{CKMfitter} J. Charles \etal ~Eur. Phys. J. {\bf C41}, 1 (2005); see also
    \url{http://www.slac.stanford.edu/xorg/ckmfitter/ckm\_intro.html}.

\bibitem{akl} R. Aleksan \etal , Phys. Rev. Lett. {\bf 73}, (1994) 18; J. P. Silva and
    L. Wolfenstein, Phys. Rev. D{\bf 55}, 5331 (1997).



\bibitem{Zwicky} M.~A.~Srednicki,  ``Particle Physics and Cosmology: Dark Matter,"
    Amsterdam, Netherlands, North-Holland (1990).

\bibitem{Trodden} D. Hunter, G. D. Starkman and M. Trodden, Phys. Rev. {\bf D66},
 043511 (2002).

\bibitem{ExtraD} J. Hewett and  M. Spiropulu, Ann. Rev. Nucl. Part. Sci. {\bf 52}, 397
    (2002).

\bibitem{Sakh} A.~Sakharov, JETP, \textbf{5}, 32 (1967).

\bibitem{Gavela}
M. B. Gavela, P Hernandez, J. Orloff and O. Pene, Modern Phys. Lett. A {\bf 9}, 795 (1994), and references contained therein.

\bibitem{Lisa} \url{http://www.esi-topics.com/brane/interviews/DrLisaRandall.html}.

\bibitem{Lykken}
    \url{http://www.slac.stanford.edu/econf/C040802/lec\_notes/Lykken/default.htm}.

\bibitem{UTfit} See \url{http://www.utfit.org/}.

\bibitem{Albrecht1983} H. Albrecht \etal ~(ARGUS), Phys. Lett. B
\textbf{192}, 245 (1987).

\bibitem{Albajar1987} C. Albajar \etal~(UA1), Phys. Lett. B \textbf{186}, 247 (1987),
    erratum-ibid B {\bf 197} 565, (1987).

\bibitem{Akers1995} R. Akers \etal~(OPAL), Z. Phys. C \textbf{66}, 555 (1995).

\bibitem{HFAG} E. Barberio \etal~(HFAG), ``Averages of b-hadron and c-hadron
    Properties at the End of 2007," arXiv:0808.1297 [hep-ex].

\bibitem{Gaillard1974} M. Gaillard and B. Lee, Phys. Rev. D \textbf{10}, 897 (1974).


\bibitem{CDF-Bs} A. Abulencia, \etal ~(CDF), Phys. Rev. Lett. {\bf 97}, 062003 (2006)
    [arXiv:hep-ex/0606027v1], ibid {\bf 97}, 214003 (2006) [arXiv:hep-ex/0609040v1].

\bibitem{D0-Bs} V. Abazov \etal~(D0), Phys. Rev. Lett. {\bf 97}, 021802 (2006)
    [arXiv:hep-ex/0603029v2].

\bibitem{MOSER} H.~G.~Moser and A.~Roussarie,
  Nucl.~Instrum.~Methods Phys.~Res., Sect.~A {\bf 384}, 491 (1997).

\bibitem{Follana} E. Follana, C. T. H. Davies, G. P. Lepage, and J. Shigemitsu, Phys.
    Rev. Lett. {\bf 100}, 062002 (2008,) arXiv:0706.1726v2 [hep-lat].

\bibitem{CLEOfDp} B. I. Eisenstein \etal~(CLEO), Phys. Rev. D {\bf 78}, 052003
    (2008), arXiv:0806.2112v3 [hep-ex].

\bibitem{Rosner-Stone} J. Rosner and S. Stone, ``Decay Constants of Charged
    Pseudoscalar Mesons," review for the Particle Data Group \cite{PDG},
    arXiv:0802.1043v3 [hep-ex]. For updated results see P.U.E. Onyisi \etal (CLEO),
arXiv:0901.1147 [hep-ex], and J. P. Alexander \etal (CLEO),
arXiv:0901.1216 [hep-ex].

\bibitem{Tantalo} N. Tantalo, ``Lattice calculations for $B$ and $K$ mixing,"
    [arXiv:hep-ph/0703241v1].



\bibitem{Beneke}
M. Beneke, G. Buchalla, A. Lenz and U. Nierste, Phys. Lett. B {\bf 576}, 173 (2003) [hep-ph/0307344].

\bibitem{sin2beta} B. Aubert \etal~(BABAR), arXiv:0808.1903 [hep-ex]; K.~F. Chen
    \etal~(BELLE) Phys. Rev. Lett. {\bf 98}, 031802 (2007) 	 arXiv:hep-ex/0608039v4
    [hep-ex].

\bibitem{neural}
P. C. Bhat, and H. B. Prosper, ``Multivariate Methods in
High Energy Physics: The Neural Network Revolution, World
Scientific, ISBN-13: 978-9810243470 (2005); see also
\url{http://neuralnets.web.cern.ch/NeuralNets/nnwinhep.html}.


\bibitem{psiKstar} R. Itoh \etal~(BELLE), Phys. Rev. Lett. {\bf 95}, 091601 (2005)
    [hep-ex/0504030]; B. Aubert \etal~{BABAR} Phys. Rev. D {\bf 71}, 032005 (2005)
    [hep-ex/0411016].

\bibitem{alpha-pipi} B. Aubert, \etal~(BABAR), arXiv:0807.4226v2 [hep-ex]; H. Ishino
    \etal~(BELLE), Phys. Rev. Lett. {\bf 98}, 211801 (2007) [hep-ex/0608035].

\bibitem{superweak}
L. Wolfenstein Phys. Lett. {\bf 13}, 562 (1964).

\bibitem{GL} M. Gronau and D. London, Phys. Rev. Lett. {\bf 65}, 3381 (1990).

\bibitem{GQ} Y. Grossman and H. Quinn, Phys. Rev. D {\bf 58}, 017504 (1998). See also
    J. Charles, Phys. Rev. D {\bf 59}, 054007 (1999) [hep-ph/9806468]; M. Gronau, D.
    London, N. Sinha and R. Sinha, Phys. Lett. B {\bf 514}, 315 (2001)
    [hep-ph/0105308].


\bibitem{transversity} I. Dunietz, H. R. Quinn, A. Snyder, W. Toki and H. J. Lipkin,
    Phys. Rev. D {\bf 43}, 2193 (1991).

\bibitem{BaBar-rhorho} B. Aubert \etal~(BABAR) Phys. Rev. D {\bf 76}, 052007 (2007)
    arXiv:0705.2157 [hep-ex].

\bibitem{Belle-rhorho} A. Somov \etal~{BELLE}, Phys. Rev. Lett. {\bf 96}, 171801
    (2006) [arXiv:hep-ex/0601024].

\bibitem{Falk-rhorho} A. F. Falk, Z. Ligeti, Y. Nir, and H. Quinn, Phys. Rev. D {\bf
    69}, 011502 (2004) [hep-ph/0310242].

\bibitem{CKMfit} \url{http://ckmfitter.in2p3.fr/}.


\bibitem{SQ} A. E. Snyder and H. Quinn, Phys. Rev. D {\bf 48}, 2139 (1993).

\bibitem{BaBar-rhopi} B. Aubert \etal~(BABAR), [arXiv:hep-ex/0703008v2].

\bibitem{Belle-rhopi} A. Kusaka \etal~(BELLE), Phys. Rev. Lett. {\bf 98}, 221602 (2007)
    [arXiv:hep-ex/0701015].

\bibitem{isospin-p} H. J. Lipkin, Y. Nir, H. R. Quinn, and A. E. Snyder, Phys. Rev. D
    {\bf 44}, 1454 (1991); M. Gronau, Phys. Lett. B {\bf 265}, 389 (1991).

\bibitem{lhcb-alpha} P. Robbe (LHCb), Nucl. Phys. Proc. Suppl. {\bf 170}, 46 (2007).

\bibitem{controversy} J. Charles, A. Hocker, H. Lacker, F.R. Le Diberder, S.
    T'Jampens, [arXiv:hep-ph/0607246v1].

\bibitem{GW} M. Gronau and D. Wyler, Phys. Lett. B {\bf 265}, 172 (1991); see also M.
    Gronau and D. London, Phys. Lett. B {\bf 253}, 483 (1991).

\bibitem{BG} A. Bondar and T. Gerson, Phys. Rev. D {\bf 70}, 091503 (2004)
    [arXiv:hep-ph/0409281].

\bibitem{GLW-gamma} B. Aubert \etal~{BABAR}, Phys. Rev. D {\bf 77} 111102 (2008)
    arXiv:0802.4052 [hep-ex]; K. Abe \etal~(BELLE), Phys. Rev. D {\bf 73}, 051106
    (2006) [hep-ex/0601032]; K. Gibson~(CDF), arXiv:0809.4809; B. Aubert
    \etal~{BABAR},  	 arXiv:0807.2408v2 [hep-ex]; K. Abe \etal~(BELLE), Phys. Rev.
    D {\bf 73}, 051106 (2006) [hep-ex/0601032]; G. Marchiori~(BABAR) arXiv:0810.0502.

\bibitem{ADS} D. Atwood, I. Dunietz and A. Soni, Phys. Rev. Lett. {\bf 78}, 3257
    (1997) [hep-ph/9612433].

\bibitem{ADS-gamma} Y. Horii \etal~(BELLE), arXiv:0804.2063v1 [hep-ex]; B. Aubert
    \etal~(BABAR) Phys. Rev. D {\bf 72}, 032004 (2005) [hep-ex/0504047].

\bibitem{Dalitz} R.H. Dalitz, Phys. Rev. {\bf 94}, 1046 (1954).

\bibitem{Giri} A. Giri, Y. Grossman, A. Soffer, and J. Zupan, Phys. Rev. D {\bf 68},
    054018 (2003) [arXiv:hep-ph/0303187].


\bibitem{babar-dalitz} B. Aubert \etal~(BABAR), Phys. Rev. D {\bf 78}, 034023 (2008) 	
    arXiv:0804.2089 [hep-ex], they also include the mode $D^0\to K_S K^+K^-$.

\bibitem{belle-dalitz} K. Abe \etal~(BELLE), 	arXiv:0803.3375 [hep-ex];

\bibitem{Bondar} A. Bondar and Poluektov, Eur. Phys. J. {\bf C47}, 347 (2006); A.
    Bondar, A. Poluektov, arXiv:0801.0840 [hep-ex].

\bibitem{Naik}
 P. Naik~(CLEO), arXiv:0810.3666 [hep-ex].

\bibitem{Rosner1} J. L. Rosner, Phys. Rev. D {\bf 42}, 3732 (1990).

\bibitem{Rosner2} A. S. Dighe, I. Dunietz, H. J. Lipkin and J. L. Rosner, Phys. Lett.
    B {\bf 369}, 144 (1996) [hep-ph/9511363].

\bibitem{trans-fig} T. Kuhr (CDF), arXiv:0710.1789 [hep-ex]).

\bibitem{DDF} A. S. Dighe, I. Dunietz and R. Fleischer, Eur. Phys. J. C {\bf 6}, 647
    (1999).

\bibitem{CDF-chi}
T. Aaltonen \etal (CDF), Phys. Rev. Lett. {\bf 100}, 161802 (2008) arXiv:0712.2397v1 [hep-ex];
updated results in D. Tonelli \etal ~(CDF), arXiv:0810.3229v2 [hep-ex].

\bibitem{D0-chi}
V. M. Abazov \etal~(D0), arXiv:0802.2255v1 [hep-ex].

\bibitem{Charles}
J. Charles (CKM Fitter) \url{http://www.slac.stanford.edu/xorg/ckmfitter/TalkBuffer/StatusOfTheCkmMatrixCapri2008.pdf},
see also O. Deschamps (CKM Fitter) 	arXiv:0810.3139v1 [hep-ph].

\bibitem{Blouw}
J. Blouw (LHCb), arXiv:0710.5124v2 [hep-ex].

\bibitem{Stone-Zhang}
S. Stone and L. Zhang, arXiv:0812.2832v3 [hep-ph].

\bibitem{BaBar-psiKstar}
B. Aubertet \etal~(BABAR), Phys. Rev. D{\bf 76}, 031102 (2007)
arXiv:0704.0522v2 [hep-ex].

\bibitem{BaBar-Kpi}
B. Aubert \etal (BABAR), Phys. Rev. Lett. {\bf 93}, 131801 (2004)
[arXiv:hep-ex/0407057].

\bibitem{BaBarlate-Kpi} B. Aubert \etal,~
(BABAR), arXiv:0807.4226 [hep-ex].

\bibitem{Belle-Kpi}
S.-W. Lin \etal (BELLE), Nature {\bf 452}, 332 (2008).

\bibitem{Peskin}
M. Peskin, Nature {\bf 452}, 293 (2008).

\bibitem{Jarlskog} C. Jarlskog, Phys. Rev. D {\bf 35}, 1685 (1987).

\bibitem{BAU} M. Trodden, Rev. Mod. Phys. {\bf 71}, 1463, (1999).


\bibitem{pole_mass}
R.~Tarrach, Nucl. Phys. B {\bf 183}, 384 (1981);
A.~S.~Kronfeld, Phys. Rev. D {\bf 58}, 051501 (1998);
P.~Gambino and P.~A.~Grassi, Phys. Rev. D {\bf 62}, 076002 (2000).

\bibitem{infrared}
I.~I.~Bigi, M.~A.~Shifman, N.~G.~Uraltsev and A.~I.~Vainshtein, Phys. Rev. D {\bf 50}, 2234 (1994);
M.~Beneke and V.~M.~Braun, Nucl. Phys. B {\bf 426}, 301 (1994).

\bibitem{msbar}
N.~Gray, D.~J.~Broadhurst, W.~Grafe and K.~Schilcher, Z.~Phys. C {\bf 48}, 673 (1990);
K.~Melnikov and T. v. Ritbergen, Phys. Lett. B {\bf 482}, 99 (2000);
K.~G.~Chetyrkin and M.~Steinhauser, Nucl. Phys. B {\bf 573}, 617 (2000).

\bibitem{threshold}
A.~H.~Hoang {\it et al.}, Eur. Phys. J. C {\bf 3}, 1 (2000).
\bibitem{bigi} I.~I.~Bigi, M.~A.~Shifman and N.~G.~Uraltsev, Ann. Rev. Nuc. Part. Sci. {\bf  47}, 591 (1997).

\bibitem{ref:2}
  P.~Gambino, private communication (2006).



\bibitem{1s_scheme}
A.~H.~Hoang, Z.~Ligeti and A.~V.~Manohar, Phys. Rev. Lett. {\bf 82}, 277 (1999).
A.~H.~Hoang, Z.~Ligeti and A.~V.~Manohar, Phys. Rev. D {\bf 59}, 074017 (1999).

 \bibitem{gardi} E. Gardi for HFAG 2008 averages.
\bibitem{Lattice_mb} C.McNelle, C. Michael and G. Thompson, Physics Letters B {\bf 600}, 77 (2004).
\bibitem{penin} A. Penin and M. Steinhauser, Phys. Lett. B {\bf 538}, 335 (2002).
\bibitem{khun} J.H Kuhn et al, Nucl. Phys. B {\bf 778}, 192 (2007).
\bibitem{Hoang} A. H. Hoang,  Phys. Lett. B {\bf 483}, 94 (2000).




\bibitem{Isgur-Wise-BBook}
N. Isgur and M. B. Wise, ``Heavy Quark Symmetry," page 231, in {\bf $B$ Decays} Revised 2nd Edition, ed. S. Stone, World
Scientifc, Singapore (1994) .

\bibitem{grinstein}
C. Glenn Boyd, B. Grinstein and R.F. Lebed,
Phys.\ Lett.\ B {\bf 353}, 306 (1995).


\bibitem{Chay:1990da}
  J.~Chay, H.~Georgi and B.~Grinstein,
  Phys.\ Lett.\  B {\bf 247}, 399 (1990).


\bibitem{Laiho:2008pn}  C. Bernard et al.  Submitted to Phys. Rev. D,
arXiv:0808.2519~.


\bibitem{Adam:2002uw}
  N.~E.~Adam \etal~(CLEO),
  Phys.\ Rev.\  D {\bf 67}, 032001 (2003)
  [arXiv:hep-ex/0210040].

\bibitem{CLEO-Vcb}
 J. E. Duboscq {\it et al.} (CLEO),  Phys. Rev. Lett. {\bf 76}, 3898 (1996).

\bibitem{bellen}
I.~Adachi \etal~(BELLE),
 arXiv:0810.1657 [hep-ex].

 \bibitem{hep-ex/0602023}
 B. Aubert  {\it et al.}  (BABAR),   Phys. Rev. D {\bf 74}, 092004 (2006).

  \bibitem{hfag08}
ICHEP08 updates available online at: \url{http://www.slac.stanford.edu/xorg/hfag/semi/ichep08/home.shtml}.

\bibitem{LEP}  D.~Abbaneo {\it et al.}  [ALEPH, CDF, DELPHI, L3, OPAL, SLD],
  arXiv:hep-ex/0112028. See references therein.

\bibitem{BaBar(excl)}   B.~Aubert \etal~(BABAR),
 Phys.Rev.D {\bf 77}, 032002 (2008).
\bibitem{babar07}  	   B.~Aubert \etal~(BABAR),
 arXiv:0712.3493 [hep-ex]
\bibitem{babar08}   B.~Aubert \etal~(BABAR),
 arXiv:0809.0828 [hep-ex]
\bibitem{belle1} K.Abe, {\it et al.}  (BELLE), Phys. Lett .B {\bf 526}, 247 (2002).
\bibitem{cleods} R. A. Briere {\it et al.} (CLEO), Phys. Rev. Lett. {\bf 89}, 081803 (2002).

\bibitem{CLN}
  I. Caprini, L. Lellouch and  M. Neubert,
{Nucl. Phys. B} {\bf 530}, 153 (1998);
C.G. Boyd, B. Grinstein, R.F. Lebed, {Phys. Rev. D} {\bf56}, 6895 (1997).
\bibitem{bard}  B.~Aubert \etal~(BABAR),
 arXiv:0807.4978 [hep-ex].
\bibitem{bard1} B.~Aubert \etal~(BABAR),
arXiv:0809.0828 [hep-ex].
\bibitem{bard2}  B.~Aubert \etal~(BABAR),
arXiv:0712.3503 [hep-ex].
\bibitem{bd}  K. Abe,  {\it et al.}  (BELLE), Phys. Lett. B {\bf 526}, 258 (2002).

\bibitem{cd} J. Bartelt, {\it et al.}  (CLEO), Phys. Rev. Lett. {\bf 82}, 3746 (1999).


\bibitem{IWS} N. Isgur and M. Wise, Phys. Rev. D {\bf 93}, 819 (1991).

\bibitem{Okamoto:2005hr}
  M.~Okamoto  (Fermilab Lattice, MILC and HPQCD Collaborations),
  Int.\ J.\ Mod.\ Phys.\  A {\bf 20}, 3469 (2005).

\bibitem{Aubert:2008yv}
  B.~Aubert \etal~(BABAR),
  arXiv:0809.0828 [hep-ex].


\bibitem{cleod1d2} A. Anastassov {\it et al.}  (CLEO), Phys.\ Rev.\ Lett.\  {\bf 80}, 4127 (1998).

\bibitem{D12D0} V. Abazov {\it et al.} (D0), Phys.\ Rev.\ Lett.\  {\bf 95}, 171803 (1995).

\bibitem{b22}
  B.~Aubert \etal~(BABAR),
  arXiv:0708.1738 [hep-ex].

\bibitem{matsumoto} D.~Liventsev, T.~Matsumoto {\it et al.} (BELLE), Phys. Rev. {\bf{D 72}}, 051109(R) (2005).

\bibitem{kuzmin} K.~Abe {\it et al.} (BELLE), Phys. Rev. Lett. {\bf 94}, 221805 (2005);\\
  K.~Abe {\it et al.} (BELLE), Phys. Rev. {\bf D 69}, 112002 (2004).

\bibitem{Bauer:2004ve}
C.~W.~Bauer {\it et al.},  Phys.\ Rev.\ {\bf D  70}, 94017 (2004).

\bibitem{Buchmuller:2005zv}
  O.~Buchmuller, H.~Flacher, Phys.\ Rev.\ {\bf D  73}, 73008 (2006).

\bibitem{wilson} K.~Wilson, Phys.\ Rev.\ {\bf 179}, 1499 (1969).

\bibitem{Benson:2003kp}
  D.~Benson {\it et al.},   Nucl.\ Phys. {\bf B 665}, 367 (2003).



\bibitem{gremm-kap}M.~Gremm and A.~Kapustin, Phys. Rev. D{\bf 55}, 6924 (1997).

\bibitem{falk} A. Falk, M. Luke, and M.J. Savage, Phys. Rev. D {\bf 53}, 2491 (1996);
A.~F.~Falk and M.~Luke, Phys. Rev. D {\bf 57}, 424 (1998).

\bibitem{Gambino:2004qm}
  P.~Gambino and N.~Uraltsev,   Eur.\ Phys.\ J.\ C {\bf 34}, 181 (2004).

\bibitem{Benson:2004sg}
  D.~Benson {\it et al.}, Nucl.\ Phys.\ B {\bf 710}, 371 (2005).


\bibitem{bb} B.~Aubert {\it et al.} (BABAR),  Phys. Rev. D {\bf 69}, 111103 (2004).

\bibitem{mx} C.~Schwanda {\it et al.}  (BELLE), Phys. Rev. D {\bf 75}, 032005 (2007).

\bibitem{cl} A.H Mahmood {\it et al.}  (CLEO)  Phys. Rev. D {\bf 70}, 32002 (2004).

\bibitem{chen2001}S. Chen  \etal, Phys. Rev. Lett. 87, 251807 (2001).

\bibitem{battaglia} J.~Abdallah {\it et al.} (DELPHI), Eur. Phys. J. {\bf C 45}, 35 (2006).

\bibitem{cdfmom} D. Acosta, {\it et al.} (CDF), Phys.Rev. D {\bf 71}, 051103 (2005).

\bibitem{isgur}
N.~Isgur, Phys. Lett. B {\bf 448}, 111 (1999).

\bibitem{belle} K. Abe \etal~(BELLE), [ArXiv:hep-ex/0408139]; {\it ibid.}  [ArXiv:hep-ex/0409015].

\bibitem{prd2003} A. H. Mahmood {\it et al.}  (CLEO), Phys. Rev. Lett. {\bf 67}, 072001, (2003).


\bibitem{ref:13} A.~H\"ocker and V.~Kartvelishvili, Nucl.\ Instr.\ Meth.\ A {\bf 372} (1996) 469.





\bibitem{el} P.~Urquijo {\it et al.}  (BELLE), Phys. Rev. D{\bf 75}, 032001 (2007).

\bibitem{bellefit}K.~Abe {\it et al.}  (BELLE), [arXiv:hep-ex/0611047].

\bibitem{Abe:2005cv}
  K.~Abe {\it et al.} (BELLE), [arXiv:hep-ex/0508005].


\bibitem{CKM} P.~Urquijo, ``Measurement of the CKM elements $|V_{cb}|$ and  $|V_{ub}|$ from inclusive semileptoic B decays", PhD thesis (2008).

 \bibitem{Uraltsev:2004in}
  N.~Uraltsev, Int.\ J.\ Mod.\ Phys. A {\bf 20}, 2099 (2005).




\bibitem{christoph}  Presented by C. Schwanda,  CKM08,  Rome, Italy (2008).


\bibitem{NNLO_full}
V. Aquila, P. Gambino, G. Ridolfi, and N. Uraltsev, Nucl. Phys. B {\bf 719}, 77 (2005);
A.  Pak and A. Czarnecki, Phys. Rev. D {\bf 78}, 114015 (2008);
M. Dowling et al. Phys. Rev. D {\bf 78}, 074029 (2008). See also K. Melnikov, arXiv:0803.0951v1 [hep-ph].

\bibitem{mel} K. Melnikov, arXiv:0803.0951 [hep-ph] (2008).

\bibitem{beicherlange} T. Beicher {\it et al.}, arXiv:0710.0680 (2007).

\bibitem{dassinfer}
  B.~M.~Dassinger, T.~Mannel and S.~Turczyk,
  JHEP {\bf 0703}, 087 (2007)
  [arXiv:hep-ph/0611168].

\bibitem{Adam:2007pv}
  N.~E.~Adam {\it et al.}  (CLEO),
  Phys.\ Rev.\ Lett.\  {\bf 99}, 041802 (2007)
  [arXiv:hep-ex/0703041].



\bibitem{Aubert:2006px}
  B.~Aubert \etal~(BABAR),
  Phys.\ Rev.\ Lett.\  {\bf 98}, 091801 (2007)
  [arXiv:hep-ex/0612020].

 \bibitem{Hokuue:2006nr}
  T.~Hokuue {\it et al.}  (BELLE),
  Phys.\ Lett.\  B {\bf 648}, 139 (2007)
  [arXiv:hep-ex/0604024].

   \bibitem{Aubert:2006ry}
  B.~Aubert \etal~(BABAR),
  Phys.\ Rev.\ Lett.\  {\bf 97}, 211801 (2006)
  [arXiv:hep-ex/0607089].

\bibitem{Abe:2006gb}
  K.~Abe {\it et al.}  (BELLE),
  [arXiv:hep-ex/0610054].

\bibitem{ISGW2}
N. Isgur, D. Scora, B. Grinstein and M.B. Wise, Phys Rev. D {\bf 39}, 799 (1989).


\bibitem{Ball:2004ye}
  P.~Ball and R.~Zwicky,
  Phys.\ Rev.\  D {\bf 71}, 014015 (2005)
  [arXiv:hep-ph/0406232].

\bibitem{Becirevic:1999kt}
  D.~Becirevic and A.~B.~Kaidalov,
  Phys.\ Lett.\  B {\bf 478}, 417 (2000)
  [arXiv:hep-ph/9904490].

\bibitem{Hill:2005}
  R.~Hill,
  Phys.\ Rev.\  D {\bf 73}, 014012 (2006)
  [arXiv:hep-ph/0505129v2].



\bibitem{Bowler:1999xn}
  K.~C.~Bowler {\it et al.}  [UKQCD Collaboration],
  Phys.\ Lett.\  B {\bf 486}, 111 (2000)
  [arXiv:hep-lat/9911011].

\bibitem{Shigemitsu:2002wh}
  J.~Shigemitsu, S.~Collins, C.~T.~H.~Davies, J.~Hein, R.~R.~Horgan and G.~P.~Lepage,
  Phys.\ Rev.\  D {\bf 66}, 074506 (2002)
  [arXiv:hep-lat/0207011].

\bibitem{Okamoto:2004xg}
  M.~Okamoto {\it et al.},
  Nucl.\ Phys.\ Proc.\ Suppl.\  {\bf 140}, 461 (2005)
  [arXiv:hep-lat/0409116].


\bibitem{Dalgic:2006dt}
  E.~Dalgic, A.~Gray, M.~Wingate, C.~T.~H.~Davies, G.~P.~Lepage and J.~Shigemitsu,
  Phys.\ Rev.\  D {\bf 73}, 074502 (2006)
  [Erratum-ibid.\  D {\bf 75}, 119906 (2007)]
  [arXiv:hep-lat/0601021].

\bibitem{Arnesen:2005ez}
  M.~C.~Arnesen, B.~Grinstein, I.~Z.~Rothstein and I.~W.~Stewart,
  Phys.\ Rev.\ Lett.\  {\bf 95}, 071802 (2005)
  [arXiv:hep-ph/0504209].
%
\bibitem{Flynn:2007rs}
  J.~M.~Flynn and J.~Nieves,
  PoS {\bf LAT2007}, 352 (2007)
  arXiv:0711.3339 [hep-lat].

  \bibitem{Bailey:2008wp}
  J.~Bailey {\it et al.},
  arXiv:0811.3640 [hep-lat].

\bibitem{BaBarff}
 B.~Aubert \etal~(BABAR),
  Phys.\ Rev.\ Lett.\  {\bf 97}, 211801 (2006)
  [arXiv:hep-ex/0607089].




\bibitem{Lange:2005yw}
  B.~O.~Lange, M.~Neubert and G.~Paz,
  Phys.\ Rev.\  D {\bf 72}, 073006 (2005)
  [arXiv:hep-ph/0504071].

\bibitem{Bartelt:1993xh}
  J.~E.~Bartelt {\it et al.}  (CLEO),
  Phys.\ Rev.\ Lett.\  {\bf 71}, 4111 (1993).

\bibitem{Isgur:1992iv}
  N.~Isgur,
  J.\ Phys.\ G {\bf 18}, 1665 (1992).
\bibitem{Bigi:1993ex}
  I.~I.~Y.~Bigi, M.~A.~Shifman, N.~G.~Uraltsev and A.~I.~Vainshtein,
  Int.\ J.\ Mod.\ Phys.\  A {\bf 9}, 2467 (1994)
  [arXiv:hep-ph/9312359].

\bibitem{Neubert:94QBI}
M.~Neubert, Phys. Rev. D {\bf 49}, 3392 (1994) [arXiv:hep-ph/9311325].


\bibitem{cleo:2002} A. Bornheim {\it et al.}  CLEO Collaboration Phys. Rev. Lett. {\bf 88}, 231803 (2002).

\bibitem{Belle:2005} A. Limosani  {\it et al.}  (BELLE),  Phys. Lett. B {\bf 621}, 28 (2005).
\bibitem{BaBar:2006} B.~Aubert \etal~(BABAR),
Phys. Rev. D {\bf 73}, 012006 (2006).


 \bibitem{Bigi:1997dn}
  I.~I.~Y.~Bigi, R.~D.~Dikeman and N.~Uraltsev,
  Eur.\ Phys.\ J.\  C {\bf 4}, 453 (1998)
  [arXiv:hep-ph/9706520].

\bibitem{Bauer:2001yb}
  C.~W.~Bauer, Z.~Ligeti and M.~Luke,
  arXiv:hep-ph/0111387.


  \bibitem{Bauer:2000xf}
  C.~W.~Bauer, Z.~Ligeti and M.~E.~Luke,
  Phys.\ Lett.\  B {\bf 479}, 395 (2000)
  [arXiv:hep-ph/0002161].


  \bibitem{BaBar:improved}
  B.~Aubert \etal~(BABAR),
  Phys.\ Rev.\ Lett.\  {\bf 100}, 171802 (2008)
  arXiv:0708.3702 [hep-ex].



















\bibitem{blnp} B.O. Lange, M. Neubert and G. Paz. Phys. Rev. D {\bf 72}, 073006 (2005).

\bibitem{dge} J.R. Andersen and E. Gardi. JHEP {\bf 0601}, 097 (2006).

\bibitem{Gambino:2007rp}
  P.~Gambino, P.~Giordano, G.~Ossola and N.~Uraltsev,
  JHEP {\bf 0710}, 058 (2007)
  arXiv:0707.2493 [hep-ph].

\bibitem{Aquila:2005hq}
  V.~Aquila, P.~Gambino, G.~Ridolfi and N.~Uraltsev,
  Nucl.\ Phys.\  B {\bf 719}, 77 (2005)
  [arXiv:hep-ph/0503083].

\bibitem{Gardi:2004ia}
  E.~Gardi,
  JHEP {\bf 0404}, 049 (2004)
  [arXiv:hep-ph/0403249].

\bibitem{Aglietti:2006yb}
  U.~Aglietti, G.~Ferrera and G.~Ricciardi,
  Nucl.\ Phys.\  B {\bf 768}, 85 (2007)
  [arXiv:hep-ph/0608047].

\bibitem{Rosner:2006zz}
 J.~L.~Rosner {\it et al.}  (CLEO),
  Phys.\ Rev.\ Lett.\  {\bf 96}, 121801 (2006)
  [arXiv:hep-ex/0601027].

\bibitem{Voloshin:2001}
  M.~B.~Voloshin,
  Phys.\ Lett.\  B {\bf 515}, 74 (2001)
  [arXiv:hep-ph/0106040].


\bibitem{Adam:2006nu}
  N.~E.~Adam {\it et al.}  (CLEO),
  Phys.\ Rev.\ Lett.\  {\bf 97}, 251801 (2006)
  [arXiv:hep-ex/0604044].
\bibitem{ds-semil}
J. Yelton {\it et al.}~(CLEO), [arXiv:0903.0601].



\bibitem{Steinhauser:2008pm}
  M.~Steinhauser,
  arXiv:0809.1925 [hep-ph].


\bibitem{Gray:2005ad}
  A.~Gray {\it et al.}  [HPQCD Collaboration],
  Phys.\ Rev.\ Lett.\  {\bf 95}, 212001 (2005)
  [arXiv:hep-lat/0507015].
  \bibitem{Guo:2006nt}
  X.~H.~Guo and M.~H.~Weng,
  Eur.\ Phys.\ J.\  C {\bf 50}, 63 (2007)
  [arXiv:hep-ph/0611301].

\bibitem{Bernard:2007zz}
  C.~Bernard {\it et al.}  [Fermilab Lattice, MILC and HPQCD Collaborations],
  PoS {\bf LAT2007}, 370 (2007).
\bibitem{belle:taunuhad}  K. Ikado {\it et al.}  (BELLE),  Phys. Rev. Lett. {\bf 97}, 251802 (2006).

\bibitem{belle:taunulep} I. Adachi  (\etal) (BELLE),  ArXiv:0809.3834~.

\bibitem{babar:taunuhad}  B.~Aubert \etal~(BABAR),
 Phys. Rev. D {\bf 77}, 011107 (2008).

\bibitem{:2008gx}
   {\it et al.}  (BABAR),
  arXiv:0809.4027 [hep-ex].





  \bibitem{Trine:2008qv}
  S.~Trine,
  arXiv:0810.3633 [hep-ph].


\bibitem{new_physi_bsg} Flavour in the era of the LHC CERN, 15-17 May, 2006, Yellow Reports (CERN-2007-004)  and reference therein.

\bibitem{BBL} G. Buchalla, A.J. Buras, M. E. Lautenbacher, Rev. Mod. Phys. {\bf 68}, 1125 (1996).

\bibitem{cpasy} D.Atwood, M. Gronau, A. Soni,  Phys. Rev. Lett. {\bf 79}, 185 (1997);
B. Grinstein, Y. Grossman, Z. Ligeti, D. Pirjol Phys. Rev. D {\bf 71}, 011504 (2005).

\bibitem{Misiak} M. Misiak et al.,  Phys. Rev. Lett. {\bf 98}, 022002 (2007).

\bibitem{matt} T. Becher, M. Neubert, Phys. Rev. Lett. {\bf 98}, 022003 (2007).

\bibitem{LNP} S. J. Lee, M. Neubert and G. Paz, Phys. Rev. D {\bf 75}, 114005 (2007) [arXiv:hep-ph/0609224].

\bibitem{tony} A. Limosani, in a private communication performed the average for this article.

\bibitem{misiaktype2}
  P.~Gambino and M.~Misiak,
  Nucl.\ Phys.\  B {\bf 611}, 338 (2001)
  [arXiv:hep-ph/0104034].



 \bibitem{cleoi} S. Chen {\it et al.} (CLEO), Phys. Rev. Lett. {\bf 87}, 251807 (2001).

 \bibitem{belles} K. Abe  {\it et al.}  (BELLE), Phys. Lett. B {\bf 511}, 151 (2001).

 \bibitem{babars}  B.~Aubert \etal~(BABAR),
 Phys. Rev. D  {\bf 72}, 052004 (2005).

\bibitem{babari}  B.~Aubert \etal~(BABAR),
Phys. Rev. Lett. {\bf 97}, 171803 (2006).

\bibitem{babarf}  B.~Aubert \etal~(BABAR),
 Phys. Rev. D {\bf 77}, 051103 (2008).

\bibitem{bellei} A. Limosani  {\it et al.}  (BELLE),  presented at Moriond E.W. (2008).


\bibitem{thurt} T. Hurt, E. Lunghi, W. Porod, Nucl. Phys. B {\bf 704}, 56 (2005).



\bibitem{pol} M. Gronau, Y. Grossman, D. Pirjol, A. Ryd,  Phys. Lett. {\bf 88}, 051802 (2002);
M. Gronau and D. Pirjol Phys. Rev. D {\bf 66}, 054008 (2002).

\bibitem{exppol} B. Aubert {\it et al.}, (BABAR), Phys. Rev. D {\bf 72}, 051103 (2004);
M. Nakao \etal~(BELLE), Phys. Rev. D {\bf 69}, 112001 (2004).


\bibitem{311}
 B.~Aubert \etal~(BABAR),
Phys.\ Rev.\  D  {\bf 70}, 112006 (2004).

\bibitem{46}
M. Nakao  {\it et al.}  (BELLE),   
Phys.\ Rev.\  D  {\bf 69}, 112001 (2004).

\bibitem{72} M. Beneke et al., JHEP {\bf 0506}, 071 (2005).


  \bibitem{2}
 B.~Aubert \etal~(BABAR),
arXiv:0805.4796 (submitted to PRL).  

\bibitem{51}
S. Nishida  \etal (BELLE),    
Phys.\ Rev.\ Lett. {\bf 93}, 031803 (2004).


 \bibitem{74}
T. E. Coan \etal~(CLEO),  
Phys.\ Rev.\ Lett. {\bf 86}, 5661 (2001).









\bibitem{babarmm}
 B.~Aubert \etal~(BABAR),
Babar Collaboration Phys. Rev. D {bf 77}, 032007 (2008).

\bibitem{cdfmm} T. Aaltonen {\it et al} (CDF), Phys. Rev. Lett. {\bf 100}, 101802 (2008).

\bibitem{d0mms}    V. Abazov et al . (D0),  Phys. Rev. D {\bf 76}, 092001 (2007).

\bibitem{babaree}  B.~Aubert \etal~(BABAR),
 Phys. Rev. D {\bf 77}, 032007 (2008).

 \bibitem{belleee}  M.C. Chang et al BELLE collaboration Phys. Rev. D {\bf 68}, 111101 (2003).

\bibitem{cdee}
R. F. Harr, (CDF), ``Rare Decays of $B$ and $D$ Hadrons at CDF,"
arXiv:0810.3444 [hep-ex].





%
\bibitem{3}
 B.~Aubert \etal~(BABAR),
 Phys.\ Rev.\  D  {\bf 73}, 092001 (2006).  

\bibitem{31}
I. Adachi, {\it et al.}  (Belle),
 arXiv:0810.0335 [hep-ex].

\bibitem{65}
A. Aaltonen \etal~(CDF), arXiv:0804.3908, (2008). 

\bibitem{bellefisrt} M. Iwasaki \etal~(BELLE, Phys. Rev. D {\bf 72}, 092005 (2005).


\bibitem{second}   B.~Aubert \etal~(BABAR),
Phys. Rev. Lett. {\bf 93}, 081802 (2004).

 \bibitem{alilungji} A. Ali, E. Lunghi, C. Greub, M. Walker, Phys. Lett. B {\bf 66}, 034002 (2002).


\bibitem{12}
 B.~Aubert \etal~(BABAR),
 Phys.\ Rev.\  D  {\bf 73}, 092001 (2006).  


\bibitem{14}
 B.~Aubert \etal~(BABAR),
  Phys.\ Rev.\ Lett. {\bf 93}, 081802 (2004).  



 \bibitem{a910}
  A.~Ishikawa \etal,
  Phys.\ Rev.\ Lett.\  {\bf 96}, 251801 (2006)
  [arXiv:hep-ex/0603018].






\bibitem{UTfit-CKM}
M. Pierini,``Update on the unitarity triangle (UTFit)," presented at 34th Int. Conf. on High Energy Physics (ICHEP 2008), Philadelphia, PA, July 2008.

\bibitem{MB}
E. Blucher and W. B. Marciano, ``$V_{ud}$, $V_{us}$, THE CABIBBO ANGLE,
AND CKM UNITARITY,"  in the Particle Data Group reviews \cite{PDG}.

\bibitem{NMFV}
This type of analysis was done with somewhat different assumptions earlier by
K. Agashe, M. Papucci, G. Perez and D. Pirjol [hep-ph/0509117].

\bibitem{ASL}
Measurement of $A_{SL}$ for $B_s$ mesons are given by
V. Abazov et al. (D0), Phys. Rev. Lett. {\bf 98}, 151801 (2007); CDF Note 9015 07-08-16 available at
\url{http://www-cdf.fnal.gov/physics/new/bottom/bottom.html}.
See also S. Laplace, Z. Ligeti, Y. Nir, and G. Perez, Phys. Rev. D {\bf 65}, 094040, (2002) [hep-ph/0202010] and references contained therein.

\bibitem{LenzN}
There are considerable theoretical uncertainties,
see A. Lenz and U. Nierste, JHEP {\bf 06}, 72 (2007) [hep-ph/0612167].



\end{thebibliography}
\end{document}